\documentclass[aps,prc,onecolumn,groupedaddress]{revtex4}
\pdfoutput=1
\usepackage{epsfig}
\usepackage{xcolor}

\begin{document}

\title{Instanton - motivated study of spontaneous fission 
 of odd-A nuclei }
\author{W.~Brodzi\'nski, J.~Skalski}

\affiliation{National Centre for Nuclear Research, Pasteura 7,
PL-02-093 Warsaw, Poland}

\date{\today}

\begin{abstract}
{\noindent Using the idea of the instanton approach to quantum tunneling 
 we try to obtain a method of calculating spontaneous fission rates for  
 nuclei with the odd number of neutrons or protons. This problem  
 has its origin in the failure of the adiabatic cranking approximation 
 which serves as the basis in calculations of fission probabilities. 
 Selfconsistent instanton equations, with and without pairing,  
 are reviewed and then simplified to non-selfconsistent versions with 
 phenomenological single-particle potential and seniority pairing interaction. 
 Solutions of instanton-like equations without pairing 
 and actions they produce are studied for the Woods-Saxon 
 potential along realistic fission trajectories. Actions for unpaired 
 particles are combined with cranking actions for even-even cores and 
 fission hindrance for odd-$A$ nuclei is studied in such a hybrid model. 
 With the assumed equal mass parameters for neighbouring odd-A and even-even 
 nuclei, the model shows that freezing the $K^{\pi}$ configuration 
 leads to a large overestimate of the fission hindrance factors. Actions 
 with adiabatic configurations mostly show not enough hindrance; instanton-like
 actions for blocked nucleons correct this, but not sufficiently. 
  }
\end{abstract}
\maketitle

\section{Introduction}

  Nuclear fission is thought to be a collective process, 
 classically envisioned in analogy 
  to fragmentation of a liquid drop. In reactions induced by neutrons and 
  light or heavy ions, fission is one of many possible deexcitation channels 
   of a formed compound nucleus. 
  On the other hand, spontaneous fission is a decay of the nuclear ground state 
 (g.s.) which exhibits its meta-stability and involves quantum  
 tunneling through a potential barrier. In a theoretical approach, the fission 
 barrier follows from a model of the shape-dependent nuclear energy. In 
 practical terms, it is calculated either from a selfconsistent mean-field 
 functional or a microscopic-macroscopic model, as a landscape formed by the 
 lowest energies $E({\bf q})$ at fixed values of a few arbitrarily chosen 
 coordinates ${\bf q}=(q_1,...,q_i,...)$ (for simplicity assumend 
 dimensionless) describing nuclear shape. 
  The obscure part of the current approach relates to a) the likely 
  insufficiency of included coordinates and b) a description of tunneling 
  dynamics, essentially shaped after the Gamow method, but without a clear 
  understanding of mass parameters and conjugate momenta entering the formula 
  for decay rate.  

 The experimentally well established presence of pairing correlations in nuclei
 gives rationale for using cranking \cite{Inglis,Funny} or adiabatic 
 Time-Dependent Hartree-Fock(-Bogolyubov) - ATDHF(B) - approximation 
 \cite{BV,GQ,DS} in the 
 description of fission in even-even (e-e) nuclei. Indeed, as the lowest 
 two-quasiparticle excitation in such nuclei has  energy of at least 
 twice the pairing gap $2\Delta$, which in heavy nuclei amounts to more than 
 1 MeV,  
 one can, for collective velocities $|\hbar {\dot {\bf q}}|$ 
 reasonably smaller than that, solve the time-dependent Schr\"odinger (or 
 mean-field) equation to the first order in ${\dot {\bf q}}$ 
 and obtain kinetic energy of shape changes: 
 $\frac{1}{2}\sum_{i j} B_{q_i q_j}({\bf q}) {\dot q}_i {\dot q}_j$, with 
 cranking (or ATDHFB) mass parameters $B_{q_i q_j}({\bf q})$. 
 Then one can apply the Jacobi variational principle to the imaginary 
 under-the-barrier motion in order to find the quasiclassical tunneling path 
 ${\bf q}(\tau)$ by minimizing action: 
\begin{equation}
S\left[{\bf q}(\tau)\right]=\int_{{\bf q}_{ini}}^{{\bf q}_{fin}} 
 \sum_{i}p_{i}dq_{i}=\int_{q_{ini}}^{q_{fin}}
 \sqrt{2B_{q q}({\bf q}(\tau))[E({\bf q}(\tau))-E_0]}\,dq . 
\label{scrank}
\end{equation}
 Here, $p_i=\sum_j B_{q_i q_j}({\bf q})\dot{q}_j$ are the conjugate momenta; 
 $q$ (without index) is an effective coordinate along a path, usually the one 
 of $q_i$ that controls elongation of the nucleus;
 $B_{q q}=\sum_{k l}B_{q_k q_l}\frac{dq_{k}}{dq}\frac{dq_{l}}{dq}$ 
 is the effective mass parametr along the fission path with respect to $q$.  
 The Jacobi principle requires that a) 
 ${\bf q}_{ini}$ and ${\bf q}_{fin}$ - the initial and final points of the path 
 through a barrier - be {\it fixed} for all tunneling paths and b) 
 on each trial path, $E({\bf q})-\frac{1}{2}\sum_{i j} B_{q_i q_j}({\bf q}) 
 {\dot q}_i {\dot q}_j$  (the potential minus kinetic energy) be constant and 
 equal to $E_0=E({\bf q}_{ini})= E({\bf q}_{fin})$, usually chosen as 
 $E_{g.s.}+E_{zp}$ - the g.s. energy augmented by the zero-point energy of 
  oscillations around the g.s. minimum in direction of fission, 
   $E_{zp}=\frac{1}{2} \hbar\omega_0$.  
 The spontaneous fission rate is given to the leading order by: 
 $(\frac{\omega_0}{2\pi}) e^{-2S_{min}/\hbar}$, with $S_{min}$ - the minimal 
 action. By the first equality in (\ref{scrank}), $S$ equals the 
 integral of twice the collective kinetic energy, $B_{q q} {\dot q}^2$,  
 with $(\hbar {\dot q})^2=\frac{2[E(q)-E_0]}{B_{q q}}$, over the time of 
 passing the barrier.
 Estimating {\it a posteriori} collective velocities of the fictitious 
 under-barrier motion for heavy nuclei, with typical cranking mass parameter 
 for the Woods-Saxon potential, $B_{q q}\gtrsim 200 \hbar^2$/MeV, and the 
 fission barrier $\lesssim 7$ MeV, one obtains $\hbar{\dot q}\lesssim 
 0.25$ MeV, so the error of the cranking approximation might be believed 
 moderate. 
  
 Situation changes rather dramatically for odd-$Z$ or/and odd-$N$ nuclei. 
 For odd number of particles, their contribution to the cranking 
 mass parameter $B_{q_i q_j}$, derived {\it as if the adiabatic approximation 
 were legitimate}, reads:  
\begin{eqnarray}
\label{cranking}
  B_{q_{i} q_{j}} & = & 2\hbar^{2}
 \Bigg[\sum_{\mu,\nu\neq\nu_{0}}
 \frac{\langle \mu\mid\frac{\partial {\hat h}}{\partial q_{i}}\mid \nu\rangle 
 \langle \nu\mid\frac{\partial {\hat h}}{\partial q_{j}}\mid\mu\rangle}
 {\left(E_{\mu}+E_{\nu}\right)^3}
 \left(u_{\mu}v_{\nu}+u_{\nu}v_{\mu}\right)^2  \\   \nonumber 
 & + &\frac{1}{8}\sum_{\nu\neq\nu_{0}}\frac{\left(\tilde{\epsilon}_{\nu}
  \frac{\partial \Delta}{\partial q_{i}} -\Delta \frac{\partial 
 \tilde{\epsilon}_{\nu}}{\partial q_{i}}\right) 
\left(\tilde{\epsilon}_{\nu}\frac{\partial \Delta}{\partial q_{j}}-
 \Delta \frac{\partial \tilde{\epsilon}_{\nu}}{\partial q_{j}}\right)}
 {E_{\nu}^5}\Bigg]   \\  \nonumber 
& + & 2\hbar^{2} \sum_{\nu\neq\nu_{0}}\frac{\langle \nu\mid
 \frac{\partial {\hat h}}{\partial q_{i}}\mid\nu_{0}\rangle \langle \nu_{0}\mid
\frac{\partial {\hat h}}{\partial q_{j}}\mid\nu\rangle}
{\left(E_{\nu}-E_{\nu_{0}}\right)^{3}}
\left(u_{\nu}u_{\nu_{0}}-v_{\nu}v_{\nu_{0}}\right)^{2} . 
 \\  \nonumber 
\end{eqnarray}
 Here, the odd nucleon occupies the orbital $\nu_0$ in the g.s.; 
 ${\hat h}$ is the mean-field single - particle (s.p.) Hamiltonian, 
 $\epsilon_{\mu}$ are its eigenenergies, 
 $\tilde{\epsilon}_{\nu}=\epsilon_{\nu}-\lambda$, 
 $E_{\mu}=\sqrt{\tilde{\epsilon}_{\mu}^2+\Delta^2}$, $u$ and $v$ are 
 the usual BCS amplitudes. A common pairing gap $\Delta$ and Fermi energy 
 $\lambda$ were assumed for the g.s. 
 and its two-quasiparticle excitations: those with the odd particle in the 
 state $\nu_0$ which give contribution in the square bracket that has the 
 same form as the mass parameter for an e-e nucleus, and those with 
 the odd particle in the state $\nu\neq\nu_0$ and the orbital $\nu_0$ paired,  
 given by the last term of the formula. The latter 
  becomes nearly singular, $\sim (E_{\nu_0}-E_{\nu})^{-3}$, at close avoided 
 level crossings where $E_{\nu_0}-E_{\nu}$ can be of the order of keV or less. 
 This invalidates the very assumption 
 underlying the cranking formula, except for ridiculously small collective 
 velocities. But there is still another deficiency: a departure from the 
 symmetry preserved on a part of the fission trajectory often produces a 
 negative contribution to the inertia parameter whose magnitude would depend 
 on the proximity of the relevant crossing of levels of different 
 symmetry classes. 
 Although  some calculations of fission half-lives for odd nuclei with the 
 cranking mass parameters (\ref{cranking}) were done in the past, 
 e.g. \cite{Lojew}, the above-mentioned problems make the precise minimization 
 of action (\ref{scrank}) for those nuclei both questionable and practically 
 very difficult - a good illustration of near-singular cranking mass 
 parameter [calculated with a formula more refined than (\ref{cranking})] in 
 the odd nucleus is provided in \cite{Mirea2019} 
 (the middle panel of Fig. 4 there) \cite{foot1}. 

  The well known experimental evidence, reviewed recently in \cite{Hess}, 
 shows that the spontaneous fission rates of odd nuclei are three 
 to five orders of magnitude smaller than those of their e-e neighbours. 
 Although the  explanation usually invokes the specialization energy - an 
 increase in the fission barrier by the blocking of one level by a single 
 nucleon - a quantitative understanding is lacking at present. In particular,
 the combination of axial symmetry of the nuclear deformation and very 
 different densities of s.p. levels with low- and high-$\Omega$ quantum numbers 
 ($\Omega$ being the projection of the s.p. angular momentum on the symmetry 
 axis of a nucleus) could suggest a higher specialization energy, and thus 
 smaller fission rate, for configurations based on high-$\Omega$ orbitals, but 
 the data \cite{Hess} contradict this. 
 
  While estimates of fission half-lives rely on the assumption of 
 nearly adiabatic motion, doubtful for odd-$A$ nuclei, the real-time solutions 
 of Schr\"odinger-like dynamics are regular for any velocity profile 
 ${\dot {\bf q}}$ and any avoided crossings. In general, they lead to a 
 population of levels above the Fermi energy. Analogous possibility must exist 
 in the fictitious imaginary-time motion, pertinent to quantum tunneling.  
 In this light, a consideration of non-adiabatic tunneling - with fission paths
 formed at least in part by non-adiabatic configurations - presents itself 
 as an interesting subject. 
 Beyond-cranking effects could provide corrections to the standard cranking 
 spontaneous fission rates in e-e nuclei and can be crucial   
 for spontaneous fission of odd-$A$ nuclei and high-$K$ isomers 

 In this paper, we present an attempt towards replacing the adiabatic
 cranking approximation by a scheme including non-adiabatic 
 fission paths, motivated by the instanton method 
 \cite{Coleman,LNP,Neg1,PudNeg,Neg2}. 
 Instantons are solutions with the infinite period to time-dependent mean-field 
 equations in imaginary time $\tau=it$, with the nuclear g.s. wave 
 function as the boundary value. They arise from the saddle-point approximation
  to the path integral representation of the propagator and give the leading 
 contribution to spontaneous fission rate of the form: 
 $A_{inst}\exp(-S_{inst}/\hbar)$. Here, $S_{inst}$ - instanton action, is the 
 counterpart of $2 S\left[{\bf q}(\tau)\right]$ in (\ref{scrank}), while  
 the prefactor $A_{inst}$ -  the ratio of determinants including frequencies 
 of quadratic fluctuations around the instanton and the g.s. - for review see
 e.g.  \cite{ChT,Ander,tunnsplit} -  will not be considered it in the following.
 The instanton with the smallest action (there can be more than one as the 
 instanton equation determines local minima of action) 
 gives fission half-life {\it without the necessity of defining mass 
 parameters}. 
 The resulting fission path involves all degrees of freedom of the 
 mean-field state, not only shape parameters.  

 The difficulty in solving for a selfconsistent instanton including pairing 
 is beyond that of solving real-time TDHFB equations: the generically 
 exponential $\tau$-dependence of the HFB $Z$ matrix \cite{Ring}, introducing 
 components differing by orders of magnitude, has to be found from equations 
 {\it non-local in} $\tau$ (see Sect. \ref{sec:metinst3}). 
 Here, we treat the selfconsistent theory as a motivation, and solve 
 imaginary-time-dependent Schr\"odinger equation (iTDSE) with the 
 phenomenological Woods-Saxon (W-S) potential to calculate action along various 
 chosen paths. We use  
 micro-macro energy for $E({\bf q})$. Since we reject cranking mass 
 parameters for odd-$A$ nuclei, we have to provide ${\dot {\bf q}}$ without 
 them. To this aim we use cranking mass parameters of the neighbouring e-e 
 nucleus. With this prescription, we can calculate manifestly 
 beyond-cranking actions and study their behaviour. 
 Although we formulate equations with pairing, in the present paper we 
 present iTDSE instanton-like solutions without it. To the best of our 
 knowledge, such solutions and their actions are discussed for the first time. 
 Then, we combine instanton-like solutions for the odd nucleon with the 
 cranking action with pairing for the e-e core in a hybrid model to study 
 fission hindrance in odd-$A$ nuclei. Within this model we calculate and 
 compare fission half-lives obtained with and without constraining the 
 $\Omega^{\pi}$ (with $\pi$ - parity) g.s. configuration. 
   
 The presented approach cannot be as yet a basis for the systematic 
 minimization of action over fission paths. Moreover, it differs from the 
 instanton method by 
 ignoring the anti-hermitean part of the imaginary-time mean-field. We think, 
 however, that it presents some features of the instanton method and may be 
 useful for developing either a more refined non-selfconsistent method  
 or ways to implement the selfconsistent instanton treatment of 
 spontaneous fission half-lives, including odd-$A$ nuclei and 
 high-$K$ isomers.

  The paper is organized as follows: in sect. II we briefly describe 
 the instanton formalism with and without pairing, specifying a simplification 
 of each of them
 to a non-selfconsistent version with the phenomenological s.p. potential. 
 To provide an illustration of imaginary-time solutions, in sect. III we 
  discuss the two-level model, in particular the dependence of 
 action on the interaction between levels and the collective velocity. 
 Properties of solutions and actions obtained from the iTDSE with the 
 realistic W-S potential are described in sect. IV, including 
 an example of the action calculation along the path through non-axial 
 deformations. Sect. V contains a study of the fission hindrance in odd nuclei 
 made within a hybrid model utilizing adiabatic cranking action for the 
 e-e core and the iTDSE action without pairing for the odd nucleon. 
 This approach is meant to mimic a model with pairing which we have not solved 
 yet. As a byproduct, we study the effect of freezing the configuration along 
 the path of axially-symmetric deformations on the fission rate. 
 This is done under the assumption that the collective velocity along a given 
 path in odd-$A$ nucleus is as if it had the mass parameter of the e-e 
 neighbour; stated otherwise, the difference in ${\dot q}$ between the odd-$A$ 
 nucleus and its e-e $A-1$ neighbour comes solely from their different fission 
 barriers. Summary and conclusions are given 
 in sect. VI. In appendices we derive expressions for the Floquet exponent and 
 action for periodic solutions within the cranking approximation (Appendix A), 
 describe the method of solution of the iTDSE (Appendix B), tests of the 
 reliability of the calculated actions (Appendix C) and the problem of 
 calculating action along paths through non-axial shapes (Appendix D).

  \section{Instanton-motivated approach} 



 The instanton approach to nuclear fission was formulated in the mean-field 
 setting in \cite{LNP,Neg1,JS1,JS,JS2}. After reviewing the selfconsistent 
 formulation without pairing in Subsect. A, in Subsect. B, we formulate the 
 non-selfconsistent version with the phenomenological nuclear potential, the 
 solutions to which we present in this work. 
 For completeness, as the pairing interaction is crucial to nuclear 
 fission, we review also the selfconsistent equations with pairing in 
 Subsect. C, and formulate the model with the phenomenological potential   
 and the monopole pairing with the selfconsistent pairing gap in 
 Subsect. D. 

  \subsection{Instantons of Hartree-Fock equations \label{sec:metinst1}}    

 A transition to imaginary time, $t\rightarrow -i\tau$, transforms 
 TDHF equations for s.p. amplitudes $\psi_k(t)$ into imaginary-TDHF (iTDHF) 
 equations for amplitudes $\phi_k(x,\tau) = \psi_k(x,-i\tau)$, 
 with the complex-conjugate amplitudes $\psi^*_k(t)$ becoming 
 $\psi^{*}_k(x,-i\tau)=\phi^{*}_k(x,-\tau)$, so that the scalar products 
 $\langle\psi_k(t)|\psi_l(t)\rangle$ transform to 
 $\langle\phi_k(-\tau)|\phi_l(\tau)\rangle$.   
 Mean-field solutions dominating the quasiclassical tunneling rate are periodic 
 \cite{LNP,Neg1}, hence the iTDHF equations acquire the additional terms 
 $\zeta_{k}\phi_k$, with $\zeta_{k}$ - Floquet exponents with the dimension 
 of energy, which ensure periodicity: 
\begin{equation}
\hbar \frac{\partial \phi_{k}({\tau})}{\partial\tau}=-(\hat{h}(\tau)-
 \zeta_{k})\phi_{k}(\tau).
\label{eq:itdhf}
\end{equation}
  The mean-field hamiltonian 
 $\hat{h}(\tau)=\hat{h}[\phi^{*}(-\tau),\phi(\tau)]$ is defined by: 
 $\hat{h}(\tau)\phi_{k}(\tau)=\delta\mathcal{H}/\delta\phi^{*}_{k}(-\tau)$, 
 where $\mathcal{H}(\tau)$ is the energy overlap 
 $\langle\Phi(-\tau)|\hat{H}|\Phi(\tau)\rangle$, playing the same role as 
 energy in the usual TDHF,  
\begin{equation}
\mathcal{H}(\tau)=\mathcal{H}[\phi^{*}(-\tau),\phi(\tau)]=
 \int d^3x \left\{\sum_{k\ occ} \frac{\hbar^2}{2m}
\nabla\phi_k^*(-\tau)\nabla\phi_k(\tau)+\mathcal{V}[\phi^{*}(-\tau),\phi(\tau)]
\right\}, 
\label{eq:tdhfener}
\end{equation}
 with $\mid\Phi(\tau)\rangle$ - the Slater determinant built of occupied 
 orbitals $\{\phi_{k}(\tau)\}$, and $\mathcal{V}$ - a two-body interaction 
 energy density composed as in the HF, but with $\phi_k(\tau)$ in place of 
 $\psi_k(t)$, and $\phi^*_k(-\tau)$ in place of $\psi^*_k(t)$. 
  The instanton solving (\ref{eq:itdhf}) that describes quantum tunneling,  
   called bounce, has to fulfil specific consditions: 
  amplitudes at the boundary are equal to static Hartree-Fock (HF) solutions 
 at the metastable state (m.s.) minimum, 
 $\phi_k(-T/2)=\phi_k(T/2)=\psi_k^{HF}$, with HF energy $E_{m.s.}$, while 
  the states $\phi_k(\tau=0)$ form a normalized Hartree-Fock state with 
 the same energy $E_{m.s.}$ at the outer slope of the barrier, that 
 corresponds to the exit point from the barrier ${\bf q}_{fin}$ in 
 Eq. (\ref{scrank}). 
 An infinite period $T$ corresponds to a decay from the m.s. - 
 evolution becomes infinitely slow close to the m.s. minimum. 
 Hence, $\partial\phi_k/\partial\tau$ become zero as 
 $\tau\rightarrow\pm\infty$, and Eq. (\ref{eq:itdhf}) reduce there to the 
 static HF equations. So, in the selfconsitent theory, the Floquet exponents 
 are equal to s.p. energies at the m.s. state. 

 Both, energy overlaps $\mathcal{H}(\tau)$ and the mean-field Hamiltonian 
 $\hat{h}(\tau)$, depend on $\phi_k(\tau)$ and $\phi_k(-\tau)$, so Eq. 
 (\ref{eq:itdhf}) are {\it nonlocal in} $\tau$ and one cannot solve them as an 
 initial value problem.
 Together with the periodicity condition, this makes iTDHF equations a kind of 
 a nonlinear boundary value problem in four dimensions. 

 Eq. (\ref{eq:itdhf}) conserve energy overlap $\mathcal{H}(\tau)$, diagonal 
 overlaps of solutions, and give the exponential $\tau$-dependence to their 
 non-diagonal overlaps. As the HF solutions at the boundary are orthonormal, 
 so remain the bounce solutions: 
\begin{equation}
\langle\phi_{i}(-\tau)|\phi_{j}(\tau)\rangle = \delta_{ij}.
\label{eq:ortonorm}
\end{equation} 

 From $\hat{H}^{\dagger}=\hat{H}$, one has 
 $\mathcal{H}(-\tau)=\mathcal{H}^*(\tau)$, and 
 the mean field hamiltonian $\hat{h}(\tau)$ is in general not hermitean, but 
 fulfils the condition: $\hat{h}(-\tau)=\hat{h}^{\dagger}(\tau)$. It may be 
 presented as a sum of its hermitean and antihermitean parts,  
 $\hat{h}(\tau)=\hat{h}_R(\tau)+\hat{h}_A(\tau)$, with: 
$\hat{h}_R(-\tau)=\hat{h}_R(\tau)=\hat{h}_R^{\dagger}(\tau)$ and 
 $\hat{h}_A(-\tau)=-\hat{h}_A(\tau)=\hat{h}_A^{\dagger}(\tau)$;  
 the $\tau$-odd, antihermitian part $\hat{h}_A$ comes from  
$\tau$-odd parts of densities building energy overlap $\mathcal{H}(\tau)$. 
 In tunneling, at least one  $\tau$-odd density is provided by the current 
 density $\mathbf{j}$, in imaginary time given by:
 $\mathbf{j}(\tau)=\sum_k[\phi_k(\tau)\nabla\phi_k^*(-\tau)-
 \phi_k^*(-\tau)\nabla\phi_k(\tau)]/2$, \cite{JS}, fulfiling: 
$\mathbf{j}(-\tau)=-\mathbf{j}^*(\tau)$. 
 Decomposing amplitudes 
 into $\tau$-even and $\tau$-odd parts, $\phi_k(\tau)=\varphi_k(\tau)-
\xi_k(\tau)$, $\phi_k(-\tau)=\varphi_k(\tau)+\xi_k(\tau)$, one has:   
  \begin{equation}
  {\bf j} = \sum_{k\ occ}
  \left[\Re(\varphi^*_k\nabla \xi_k-\xi^*_k\nabla \varphi_k)+
   i\Im(\xi^*_k\nabla \xi_k-\varphi^*_k\nabla \varphi_k)\right] . 
  \end{equation}
 One can see that, even if $\phi_k$ are purely real, 
 the $\tau$-odd components $\xi_k$ in the first part of this 
 expression generate the $\tau$-odd antihermitean mean field ${\hat h}_A$. 
 For small collective velocities, 
 the $\tau$-odd mean field ${\hat h}_A$ is a direct analogy in the 
 imaginary-time formalism of the Thouless-Valatin potential of the 
 ATDHF method in real time \cite{ThouVal}.

 After finding iTDHF solutions one can calculate action. Since in the 
 mean-field theory with a Slater determinant $\Psi(t)$,   
 $\langle \Psi(t)\mid i \hbar \partial_t - {\hat H} \mid \Psi(t)\rangle$ 
 plays a role of Lagrangian, action $\int dt
 \langle \Psi(t)\mid i \hbar \partial_t \mid \Psi(t)\rangle$ 
 in the imaginary-time version becomes \cite{LNP,Neg1}: 
\begin{equation}
S=\hbar\int_{-T/2}^{T/2}d\tau \sum \limits_{i=1}^{N} \left\langle 
\phi_{i}(-\tau)\big|\partial_{\tau} \phi_{i}(\tau)\right\rangle = 
\int_{-T/2}^{T/2}d\tau \sum \limits_{i=1}^{N} \left\langle 
\phi_{i}(-\tau)\big| \zeta_{i}-\hat{h}(\tau) \big| 
\phi_{i}(\tau)\right\rangle ,
\label{eq:sactgen}
\end{equation}
 where the summation runs over the occupied s.p. states. 

  Contrary to the unfortunate and erroneous statement in \cite{JS} [in the 
 paragraph containing the formula (14) there], repeated in \cite{JS2}  
 [after the formula (7) there], this expression is obviously composed 
 of changes in $\phi_i(\tau)$ {\it parallel} to $\phi_i(-\tau)$.

  \subsection{Non-selfconsistent instanton-motivated approach 
 \label{sec:metinst2}}


 In order to gain some idea about solutions of imaginary-time-dependent 
 Schr\"odinger-like equations with instanton boundary conditions and 
 resulting actions we replace the 
 mean-field hamiltonian ${\hat h}[\phi^*(-\tau),\phi(\tau)]$ by a simple one 
 with the phenomenological W-S s.p. potential. Releasing the 
 selfconsistency makes these equations linear iTDSEs and removes 
 non-locality in $\tau$, thus considerably simplifying solution. Certainly,
 we lose generality: the non-hermitean nature of the mean potential in 
 tunneling is lost, we have to resort 
 to the usual paramerization of nuclear shapes and have to externally 
  provide the collective velocity ${\dot q}(\tau)$ which in the selfconsistent 
 theory would follow from the energy constraint $\mathcal{H}(\tau)=E_{m.s.}$.
 However, we gain a possibility to study iTDSE solutions and their 
 actions for {\it manifestly non-adiabatic imaginary-time motions} along 
  trial fission paths which in current treatments of fission are commonly 
  considered realistic. 
 To have an approximate energy conservation we assume the effective 
 collective velocity 
 given by:  
 \begin{equation}
   \label{eq:qdotcoll}
    B^{even}_{q q} (q) {\dot q}^2 = 2 (E(q)- E_{m.s.})  , 
 \end{equation}
 with: 
 \begin{equation}
 d\tau= \frac{dq}{{\dot q}(\tau)}  . 
 \end{equation}
 Here, $E(q)$ is the microscopic-macroscopic energy and $B^{even}_{q q}(q)$ is 
 the adiabatic mass parameter along the fission path of the 
 {\it even - even nucleus} - the one 
 in question or the nearest neighbour in case of the odd-$A$. 
 The motivation will be given in section \ref{sec:calcHF-hybrid}.
 This whole procedure may be viewed as an attempt to simplify the 
 selfconsistent theory to a micro-macro version. 
 
 As a result, the phenomenological s.p. Hamiltonian ${\hat h}(\tau)$ is: 
\begin{equation}
 {\hat h}(q(\tau))= -\frac{\hbar^2}{2m}\nabla^2+V(q(\tau)) , 
\label{VWS}
\end{equation}
 where $V$ is the phenomenological s.p. potential, including Coulomb 
 repulsion for protons, depending on the collective coordinate $q$ which itself 
 depends on $\tau$. 
 In solving the equation (\ref{eq:itdhf}) with the above s.p. hamiltonian 
 along a given path we restrict to the subspace spanned by ${\cal N}$ 
 adiabatic s.p. orbitals 
 $\psi_{\mu}(q)$. 
  In this subspace, there are ${\cal N}$ bounce solutions 
 $\phi_i(\tau)$, 
 each of which tends to the s.p. orbital $\psi_i(q_{min})$ at the metastable 
 minimum as $T\rightarrow \pm \infty$. By expanding these solutions 
 onto adiabatic orbitals 
 \begin{equation} 
 \label{expan1}
  \phi_i(\tau)=\sum_{\mu} C_{\mu i}(\tau) \psi_{\mu}(q(\tau))  ,
 \end{equation} 
  we obtain the following set of equations for the square matrix of 
 the coefficients $C_{\mu i}(\tau)$: 
 \begin{equation}
 \label{equat}
\hbar\frac{\partial C_{\mu i}}{\partial\tau}+{\dot q}
 \sum_{\nu} \langle \psi_{\mu}(q(\tau)) \mid\frac{\partial \psi_{\nu}}
 {\partial q} (q(\tau))\rangle C_{\nu i} = 
  [\zeta_i-\epsilon_{\mu}(q(\tau))] C_{\mu i}  .
 \end{equation} 
 Here, $\zeta_i$, $i=1,...,{\cal N}$, are the Floquet exponents in imaginary 
 time, 
 which for the selfconsistent instanton would be eqal to the s.p. energies at 
 the metastable minimum, $\zeta_i=\epsilon_i(q_{min})$. 
 However, for a finite imaginary-time interval $[-T/2,T/2]$, $\zeta_i\ne 
 \epsilon_i(q_{min})$, although they should tend to this limit when 
 $T\rightarrow\infty$. 

 The conservation of overlaps $\langle \phi_i(-\tau)\mid \phi_j(\tau)\rangle
 = \delta_{i j}$ leads to the condition on $C_{\mu l}(\tau)$: 
 \begin{equation} 
  \label{over1}
  \sum_{\mu=1}^{{\cal N}} C^*_{\mu i}(-\tau) C_{\mu j}(\tau) = \delta_{i j}  .
 \end{equation} 
 This means that the matrix $C_{\mu i}(\tau)$ has the inverse $C^+(-\tau)$ and 
 the adiabatic states can be expanded on (all ${\cal N}$) bounce states: 
 \begin{equation} 
 \label{expan2}
  \psi_{\mu}(q(\tau))=\sum_{i=1}^{{\cal N}} C^*_{\mu i}(-\tau) \phi_{i}(\tau) = 
  \sum_{i=1}^{{\cal N}} C^*_{\mu i}(\tau) \phi_{i}(-\tau) ,
 \end{equation} 
 where in the second equality we assumed that $q(\tau)=q(-\tau)$ which 
 strictly holds for any {\it real} bounce observable: $q(\tau)=\sum_{i\ occ}
 \langle \phi_i(-\tau)\mid {\hat q}\mid \phi_i(\tau)\rangle=q^*(-\tau)$. Then,  
 the orthonormality of $\psi_{\mu}$, combined with the overlaps 
 Eq. (\ref{over1}), produces the relation: 
 \begin{equation} 
  \label{over2}
  \sum_{i=1}^{\cal N} C_{\mu i}(\tau) C^*_{\nu i}(-\tau) = \delta_{\mu \nu}  .
 \end{equation} 
 Thus, the quantity $p_{\mu i}(\tau) = C^*_{\mu i}(-\tau)C_{\mu i}(\tau)$ may 
 be considered as a quasi-occupation (it can be negative or complex in general 
 case) of the adiabatic 
 level $\mu$ in the bounce solution $i$, with $\sum_{\mu} p_{\mu i}(\tau) =1$, 
 or as the quasi-occupation of the bounce state $i$ in the adiabatic state 
 $\mu$,  
  where $\sum_{i} p_{\mu i}(\tau) = 1$. The sums over the occupied states: 
 $\sum_{i\ occ} p_{\mu i}(\tau)$ are diagonal elements $\rho_{\mu \mu}(\tau)$ 
 of the density matrix $\rho_{\mu \nu}(\tau)$ determined by the Slater states 
 $\mid\Phi(\tau)\rangle$.   
  
 From (\ref{expan1}) and (\ref{expan2}) one obtains the relation: 
 \begin{equation} 
  \label{inver}
  \phi_i(-\tau)=\sum_{j=1}^{\cal N} \left(\sum_{\mu}^{\cal N} C_{\mu i} (-\tau) 
 C^*_{\mu j}(-\tau)\right) \phi_j(\tau) = \sum_j^{\cal N} 
 \left(C^+(-\tau)C(-\tau)\right)_{j i} \phi_j(\tau)  , 
 \end{equation} 
  where the matrix $C^+(-\tau)C(-\tau)$ is hermitean and positive. 
  One can define: $C^+(-\tau)C(-\tau)=\exp(2{\hat {\cal S}}(\tau))^T$, so that 
  ${\hat {\cal S}}(\tau)$ is $\tau$-odd and hermitean and: 
 \begin{equation} 
  \label{Soper}
  \phi_i(-\tau) = \exp({\hat {\cal S}}(\tau))\psi_{0 i}(\tau) ,\hspace{3mm} 
  \phi_i(\tau) = \exp(-{\hat {\cal S}}(\tau))\psi_{0 i}(\tau) , 
 \end{equation} 
 where the states $\psi_{0 i}(\tau)$ are $\tau$-even and orthonormal, so they 
 could be considered as some "mean" TDHF orbitals related to the bounce 
 solutions $\phi_i(\tau)$ \cite{JS}. 


  Action is equal to the sum over the occupied iTDHF solutions: 
 \begin{equation}
  S = \Re \sum_{i\ occ} \int_{-T/2}^{T/2}
 \langle \phi_i(-\tau)\mid \zeta_i-{\hat h} \mid \phi_i(\tau) \rangle = 
    \int_{-T/2}^{T/2}\sum_{i\ occ} \sum_{\mu=1}^{\cal N}  
   [\zeta_i-\epsilon_{\mu}(q(\tau))]C^*_{\mu i}(-\tau)C_{\mu i}(\tau) d\tau ,
 \end{equation}
 so, using the quasi-occupations $p_{\mu i}$, it can be written as: 
 \begin{equation}
 \label{eq:stot}
  S = \int_{-T/2}^{T/2}\sum_{i\ occ} \sum_{\mu=1}^{\cal N}  
   [\zeta_i-\epsilon_{\mu}(q(\tau))] p_{\mu i}(\tau) d\tau .
 \end{equation}
  From this, the sum of actions for all individual s.p. bounce states is 
  the integral of a difference between two sums: of all Floquet exponents 
  and all adiabatic s.p. energies: $\sum_{i=1}^{\cal N}(\zeta_i - \epsilon_i)$. 
   It can be shown that this integral vanishes \cite{foot2}, so the sum of all
  actions is zero.
  
  When the collective motion is nearly adiabatic, one recovers from this 
  formalism action (\ref{scrank}) with the 
 cranking mass parameter and, ususally not mentioned, related formula 
 for the Floquet exponent - see Appendix \ref{app:crankmet}.   


\subsection{Instantons with pairing interaction \label{sec:metinst3}}

 In the presence of pairing interaction a proper mean-field formalism is 
 the imaginary-time-dependent HFB (iTDHFB) method. The Bogolyubov 
 transformation from the fixed, {\it independent of time} creation operators  
 $a_{\mu}^{\dagger}$ to {\it time-dependent} quasiparticle creation  
 operators $\alpha_i^{\dagger}(t)$, after passing to  
 imaginary time $t\rightarrow -i\tau$, can be written \cite{JS}:
\begin{eqnarray}
\alpha_i^\dagger(\tau)&=&\sum_{\mu}(A_{\mu i}(\tau)a_{\mu}^{\dag}+B_{\mu i}(\tau)a_{\mu}), \nonumber \\
\alpha_i(-\tau)&=&\sum_{\mu}(A_{\mu i}^{*}(-\tau)a_{\mu}+B_{\mu i}^{*}(-\tau)a_{\mu}^{\dag}),
\label{eq:transfBog}
\end{eqnarray}
 where amplitudes $A_{\mu i}(t)$ i $B_{\mu i}(t)$ became  
  functions of $\tau$, and their complex conjugate 
 $A_{\mu i}^*(t)$ and $B_{\mu i}^*(t)$ depend now on $-\tau$. The unitarity of  
 the Bogolyubov trnsformation in real time translates to the following 
 condition in imaginary time:  
\begin{equation}
\left(
\begin{array}{cc}
A^T(\tau), & B^T(\tau) \\
B^{\dagger}(-\tau), & A^{\dagger}(-\tau) \\
\end{array}
\right)^{-1} =
\left(
\begin{array}{cc}
A^*(-\tau), & B(\tau) \\
B^*(-\tau), & A(\tau) \\
\end{array}
\right).
\label{eq:Bogcond}
\end{equation}
 The hamiltonian overlap 
 $\langle\Phi(\tau)\mid {\hat H}\mid\Phi(-\tau)\rangle$ can be 
  expressed by the following contractions:
 \begin{eqnarray}
  \langle\Phi(\tau)\mid a^+_{\nu} a_{\mu}\mid\Phi(-\tau)\rangle
  &=\rho_{\mu \nu}(\tau)=& (B^*(-\tau) B^T(\tau))_{\mu \nu}   , \\
\nonumber
  \langle\Phi(\tau)\mid a_{\nu} a_{\mu}\mid\Phi(-\tau)\rangle
  &=\kappa_{\mu \nu}(\tau)=& (B^*(-\tau) A^T(\tau))_{\mu \nu}  , \\
\nonumber
  \langle\Phi(\tau)\mid a^+_{\nu} a^+_{\mu}\mid\Phi(-\tau)\rangle
  &={\tilde \kappa}_{\mu \nu}(\tau)=& (A^*(-\tau) B^T(\tau))_{\mu \nu}  , \\
\nonumber
\label{eq:hfbcontrac}
 \end{eqnarray}
  which, due to conditions (\ref{eq:Bogcond}), have the following properties 
 when regarded as matrices:
 \begin{eqnarray}
 \label{contr}
  \rho(-\tau)&=& \rho^+(\tau) ,  \\
\nonumber
  \kappa^T(\tau)&=& -\kappa(\tau) ,  \\
\nonumber
  {\tilde \kappa}(\tau)&=& \kappa^+(-\tau) .  \\
\nonumber
 \end{eqnarray}
 Using those and proceeding as in the derivation of the TDHFB equations we 
 arrive at imaginary-TDHFB (iTDHFB) equations written symbolically (where only 
 the second index of the amplitudes is explicit):
 \begin{equation}
\hbar \partial_{\tau}\left(
\begin{array}{c}
A_k(\tau) \\
B_k(\tau)  \\
\end{array}\right) +
\left(
\begin{array}{cc}
{\hat h}(\tau)-\lambda  ,&  {\hat \Delta}(\tau)  \\
-{\hat \Delta}^*(-\tau) ,&  -({\hat h}^*(-\tau)-\lambda) \\
\end{array} \right)
\left(
\begin{array}{c}
A_k(\tau) \\
B_k(\tau)  \\
\end{array}
\right) = \zeta_k
\left(
\begin{array}{c}
A_k(\tau) \\
B_k(\tau)  \\
\end{array}
\right) .
\label{eq:itdhfb}
\end{equation}
 Here, for a given two-body interaction 
$\frac{1}{2}\sum_{\mu \nu \gamma \delta} v_{\mu \nu \gamma \delta} 
 a_{\mu}^{\dagger} a_{\nu}^{\dagger} a_{\delta} a_{\gamma}$, the 
self-consistent potential:
 $\Gamma_{\mu \nu}(\tau)= \sum_{\gamma \delta}(v_{\mu \gamma \nu \delta}
 -v_{\mu \gamma \delta \nu})\rho_{\delta \gamma}(\tau)$
 and the pairing potential:
$\Delta_{\mu \nu}(\tau)=\sum_{\gamma \delta} 
 v_{\mu \nu \gamma \delta}
\kappa_{\gamma \delta}(\tau)$ have the properties: ${\hat \Gamma}(-\tau)=
 {\hat \Gamma}^+(\tau)$, and ${\hat \Delta}^T(\tau)=-{\hat \Delta}(\tau)$.
 The same properties hold for the mean fields with additional rearrangement 
 terms that follow from a density functional.
 These ensure the property ${\hat h}(-\tau)={\hat h}^+(\tau)$ of the
 mean-field Hamiltonian (${\hat t}$ - kinetic energy) ${\hat h}(\tau)=
  {\hat t}+{\hat \Gamma}(\tau)$, and the same property,
 ${\hat {\bf h}}(-\tau)={\hat {\bf h}}^+(\tau)$ of the total HFB
 mean-field Hamiltonian ${\hat {\bf h}}(\tau)$ given by the matrix in
 Eqs.(\ref{eq:itdhfb}).
 As a result of this, the equations (\ref{eq:itdhfb}) conserve both energy 
 overlap $\langle \Phi(\tau)\mid{\hat H}\mid\Phi(-\tau)\rangle$ 
 and all relations (\ref{eq:Bogcond}). The terms
 with constants $\zeta_k$ on the r.h.s. fix the periodicity of solutions and
  these constants are equal to the quasi-particle energies at the 
 HFB m.s.
 The bounce solution to Eqs.(\ref{eq:itdhfb}) has to be periodic and provide
 a path in the space of imaginary-time quasiparticle vacua which connects the 
 HFB m.s. $\mid\Phi(\pm T/2)\rangle=\mid \Psi_{gs}\rangle$ with some 
 HFB state $\mid \Phi(\tau=0)\rangle$ at the same energy beyond the barrier.

 One has to emphasize that in Eq. (\ref{eq:itdhfb}) appears the Fermi energy 
  $\lambda$ (this term is missing in \cite{JS}). It does not have to appear 
 in an initial value problem, as TDHFB equations preserve the expectation 
 value of the particle number $Tr(\rho)$, both in real \cite{Bulgac} and in 
 imaginary time. Here we look for a solution to the boundary value problem. 
 Without $\lambda$, $Tr(\rho)$ would be incorrect at the boundary and one has 
 to enforce its proper value. In particular, the solution has to tend to 
 the metastable HFB state $|\Phi(\pm T/2)\rangle$ at the boundaries as 
 $\tau\rightarrow \pm T/2$, and that fixes the value of $\lambda$.  

  Eq. (\ref{eq:itdhfb}) have the property analogous to that of the HFB 
 equations, that if $(A_{\mu i}(\tau),B_{\mu i}(\tau))$ is a periodic solution 
 with the Floquet exponent $\zeta_i$, then 
 $(B^*_{\mu i}(-\tau),A^*_{\mu i}(-\tau))$ is also a solution with the Floquet 
 exponent $-\zeta_i$. So, it suffices to find half of solutions. 
 The proper state $\mid\Phi(\tau)\rangle$ should contain exactly one of each 
 pair of two solutions with $\zeta_i$ and $-\zeta_i$ which then corresponds 
 to $\alpha_i(\tau)$. For ground states of e-e nuclei, it is natural 
 to choose the solutions with $\zeta_i>0$ as $\alpha_i^{\dagger}$ since 
 in the limit $\tau\rightarrow\pm T/2$ they correspond to positive 
 energies of quasiparticles. 
 Thus the state $\mid\Phi(-\tau)\rangle$ should be composed 
of solutions with $\zeta_i$ which at $\tau\rightarrow\pm T/2$ correspond to 
 negative quasiparticle energies. This means that in Eq. (\ref{eq:hfbcontrac}) 
 for the density matrix, $A_{\mu i}(\tau)$ and $B_{\mu i}(\tau)$ correspond at 
 $\tau\rightarrow\pm T/2$ to all positive $\zeta_i$.
 As the boundary condition fixes the correspondence with the 
 initial HFB state, the construction of matrices $\rho$ and  $\kappa$ for odd 
 nuclei is analogous to that in the HFB method \cite{Ring}: one of the 
 solutions ($A(\tau)$, $B(\tau))$ with positive $\zeta_i$ is replaced by  
 ($B^*(-\tau)$, $A^*(-\tau)$) with $-\zeta_i$.

 Decay rate is determined by instanton action which for a 
 state $|\Phi(\tau)\rangle$ can be presented in terms of the amplitudes  
 $A$ and $B$ \cite{JS}: 
{\setlength\arraycolsep{2pt}
\begin{eqnarray}
S/\hbar & = & \int_{-T/2}^{T/2} d\tau 
 \langle \Phi(\tau)|\partial_{\tau}\Phi(-\tau)\rangle \nonumber \\
  & = &\frac{1}{2}\int_{-T/2}^{T/2} d\tau \, Tr[\partial_{\tau}A^{\dagger}(-\tau)A(\tau)+\partial_{\tau}B^{\dagger}(-\tau)B(\tau)] \nonumber \\
  & = &-\frac{1}{2}\int_{-T/2}^{T/2} d\tau \, Tr[A^{\dagger}(-\tau)\partial_{\tau}A(\tau)+B^{\dagger}(-\tau)\partial_{\tau}B(\tau)].
  \label{eq:Sactpair}
\end{eqnarray}}
  Substituting $\partial_{\tau}A_{\mu i}(\tau)$ and 
 $\partial_{\tau}B_{\mu i}(\tau)$ from the iTDHFB equation (\ref{eq:itdhfb})
 and using conditions  (\ref{eq:Bogcond}) we obtain for the action integrand:
\begin{equation} 
   -\sum_{i\ occ}\frac{\zeta_i}{2} - \frac{1}{2}\sum_{\mu \nu}\left((h_{\mu \nu}(\tau)
 -\lambda \delta_{\mu \nu})(2\rho_{\nu \mu}(\tau)-\delta_{\mu \nu}) 
 + \kappa_{\mu \nu}(\tau)\Delta_{\mu \nu}^*(-\tau) 
 + \kappa_{\mu \nu}^*(-\tau)\Delta_{\mu \nu}(\tau)\right).
\end{equation}

 
 One can cast the instanton method in a form analogous to the density matrix 
 formalism. The matrix:  
\begin{equation}
\label{rhoi}
 {\cal R}(\tau)=\left(
 \begin{array}{cc}
  \rho(\tau), & \kappa(\tau) \\
  -\kappa^*(-\tau), & I-\rho^*(-\tau)  \\
 \end{array}\right)  
\end{equation}
  satisfies the equation: 
 \begin{equation}
 \label{eq:DEQ} 
 \hbar\partial_{\tau} {\cal R}(\tau)+[{\hat {\bf h}}(\tau),{\cal
 R}(\tau)]=0 ,
 \end{equation}
 which follows directly from (\ref{eq:itdhfb},\ref{eq:Bogcond}). 
 The matrix ${\cal R}$ has the property: ${\cal R}^2(\tau)={\cal R}(\tau)$, as 
 a result of: 
 $\rho(\tau)\kappa(\tau)=\kappa(\tau)\rho^*(-\tau)$ and 
 $\rho^2(\tau)-\kappa(\tau)\kappa^*(-\tau)=\rho(\tau)$. However, being 
 non-hermitean, it does not represent any real-time HFB density matrix. 
 


\subsection{Phenomenological potential model with the selfconsistent pairing 
 gap $\Delta(\tau)$ \label{sec:metinst4}}

 The above scheme can be simplified by replacing the mean-field ${\hat h}$ 
 by the s.p. Hamiltonian with the W-S potential and using 
  the pairing interaction with the constant matrix element. The 
 $\tau$-dependent HFB transformation may be presented as a composition: 
 $a^+_{n} \rightarrow b^+_{\mu} \rightarrow \alpha^+_i$, where the first 
 transformation diagonalizes the deformation-dependent W-S hamiltonian in the 
 deformation-dependent basis $\psi_{\mu}(q)=b^+_{\mu}(q)\mid 0 \rangle$ [note 
 that now the independent of time operators $a^{\dagger}$ carry the Latin 
 indices $n, m$, not the Greek ones as in the preceding part of this 
 section, which are now reserved for eigenstates of the phenomenological 
 ${\hat h}(\tau)$]: 
 \begin{equation} 
  b^+_{\mu}(q) = \sum_n C_{n \mu} (q) a^+_n  . 
 \end{equation}
 The second transformation is a genuine HFB one: 
 \begin{equation} 
 \label{HFB}
    \alpha^+_i = \sum_{\mu} \left(A_{\mu i}(\tau) b^+_{\mu}(q(\tau))+
    B_{\mu i}(\tau) b_{\mu}(q(\tau))\right) .
 \end{equation}
 We assume the pairing interaction with the constant matrix element $G>0$ in 
 the  adiabatic basis which acts only between pairs of particles 
 in time-reversed states $\mu\bar{\mu}$. The only non-zero matrix elements 
 of this interaction are: 
 $v_{\mu{\bar \mu}\nu {\bar \nu}}=-\frac{G}{2}$, and those related 
 by the antisymmetry.  

  Since the matrix $C$ is $q$-dependent it must be differentiated in 
 the iTDHFB equation (\ref{eq:itdhfb}), so that this equation in the 
 adiabatic basis becomes symbolically:  
  \begin{equation}
\hbar \partial_{\tau}\left(
\begin{array}{c}
A_{i}(\tau) \\
B_{i}(\tau)  \\
\end{array}\right) +
\left(
\begin{array}{cc}
{\hat \epsilon}(q)+
{\hat D} 
,&  {\hat \Delta}(\tau)  \\
-{\hat \Delta}^*(-\tau) ,&  -{\hat \epsilon}(q)+  
{\hat D}^* 
\end{array} \right)
\left(
\begin{array}{c}
A_{i}(\tau) \\
B_{i}(\tau)  \\
\end{array}
\right) = \zeta_i
\left(
\begin{array}{c}
A_{i}(\tau) \\
B_{i}(\tau)  \\
\end{array}
\right)  . 
\label{eq:itdhfb1}
\end{equation}
 Here, ${\hat \epsilon}(q)$ is a diagonal matrix with elements 
 ${\hat \epsilon}_{\mu \nu}(q)=\delta_{\mu \nu}(\epsilon_{\mu}(q)-\lambda)$ 
 ($\epsilon_{\mu}$ are s.p. energies), ${\hat D}$ is the matrix of 
 adiabatic couplings, $D_{\mu \nu}(\tau)=\hbar\langle \mu\mid 
 \frac{\partial \nu}{\partial \tau}\rangle=\hbar\dot{q}
\langle \mu\mid \frac{\partial \nu}{\partial q}\rangle$, with 
 $\langle \mu \mid \frac{\partial \nu}{\partial \tau}\rangle = 
 {\dot q}(\tau)\sum_n C^*_{n \mu}(q) \partial_q C_{n \nu}(q)$, 
 and only non-zero elements of the matrix ${\hat \Delta}$ are: $\Delta_{\mu {\bar \mu}}(\tau)= 
 -\Delta_{{\bar \mu} \mu}(\tau)=-\Delta(\tau)$, where: 
\begin{equation}
\Delta(\tau)= G\sum_{\mu>0} {\bar \kappa}_{\mu {\bar \mu}} , 
\label{eq:delsamozg}
\end{equation}
 with ${\bar \kappa}$ the anomalous density in the adiabatic basis. 
 The connection between density matrices ${\bar \rho}$ and ${\bar \kappa}$
 in the adiabatic basis, and $\rho$ and $\kappa$ (with indices $m$, $n$)
 in the basis independent of time, reads: 
 \begin{eqnarray}
   \rho(\tau) & = & C(q(\tau)){\bar \rho}(\tau)C^{\dagger}(q(\tau))
  ,  \\ \nonumber
   \kappa(\tau) & = & C(q(\tau)){\bar \kappa}(\tau)C^T(q(\tau))
 ,  \\ \nonumber
 \label{eq:densad}
 \end{eqnarray}
 where: $\delta_{\mu \nu} \epsilon_{\mu}(q)=\left(C^+(q(\tau))
 {\hat h}(q(\tau))C(q(\tau))\right)_{\mu \nu}$. 

 Next, we intend to use further the Kramers degeneracy of s.p. states, 
 already used in defining the pairing interaction. This is quite natural for 
 e-e nuclei. In odd-$A$ nuclei, the odd nucleon perturbs the mean field,  
 breaking its invariance under time-reversal and the Kramers degeneracy; 
 three new time-reversal-odd densities emerge in the mean-field treatment 
 \cite{Engel}. 
 However, we will neglect this effect here as if it would be small (see 
 \cite{Koh} for the effect of time-odd terms on the HF$+$BCS barrier). 
 This means that also in odd-$A$ nuclei we assume two groups of states, 
 $\mu$ and ${\bar \mu}$, with $\epsilon_{\mu}=\epsilon_{{\bar \mu}}$, 
 $D_{{\bar \mu} {\bar \nu}}=D_{\mu \nu}^*$. There will be two sets of 
 solutions, $i$ and ${\bar i}$, with 
 $\rho_{\mu {\bar \nu}}= \rho_{{\bar \mu} \nu}= \kappa_{\mu \nu}= 
 \kappa_{{\bar \mu} {\bar \nu}}= 0$, for which 
  Eq. (\ref{eq:itdhfb1}) separates into two independent sets with 
 matrices: 
  \begin{equation}
\left(
\begin{array}{cc}
{\hat \epsilon}(q)+
{\hat D} 
,&  -\Delta(\tau)\cdot{\hat I} \\
-\Delta^*(-\tau)\cdot{\hat I} ,&  -{\hat \epsilon}(q)+  
{\hat D} 
\end{array} \right)
 \hspace{2mm}  {\rm and: } \hspace{2mm} 
\left(
\begin{array}{cc}
{\hat \epsilon}(q)+
{\hat D}^* 
,&  \Delta(\tau)\cdot{\hat I} \\
 \Delta^*(-\tau)\cdot{\hat I} ,&  -{\hat \epsilon}(q)+  
{\hat D}^* 
\end{array} \right)  , 
  \end{equation}
 with ${\hat I}$ - the block unit matrix. 
 Let the solutions with $\zeta_i>0$ of the first set be amplitudes:
 $(A_{\mu i}(\tau),B_{{\bar \mu} i}(\tau))$, and for the second set:  
 $(A_{{\bar \mu} {\bar i}}(\tau),B_{\mu {\bar i}}(\tau))$. Then 
 the solutions with $\zeta_i<0$ are:
 $(B_{{\bar \mu} i}^*(-\tau),A_{\mu i}^*(-\tau))$ - to the second set of 
 equations, and $(B_{\mu {\bar i}}^*(-\tau),A_{{\bar \mu}{\bar i}}^*(-\tau))$ -
 to the first one. 
 If, additionally, ${\hat D}={\hat D}^*$, which holds, for example, 
 for a mean field ${\hat h}$ with the axial symmetry or the one having the 
 reflexion symmetry in three perpendicular planes (like for shapes with 
 deformations: $\beta$, $\gamma$,
 $\beta_{40}$, $\beta_{42}=\beta_{4-2}$, $\beta_{44}=\beta_{4-4}$, etc, 
 cf Sec. \ref{sec:4}),  
 $\Delta$ will also be real and then, the solutions of the second set 
 of equations are: 
 $(A_{{\bar \mu} {\bar i}}(\tau),B_{\mu {\bar i}}(\tau))= 
  (A_{\mu i}(\tau),-B_{{\bar \mu} i}(\tau))$. In such a case, both 
 sets of equations produce the same sets of $\zeta_i$,
   one has: ${\bar \rho}_{{\bar \mu} {\bar \nu}}={\bar \rho}_{\mu \nu}$,
   ${\bar \kappa}_{{\bar \mu} \nu}=-{\bar \kappa}_{\mu {\bar \nu}}$ and 
   it suffices to know the half of density matrices (in the adiabatic basis) 
  which, from (\ref{eq:DEQ},\ref{eq:densad}), fulfill the equations
 (cf e.g. \cite{KoNix} for comparison with the TDHFB):
 \begin{eqnarray}
\label{eq:itdhfb-rho4}
 \hbar\partial_{\tau}{\bar \rho}_{\mu \nu}(\tau) & = & 
 (\epsilon_{\nu}(q)- \epsilon_{\mu}(q)){\bar \rho}_{\mu \nu}(\tau)
 - {\bar \kappa}_{\mu {\bar \nu}}(\tau)\Delta(-\tau) + 
   \Delta(\tau){\bar \kappa}_{\mu {\bar \nu}}(-\tau) \\ \nonumber
     & + & [{\bar \rho}(\tau),{\hat D}]_{\mu \nu}   ,  \\ \nonumber
 \hbar\partial_{\tau}{\bar \kappa}_{\mu {\bar \nu}}(\tau) & = & 
  \Delta(\tau)(\delta_{\mu \nu} - {\bar \rho}_{\mu \nu}(\tau) -  
 {\bar \rho}_{\nu \mu}(\tau))  
 - (\epsilon_{\nu}(q) + \epsilon_{\mu}(q)-2\lambda)
 {\bar \kappa}_{\mu {\bar \nu}}(\tau)   \\ \nonumber  
    & + & [{\bar \kappa}(\tau),{\hat D}]_{\mu {\bar \nu}} . \\ \nonumber
 \end{eqnarray}
 The Eq. (\ref{eq:itdhfb1}) are a counterpart of (\ref{equat})  
 for instanton-like solutions with pairing. One should notice that, in spite of 
 using a phenomenological potential in place of the selfconsistent one, 
 we could not avoid nonlocality in time - the matrix in Eq. (\ref{eq:itdhfb1})
 depends on both $\Delta(\tau)$ and $\Delta(-\tau)$, and the function 
$\Delta(\tau)$ has to be selfconsistent - it should fulfil the condition 
  (\ref{eq:delsamozg}).
 In the process of iterative solution for $\Delta (\tau)$ its value at the 
 current step would differ in general from the value $\Delta_r(\tau)$
 resulting from the integration of the Eq. (\ref{eq:itdhfb1}) in this step.
 Using the equation for densities one has: 
  \begin{equation}
 \hbar\frac{\partial\Delta_r}{\partial \tau} = G\left[(N_r-{\cal N}) - 
     2\sum_{\mu>0}(\epsilon_{\mu}(\tau)-\lambda) \kappa_{{\bar \mu} \mu}(\tau)
 \right] ,
  \end{equation}
 where $N_r=2\sum_{\mu>0}\rho_{\mu \mu}(\tau)$ is the expectation value 
 of the number of particles, not necessarily equal to the assumed one, 
 and ${\cal N}$ - the number of included doubly degenerate levels. 
 On the other hand, from these equations:
 \begin{equation}
  \hbar\frac{\partial N_r}{\partial \tau} = \frac{2}{G}\left(\Delta_r(\tau)
 \Delta^*(-\tau)-\Delta(\tau)\Delta_r^*(-\tau)\right) .
 \end{equation}
 One can see that the expectation value of the number of particles is 
 constant for a selfconsistent solution with $\Delta_r(\tau)=\Delta(\tau)$.

 Test solutions with a few adiabatic W-S levels indicate 
 that the (rather long) iterative procedure applied to Eq. (\ref{eq:itdhfb1}),
 equivalent to Eq. (\ref{eq:itdhfb-rho4}), leads to the exponential 
 dependence of $\Delta(\tau)$, which is large on the interval $[-T/2,0]$ and 
 small on $[0,T/2]$, with a mild variation of the product 
 $\Delta(\tau)\Delta(-\tau)$. This case is considerably more involved than 
 the the equation with the W-S potential alone. 

 Assuming that we have solutions to Eq. (\ref{eq:itdhfb1}),
 one can write down action (\ref{eq:Sactpair}) for an e-e nucleus: 
\begin{eqnarray}
\label{actionmod}
S & = & \int_{-T/2}^{T/2} d\tau\; \left\{-\sum_{i>0} \zeta_i -
 \sum_{\mu>0} \left((2{\bar \rho}_{\mu \mu}(\tau)-1)
(\epsilon_{\mu}(\tau)-\lambda)      
   +\Delta(\tau){\bar \kappa}_{\mu {\bar \mu}}^*(-\tau)+ 
    {\bar \kappa}_{\mu {\bar \mu}}(\tau)\Delta^*(-\tau)\right)\right\} \\ \nonumber
\end{eqnarray}
\begin{eqnarray}
  &=&\int_{-T/2}^{T/2} d\tau\; \left\{-\sum_{i>0} \zeta_i -
 \sum_{\mu>0}(2{\bar \rho}_{\mu \mu}(\tau)-1)(\epsilon_{\mu}(\tau)-\lambda)+
 2\frac{\Delta(\tau)\Delta^*(-\tau)}{G} \right\},    \\ \nonumber
\end{eqnarray}
 where the summation runs over solutions $i>0$ and states $\mu>0$, 
 and the last equality holds for the selfconsistent solution.
 For an odd nucleus, one has to exchange in densities (\ref{eq:hfbcontrac}) 
 one amplitude with positive $\zeta$ by the other one with $-\zeta$. 

 In the limit of no pairing, the positive Floquet exponents of decoupled 
 Eq. (\ref{eq:itdhfb1}) are: $\zeta_i^{NP}-\lambda$ for amplitudes $A$ of 
 empty states, and $\lambda-\zeta_i^{NP}$ for amplitudes $B$ of occupied 
 states, where $\zeta_i^{NP}$ are Floquet exponents of solutions to 
 (\ref{equat}). Density ${\bar \rho}_{\mu \mu}$, composed of amplitudes 
 of occupied states, expressed in terms of quasi-occupations $p_{\mu i}$ of 
 Sec. \ref{sec:metinst2}, is: $\sum_{i>0 , \zeta_i^{NP}<\lambda}\ p_{\mu i}$. 
 For solutions $i>0$ one has:
 $2{\bar \rho}_{\mu \mu}-1=\sum_{\zeta_i^{NP}<\lambda} p_{\mu i} 
 -\sum_{\zeta_i^{NP}>\lambda}\ p_{\mu i}$ (since $\sum_{i>0}\ p_{\mu i}=1$). 
 Hence, the sum in the integrand (\ref{actionmod}) is equal to the 
 difference $\sum_{\zeta_i^{NP}<\lambda}-\sum_{\zeta_i^{NP}>\lambda}$ of the 
 following expressions:
 $(\zeta_i^{NP}-\lambda)-\sum_{\mu>0} p_{\mu i}(\epsilon_{\mu}-\lambda)$.
 The terms with $\lambda$ vanish after summation as a consequence of: 
 $\sum_{\mu>0} p_{\mu i}=1$; one is thus left with the difference of sums of 
 actions without pairing for solutions $i>0$: (below) $-$ (above) the Fermi 
 level. We know from Sec. \ref{sec:metinst2} that those sums add to zero; 
  therefore the result is $2\times$ the sum of actions for $i>0$ occupied  
 solutions, equal to action without pairing for all 
 (i.e. $i$ and ${\bar i}$) occupied states.

\section{Two - level model \label{sec:2levmod}}

 It turns out that a main difficulty in integrating Eq. (\ref{equat}) are 
 avoided crossings with a minuscule interlevel interaction - see 
 Sec. \ref{sec:instnonax}. Here we study a dependence of bounce-like action 
 for such a crossing on the collective velocity and level slopes in a simple 
 model with two s.p.  levels - a kind of analogy with the Landau - Zener 
 problem \cite{Landau,Zener,Stuck}. The Hamiltonian is: 
\begin{equation}
\hat{h}(q(\tau))=\left(\begin{array}{cc}
E_{1}(q(\tau)) & V \\ 
V^* & E_{2}(q(\tau))
\end{array}\right),
\label{eq:hamdiab} 
\end{equation}
 where $q(\tau)$ is a time-dependent parameter, e.g. some nuclear deformation.
 We assume: $V=V^*$, $E_{1,2}=\pm\, E(q-q_{0})$, so that diagonal elements are 
 linear in $q$ and cross at $q_{0}$. 
 The states: $|\chi_{1}\rangle=(1,0)^{T}$, $|\chi_{2}\rangle=(0,1)^{T}$ we call \textit{diabatic}; the basis: 
\begin{equation}
|\psi_{1}\rangle=\left(\begin{array}{c}
\cos\frac{\theta}{2} \\ 
\sin\frac{\theta}{2}
\end{array}\right), \quad 
|\psi_{2}\rangle=\left(\begin{array}{c}
-\sin\frac{\theta}{2} \\ 
\cos\frac{\theta}{2}
\end{array}\right),
\label{eq:vectadiab}
\end{equation}
 in which ${\hat h}$ is diagonal with eigenvalues:
\begin{equation}
\epsilon_{1,2}= \mp \frac{1}{2} \sqrt{(E_{1}-E_{2})^{2}+4V^2}
\label{eq:eadiab2lvl}
\end{equation}
  we call \textit{adiabatic}. Here, $\tan\theta=\frac{2V}{E_{1}-E_{2}}$. 
 So, for $q<q_{0}$, $\theta \rightarrow 0$ and adiabatic states tend to 
 diabatic ones, $|\psi_{1,2}\rangle \rightarrow |\chi_{1,2}\rangle$. 
 At the pseudo-crossing $q_{0}$, $\theta=-\pi/2$ and the mixing of diabatic 
 states is maximal. Due to the interaction, adiabatic energies do not 
 cross but at $q_0$ approach their minimal distance 
 $\epsilon_{2}-\epsilon_{1}=2V$. For $q>q_{0}$,   
  $\theta \rightarrow -\pi$ and $|\psi_{1}\rangle \rightarrow 
 -|\chi_{2}\rangle$ (note the change of sign), $|\psi_{2}\rangle \rightarrow 
 |\chi_{1}\rangle$, so after passing the pseudo-crossing the 
 adiabatic states exchange their characteristics. 
 The coupling of adiabatic states in the iTDSE is:
\begin{equation}
\left\langle\psi_{1}\bigg|\frac{d\psi_{2}}{dq}\right\rangle=-\frac{1}{2}\frac{d\theta}{dq} =\frac{1}{2}\frac{EV}{E^{2}(q-q_{0})^{2}+V^{2}}=\frac{1}{2}\frac{\alpha}{(q-q_{0})^{2}+\alpha^{2}},
\label{eq:2levcoupl}  
\end{equation}
 where we introduced $\alpha=V/E$. It has the Lorentz shape with a maximum at 
  $q_{0}$ and the width and height regulated by $\alpha$. In the limit 
  $V \rightarrow 0$, i.e., $\alpha \rightarrow 0$, the coupling 
  element tends to the Dirac $\delta$-function. 

 To define the model we have to specify $q(\tau)$ and the resulting collective 
 velocity $\dot{q}(\tau)$. In the following we use the ansatz: 
\begin{equation}
q(\tau)=\frac{q_{fin}-q_{ini}}{\cosh(\Gamma\tau)}+q_{ini},
\label{eq:impuls}
\end{equation}
 where $q_{ini},\ q_{fin}$ are the initial and final collective deformation
 (e.g. the entrance and exit from the barrier). So defined $q(\tau)$ has 
 an impulse shape, typical for instanton, 
 which means that the motion takes place in a finite time interval around 
  $\tau=0$, while in the asymptotic region, $\tau \rightarrow \pm \infty$, 
  $q(\tau)\rightarrow q_{ini}$ with vanishingly small ${\dot q}$.
 The equation reads: 
\begin{eqnarray}
\label{eq2lvl2}
\hbar\dot{c}_{1} & = & -\epsilon_{1}c_{1}-\hbar\dot{q}\langle\psi_{1}|
\partial_{q}\psi_{2}\rangle c_{2}, \\ \nonumber 
\hbar\dot{c}_{2} & = & -\epsilon_{2}c_{2}+\hbar\dot{q}\langle\psi_{1}|
\partial_{q}\psi_{2}\rangle c_{1}.
\end{eqnarray}
 After using definitions of the model and 
 introducing dimensionless time parameter $z=\tau\frac{|E|}{\hbar}$  
 the following form of iTDSE is obtained:
\begin{eqnarray}
\label{eq2lvl2B}
\frac{d}{dz}\tilde{c}_{1} & = & 
\sqrt{(q-q_0)^2+\alpha^2}\ \tilde{c}_{1}+\frac{1}{2}
\beta\tanh(\beta z)\,(q-q_{ini})\,\frac{\alpha}{(q-q_0)^2+\alpha^2}\  
\tilde{c}_{2}, \\  \nonumber
\frac{d}{dz}\tilde{c}_{2} & = & 
-\sqrt{(q-q_0)^2+\alpha^2}\ \tilde{c}_{2}-\frac{1}{2}
 \beta\tanh(\beta z)\,(q-q_{ini})\,\frac{\alpha}{(q-q_0)^2+\alpha^2}\ 
\tilde{c}_{1},  \\ \nonumber
\end{eqnarray}
 where $\tilde{c}_{i}(z)=c_{i}(\tau)$ and $\beta=\hbar\Gamma/|E|$. The 
 following parameters were fixed: $q_{ini}=0.2212$, $q_{fin}=0.7343$ and 
 $q_0=0.55$. Then, from (\ref{eq2lvl2B}), bounce-like solutions 
 $\tilde{c}_k(z)$ and action depend on two 
 parameters: $\alpha$ and $\beta$: $S=S(\alpha,\beta)$. Pertinent to 
 difficulties of realistic calculations are the non-obvious changes in $S$ 
 for small $\alpha$ {\it and} $\beta$ - see Sec. \ref{sec:instnonax}. 
 Accordingly, other parameters were set as follows: 
 $\Gamma = 0.5 \times 10^{21}$ s$^{-1}$ (the maximal possible velocity was 
 $|{\dot q}_{max}|\approx0.128 \times 10^{21}$ s$^{-1}$), 
 $E=5, 10, 15 ...$ MeV 
 defined values of $\beta$, and $V$ covered a range of exponentially small 
 values. Solutions were obtained by the method described in 
 Appendix \ref{app:solu}, but for small $\alpha$ Eq. (\ref{eq2lvl2}) 
 was solved in the diabatic basis.

\begin{figure}[t]
	\centering
	\includegraphics[angle=-90, width=0.60\textwidth]{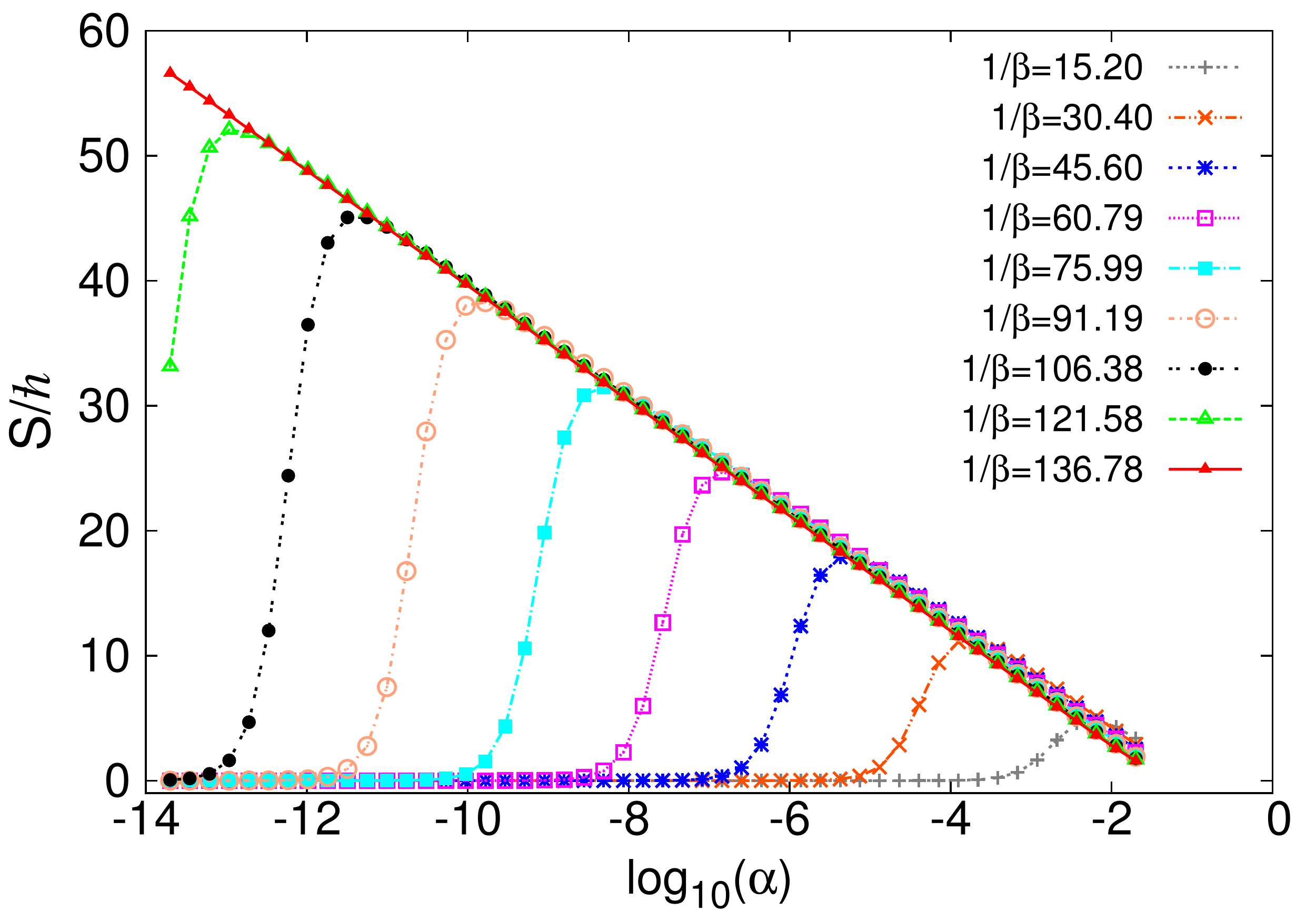}
	\caption{Action $S(\alpha)$ for various parameters $1/\beta$.}
	\label{fig:sarowbcol}
\end{figure}

  In Fig. \ref{fig:sarowbcol} the calculated action is displayed as a function 
 of the parameter $\alpha$ at fixed values of $\beta$. The parameter $\alpha$ 
 is proportional to $V$ - the strength of interaction between levels. 
 The extremal cases are when $V$ is very large or very small. In the first 
 case, levels are repelling each other and transitions between the adiabatic 
 levels are reduced - one can expect a small action (note that the adiabatic 
 limit of small $\beta/\alpha=\hbar{\dot q}/V$ is not covered in 
   Fig. \ref{fig:sarowbcol}). When $V\rightarrow 0$, the 
 transitions between diabatic levels cease, and action tends to zero again. 
 A larger action can be expected for intermediate values of $\alpha$ 
 and there has to be at least one maximum of $S$. Calculated values of 
 $S(\alpha)$ in Fig. \ref{fig:sarowbcol} show a maximum at some 
 $\alpha_{max}$, while for smaller and larger values of $\alpha$, respectively, 
 action rises from, and falls down to zero. In the covered range of $\alpha$,   
 one can observe an approximate scaling: $S(\log_{10}\alpha,\beta') \sim 
 (\beta/\beta') S((\beta'/\beta)\log_{10}\alpha,\beta)$.  
\begin{figure}[h]
	\centering
	\includegraphics[angle=-90, width=0.48\textwidth]{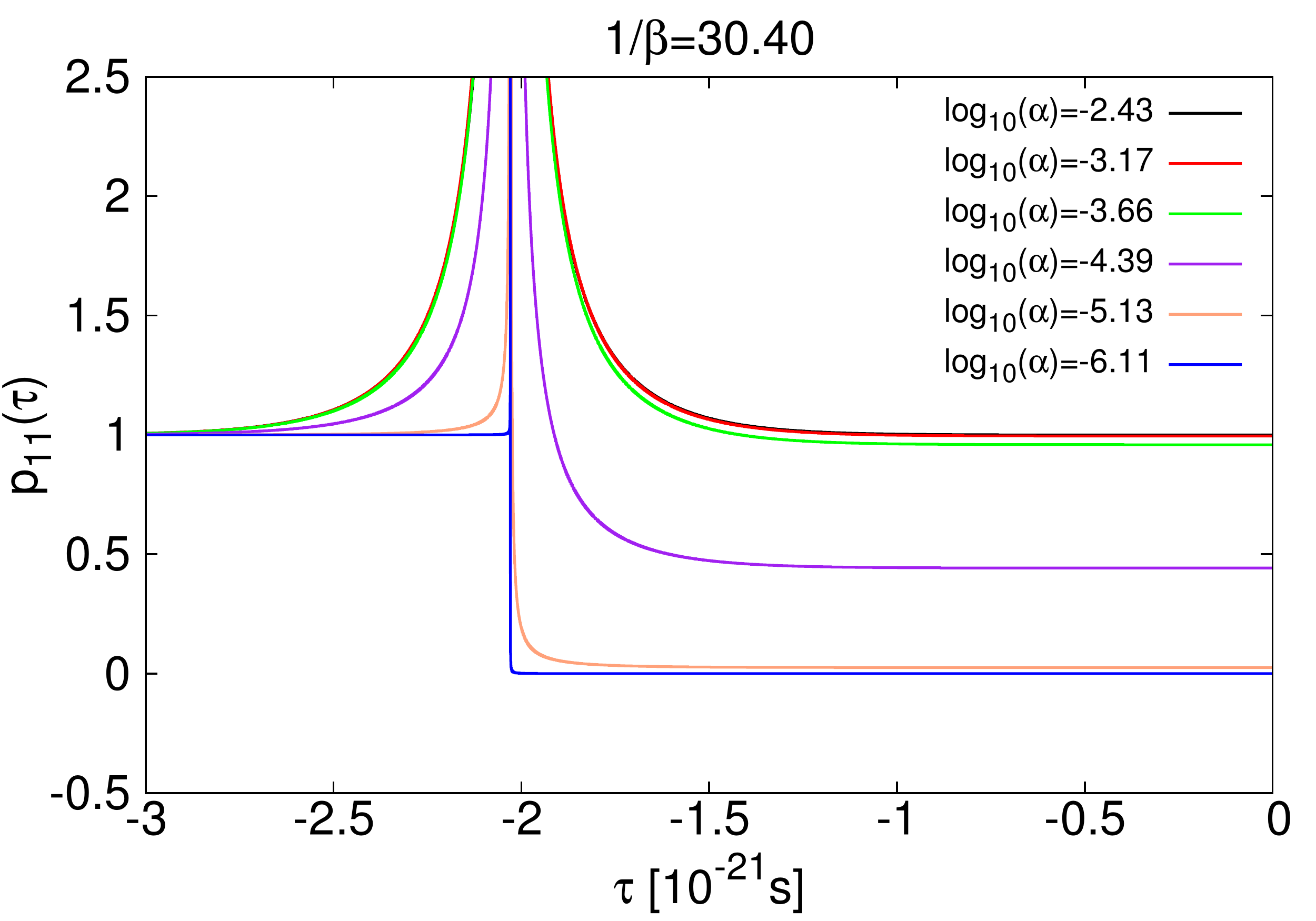}
	\includegraphics[angle=-90, width=0.48\textwidth]{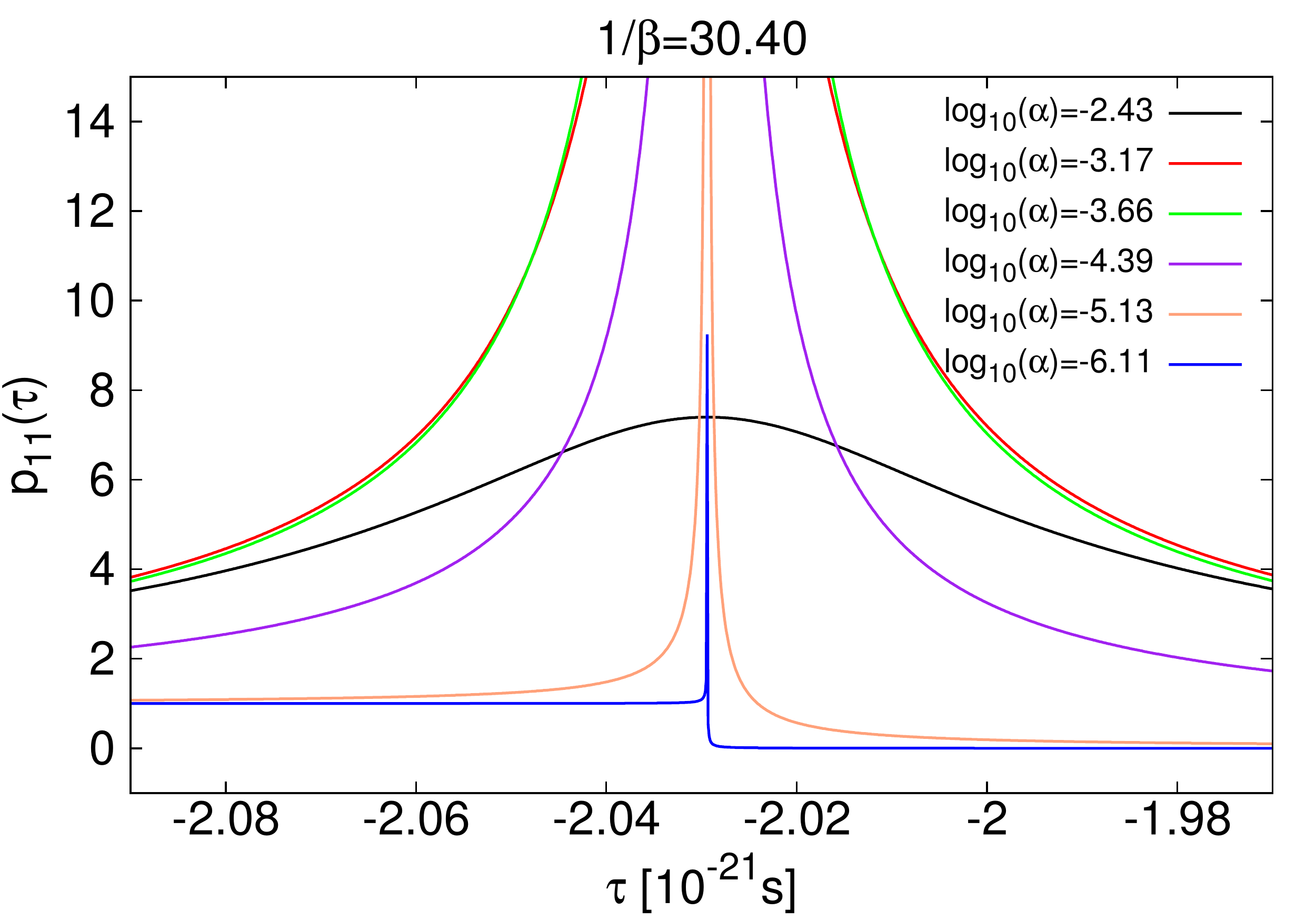}
	\caption{\textit{Left panel:} Pseudo-occupation of the lower adiabatic 
 level for solutions with various $\alpha$ at fixed $1/\beta=30.40$. 
 The corresponding $S(\alpha)$ is shown in Fig. \ref{fig:sarowbcol}. The 
  pseudocrossing occurs at $\tau_{c}\approx -2.03$. \textit{Right panel :} 
 The same in greater detail, close to $\tau_{c}$.  }
	\label{fig:p11arowbcol}
\end{figure}

For an illustration of non-adiabatic transitions, in Fig. \ref{fig:p11arowbcol}
 we show the pseudo-occupation $p_{11}(\tau)$ defined in Sect. II B [after 
  the formula (\ref{over2})]. It is displayed for the same $\alpha$ values 
 which were used to calculate $S(\alpha)$ in Fig. \ref{fig:sarowbcol},  
 for $1/\beta = 30.40$. It can be seen that for $\alpha$ 
 greater than $\alpha_{max}$ ($\log_{10}(\alpha_{max})\approx -3.95$),  
 most of the time $p_{11}$ is concentrated in the lower adiabatic state; 
 a transition to the upper adiabatic state takes place only around the 
 pseudo-crossing, while behind it the system returns to 
 the lower state, i.e. $p_{11}(\tau=0)=1$. This behaviour changes when 
 we approach the maximum of action - for $\log_{10}(\alpha)=-4.39$ - 
 the system behind the crossing remains partially excited to the upper 
 adiabatic level ($0<p_{11}(\tau=0)<1$). For still smaller 
 $\alpha<\alpha_{max}$, behind the pseudo-crossing the system 
 occupies exclusively the upper adiabatic level, till the end of the barrier
 ($p_{11}(\tau=0)=0$; $p_{21}(\tau=0)=1$). In such a case we have 
 a continuation of the diabatic state. 

\begin{figure}[t]
	\centering
	\includegraphics[angle=-90, width=0.65\textwidth]{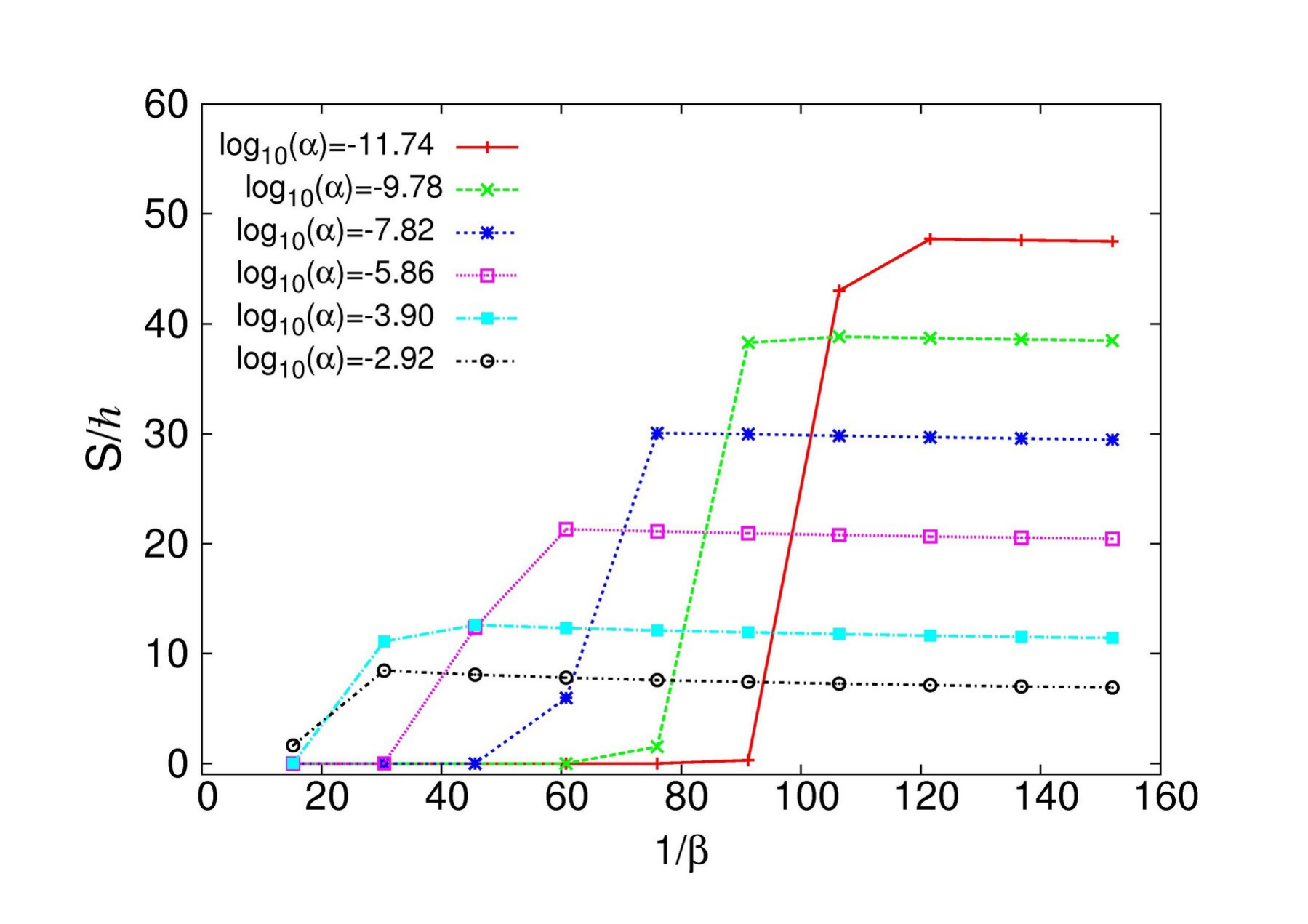}
	\caption{Action $S(1/\beta)$ for various values of $\alpha$.}
	\label{fig:sacolbrow}
\end{figure} 
\begin{figure}[!h]
	\centering
 	\includegraphics[angle=-90, width=0.48\textwidth]{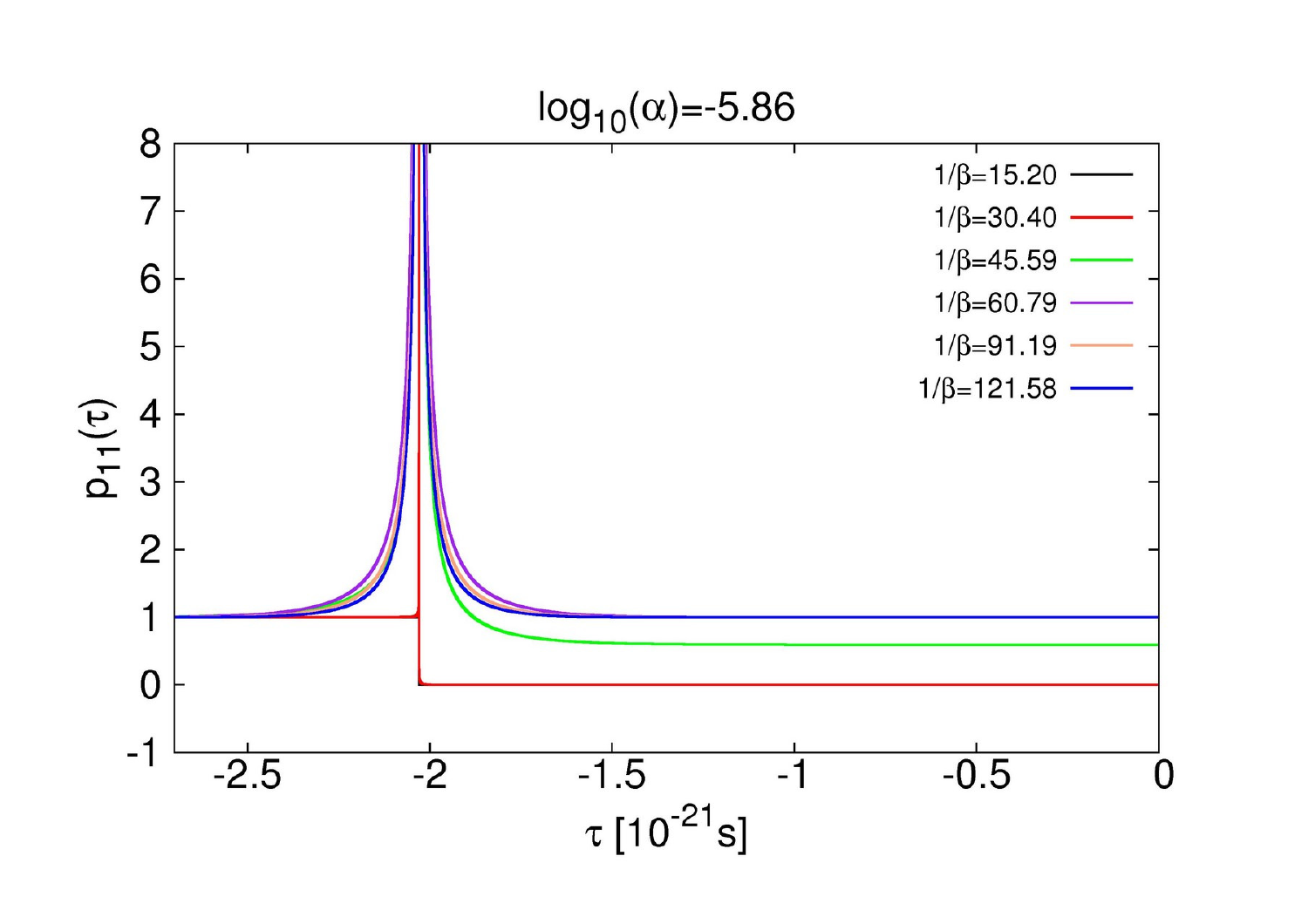}
	\includegraphics[angle=-90, width=0.48\textwidth]{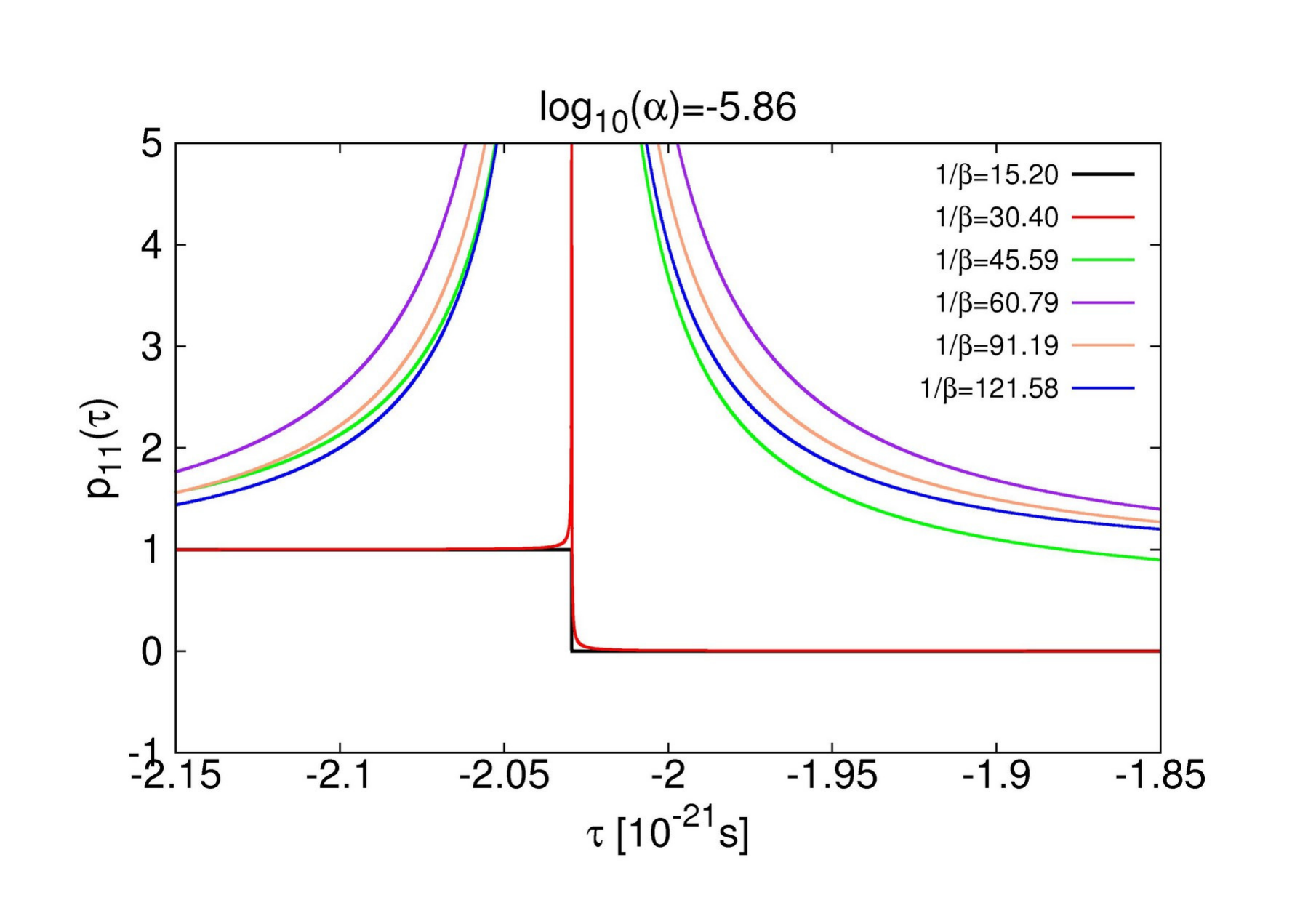}
	\caption{\textit{Left panel:} Pseudo-occupation of the lower adiabatic 
 level for solutions with various $1/\beta$ at fixed 
 $\log_{10}(\alpha)=-5.86$. The corresponding action $S(1/\beta)$ is shown 
 in Fig. \ref{fig:sacolbrow}. The pseudo-crossing occurs at 
 $\tau_{c}\approx -2.03$. \textit{Right panel:} The same in greater detail, 
 close to $\tau_{c}$.  }
	\label{fig:p11acolbrow}
\end{figure} 
   In Fig. \ref{fig:sacolbrow} is shown a plot of action as a function of 
 $1/\beta$ at the fixed $\alpha$, which corresponds to the fixed 
 matrix element $\langle\psi_{1}|\partial_{q}\psi_{2}\rangle$. 
 One can see its jump-like character: for small $1/\beta$ action is close to 
 zero, over a short interval of $1/\beta$ it rises rapidly to 
 a maximal value and then it decreases very slowly. The jump is more sharp 
 and larger for smaller values of $\alpha$, which correspond to a sharper 
 pseudo-crossing between the adiabatic levels. 
 As $1/\beta \sim 1/\Gamma \sim 1/{\dot q}_{max}$, the greater the velocity, 
 the stronger the coupling between the adiabatic levels, so for 
 sufficiently large $\dot{q}$ (small $1/\beta$) one can expect a diabatic 
 continuation (transition to an upper adiabatic level) when passing through 
 the pseudo-crossing, which means a small action. One should notice that 
 action vanishing in the limit of very large ${\dot q}$ is an artificial 
  property of the model with a finite number of states - after reaching the 
 highest one the system cannot excite anymore. 

  For smaller ${\dot q}$, after passing through the pseudo-crossing, 
 pseudo-occupations of both adiabatic states become comparable - 
 action becomes sizable. For still smaller $\dot{q}$, the pseudo-occupation 
 $p_{21}$ of the upper adiabatic state is non-zero only around the 
 pseudo-crossing, and action does not change much. This also can be seen in 
 Fig. \ref{fig:p11acolbrow} where the pseudo-occupation of the lower 
 adiabatic state is shown for the lower iTDSE solution at the fixed value 
 of $\alpha$.  
 The diabatic behaviour - a sharp fall of $p_{11}$ from 1 to 0 at the 
 pseudo-crossing (red and black lines) - gives way to an intermediate 
 situation - $0<p_{11}<1$ behind pseudo-crossing (green line) - 
 and then to the adiabatic one - $p_{11}=1$ except the close neighbourhood of 
 the pseudocrossing (all other lines). One can notice from 
 Fig. \ref{fig:sacolbrow} that a smaller $\alpha$ means a larger domain of 
 diabatic behaviour in $1/\beta$, i.e. as $\alpha$ decreases, the interval 
 of a diabatic - to - adiabatic transition shifts towards smaller 
 collective velocities (larger $1/\beta$).

 Presented solutions determine whether the evolution is diabatic, intermediate 
 or adiabatic. Since values of $\alpha$ pertinent to nuclear potential with 
 nonaxial deformation can be as small as $\sim 10^{-6}$ - $10^{-7}$, cf  
 Sec. \ref{sec:instnonax}, this simple model demonstrates a possibility 
 of large variation in action for a fixed $\alpha$, resulting from 
 the dependence on the collective velocity ${\dot q}$ at the crossing. 
 As Fig. \ref{fig:sarowbcol} suggests, {\it even for very small} $V$ 
 {\it one  can get sizabele} action. 
 In a realistic case,  with many interacting levels, it is difficult 
 to predict the effect of one pseudo-crossing on the value of action without 
 solving for the instanton-like solution.  
 
 Independent of the above results, we have checked that in the adiabatic limit 
 of small ${\dot q}/V=\beta/\alpha$, 
 the two-level model produces action which tends to the value given by the 
 formula (\ref{Scrankim}) with {\it the cranking mass parameter}, 
 see \cite{Piaski}.

 \section{Instanton-like solutions with the Woods-Saxon potential
 \label{sec:4}}

  From this point on, we shall consider instanton-like iTDSE solutions related 
  to the realistic s.p. Woods-Saxon potential within the 
 microscopic-macroscopic framework briefly described below. 
 
  Deformation enters the s.p. potential via a definition of the nuclear 
  surface by \cite{WS}: 
  \begin{eqnarray}
  \label{shape}
    R(\theta,\varphi)&=& c(\{\beta\}) R_0 \{ 1+\sum_{\lambda>1}\beta_{\lambda 0}
   Y_{\lambda 0}(\theta,\varphi)+ \nonumber \\
&& \sum_{\lambda>1, \mu>0, even}
   \beta_{\lambda \mu c} Y^c_{\lambda \mu} (\theta,\varphi)\}  ,
\end{eqnarray}
  where $c(\{\beta\})$ is the volume-fixing factor. The real-valued spherical
  harmonics $Y^c_{\lambda \mu}$, with even
  $\mu>0$, are defined in terms of the
  usual ones as: $Y^c_{\lambda \mu}=(Y_{\lambda \mu}+Y_{\lambda -\mu})/
 \sqrt{2}$. Here we restrict shapes to reflection-symmetric ones and 
 allow only for the quadrupole non-axiality $\beta_{22}$. 
 The $n_{p}=450$ lowest proton levels and $n_{n}=550$ lowest
 neutron levels from $N_{max}=19$ lowest major shells of the deformed harmonic 
 oscillator were taken into account in the diagonalization procedure.
 Eigenenergies are used to calculate the shell- and pairing corrections.  
 The macroscopic part of energy is calculated by using the Yukawa plus 
 exponential model \cite{KN}. All parameters used here, of 
  the s.p. potential, the pairing strength and the
  macroscopic energy, are equal to those used previously in the calculations
  of masses \cite{WSparmac,Qmass} and fission barriers 
 \cite{Kow,2bar,JKSs,JKSa} of heaviest nuclei, whose results are in 
 reasonable agreement with data. 
 In particular, we took the "universal set" of potential parameters and the 
 pairing strengths $G_n=(17.67-13.11\cdot I)/A$ for neutrons, 
 $G_p=(13.40+44.89\cdot I)/A$ for protons ($I=(N-Z)/A$), as adjusted 
 in \cite{WSparmac}. As always within this model, $N$ neutron and 
  $Z$ proton s.p. levels have been included when solving BCS equations.  

\begin{figure}[!b]
	\centering
	\includegraphics[width=0.65\textwidth]{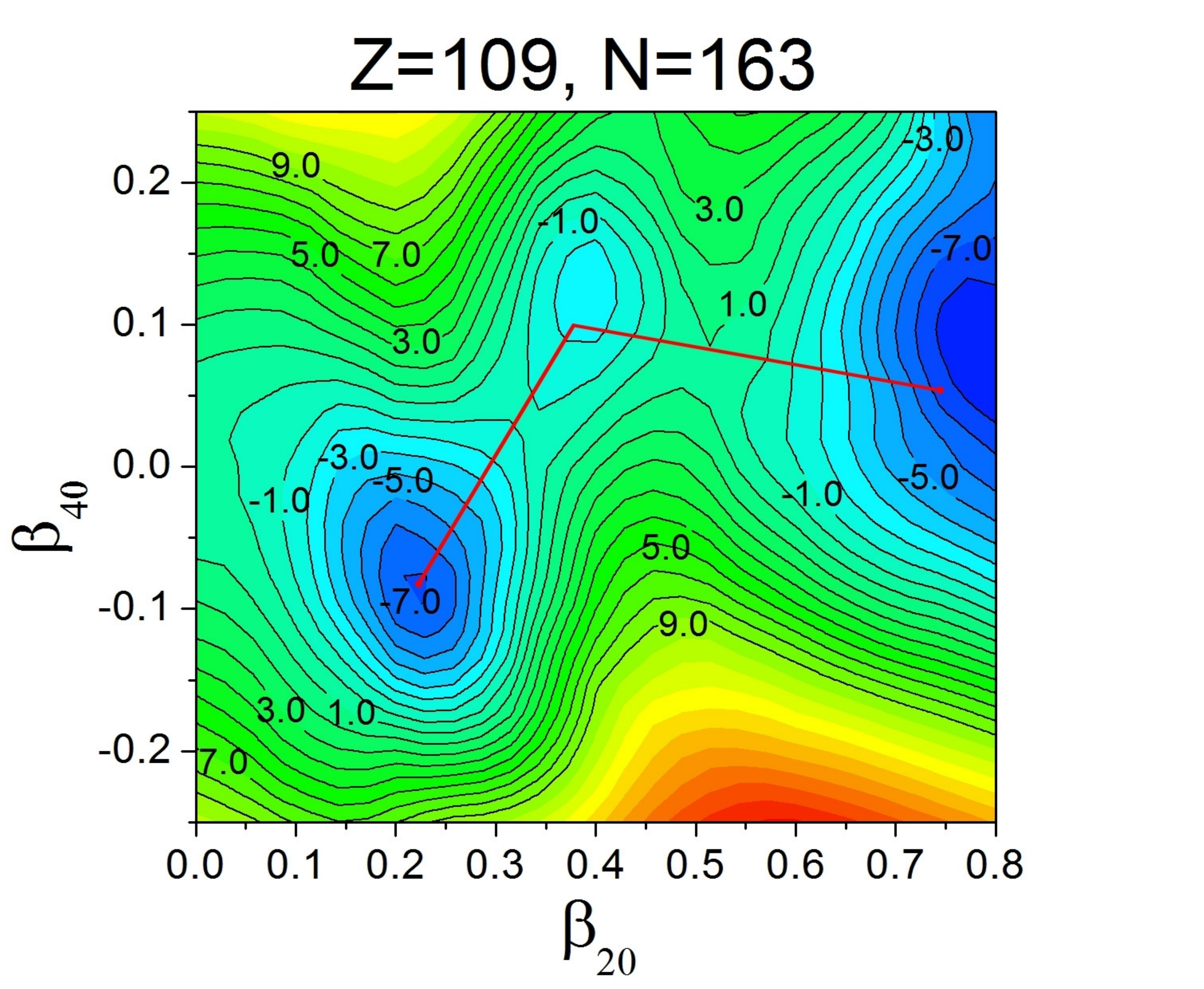}
	\caption{Energy surface of $^{272}$Mt; a chosen trajectory 
        coloured in red.}
	\label{fig:mapaadiabmt} 	
\end{figure}

\begin{figure}[h]
	\centering
	\includegraphics[angle=-90, width=0.60\textwidth]{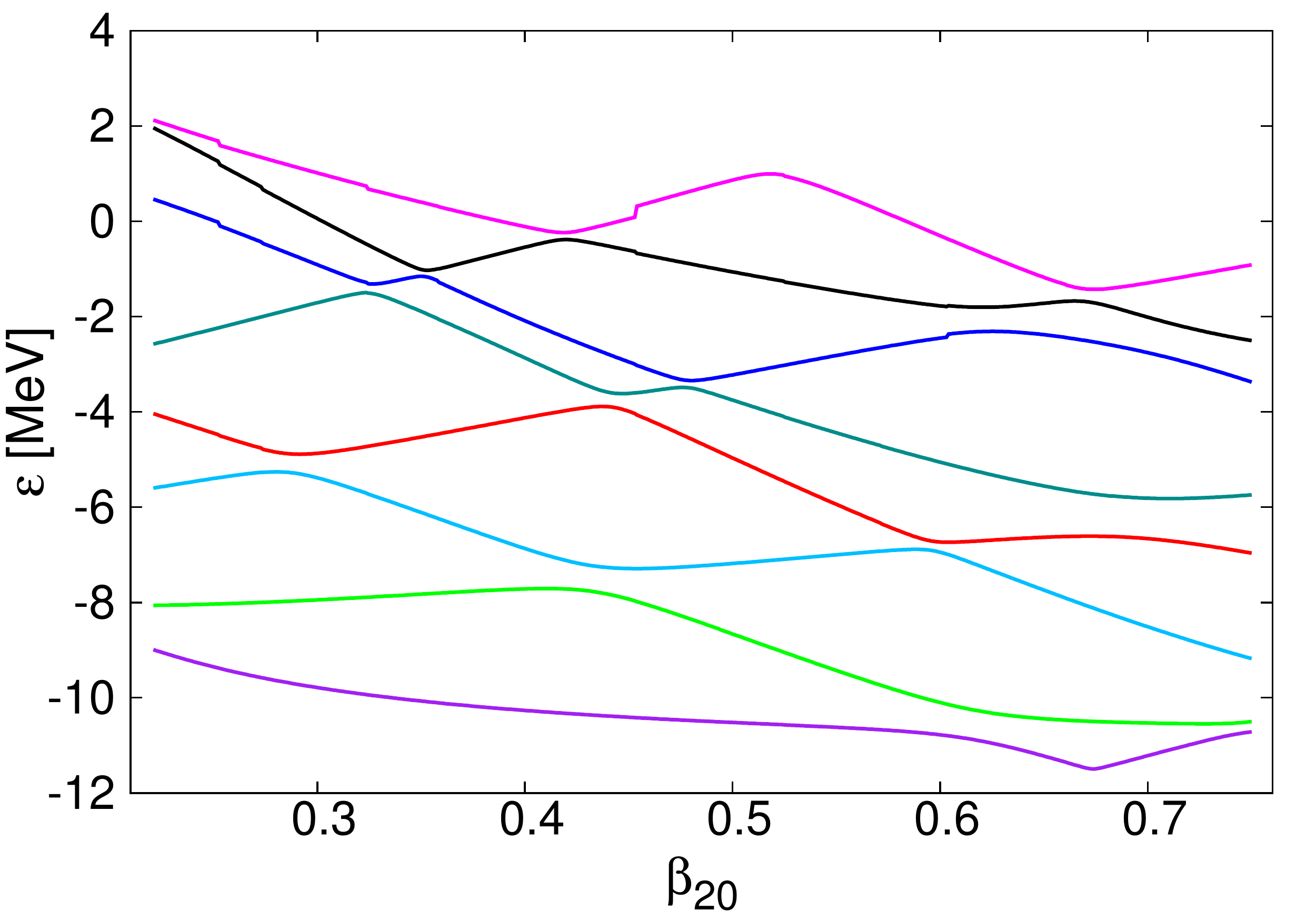}
	\caption{Neutron levels $\Omega^{\pi}=1/2^+$ around the Fermi level of 
 $^{272}$Mt along the trajectory shown in Fig. \ref{fig:mapaadiabmt}.}
	\label{fig:enerbet1-2plu}
\end{figure}

\begin{figure}[h]
	\centering
  \includegraphics[angle=-90, width=0.60\textwidth]{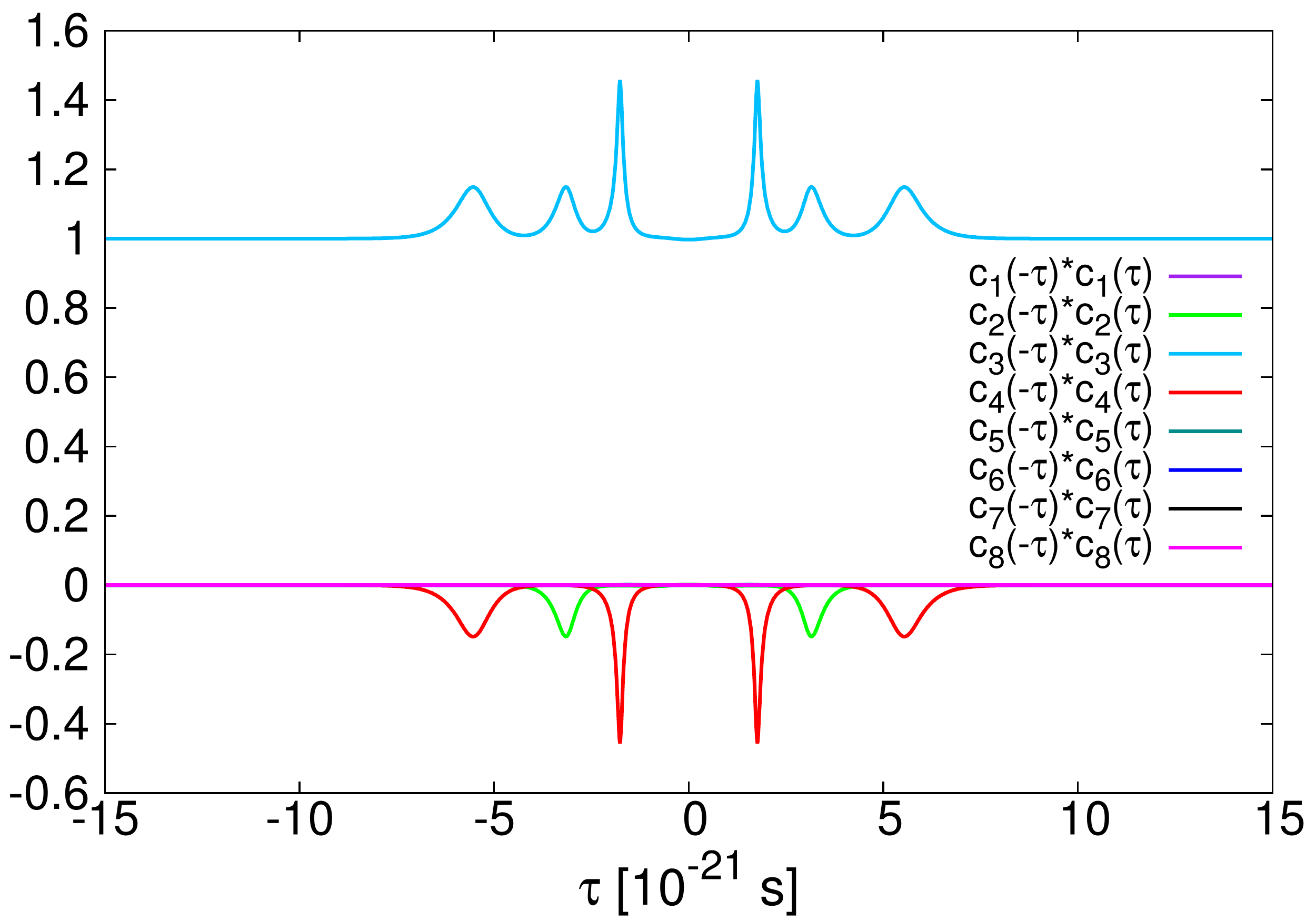}
  \includegraphics[angle=-90, width=0.60\textwidth]{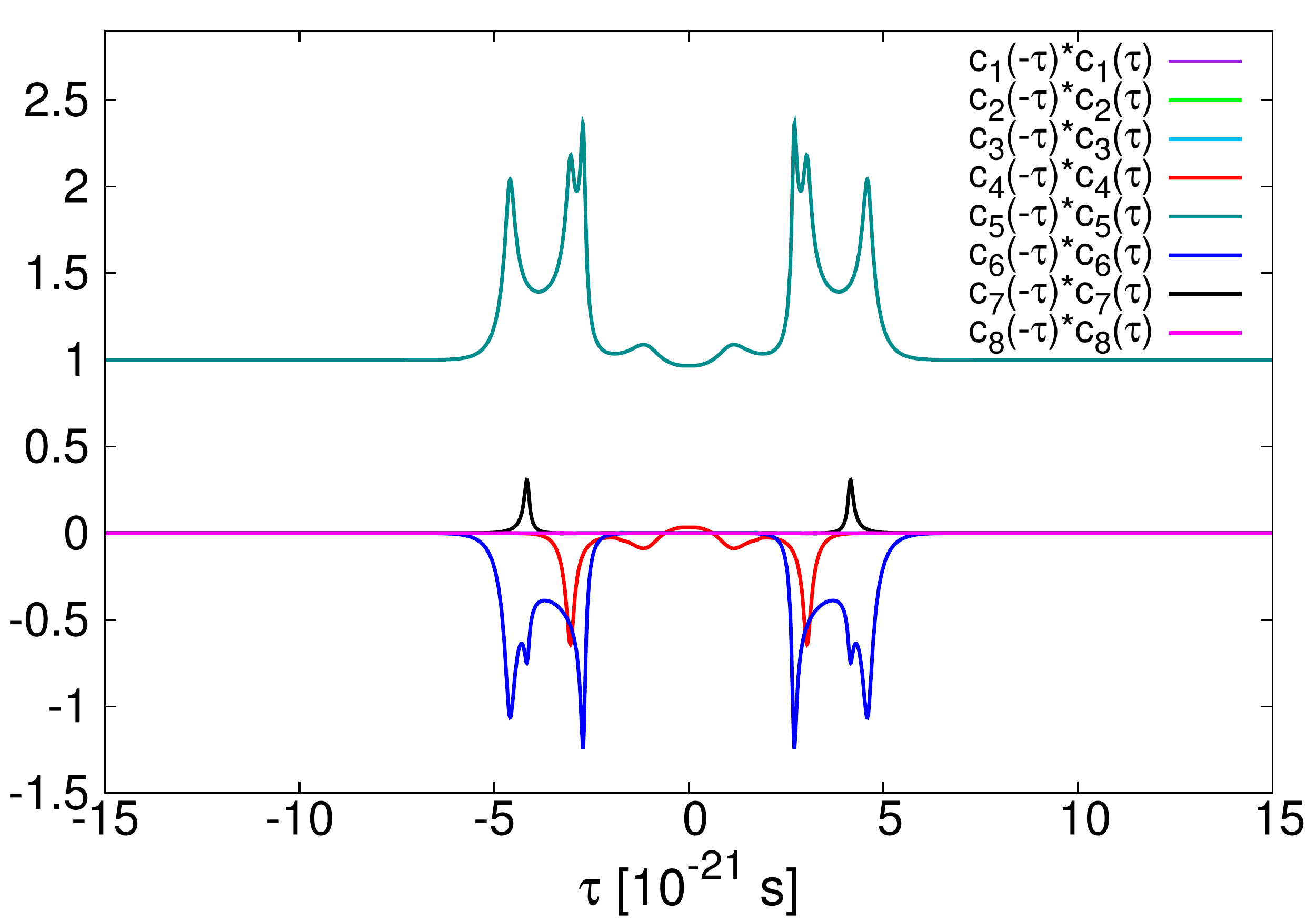}
	\caption{Pseudo-occupations of the adiabatic states for instanton - 
 like iTDSE solutions; \textit{upper panel:} for $\phi_3$, 
 \textit{lower panel:} for 
 $\phi_5$. Colours correspond to the levels of Fig. \ref{fig:enerbet1-2plu}. }
	\label{fig:obsadz8lev}
\end{figure}

 First we discuss the iTDSE solutions for axially-symmetric nuclear shapes 
 composed of multipoles with even $\lambda$.
 In this case the $\tau$-evolution of groups of states with different 
 $\Omega^{\pi}$ are indepedendent of each other. As an example we take 8 
 neutron $\Omega^{\pi}=1/2^+$ states in the W-S potential for 
 $^{272}$Mt along the axially symmetric fission path shown on the energy map in 
 Fig. \ref{fig:mapaadiabmt}. The map was obtained from the four-dimensional 
 (4D) calculation by minimizing energy of the lowest odd proton and neutron 
 configuration over $\beta_{60},\beta_{80}$ at each $\beta_{20},\beta_{40}$, 
 i.e. {\it without} keeping the $K^{\pi}$ configuration of the g.s. 
 Then, to assure 
 a continuity of the path, $\beta_{60}$ and $\beta_{80}$ were chosen continuous
  and close to those of the minimization, with energy changed by no more than 
 200-300 keV. 
 Collective velocity was calculated from Eq. (\ref{eq:qdotcoll}) by taking the 
 effective (i.e. tangent to the path) cranking mass parameter of the e-e 
 ($Z-1$,$N-1$) nucleus $^{270}$Hs.
 The adiabatic neutron levels in the basis for solving iTDSE were chosen so, 
 that in the g.s. the lower four are occupied (the fourth one singly) and 
 the upper four are empty. 
 In Fig. \ref{fig:enerbet1-2plu}, they are shown along $\beta_{20}$ which,  
 here and in the following, will play a role of the effective collective 
 coordinate $q$ along fission paths.

 The method which we used for solving the iTDSE in this and all other cases 
 reported here is described in Appendix  \ref{app:solu}. 
 We find solutions for a finite period $T$ in a finite adiabatic basis and  
 for each of them we calculate action. A natural question then is what would 
 be the limiting values of $S_i$ for occupied states when 
 $T\rightarrow \infty$ and 
 the dimension of the basis ${\cal N} \rightarrow \infty$. We tried to answer 
 this by finding actions for increased periods, and by incresing dimension of 
 the adiabatic basis and inspecting the quasi-occupation coefficients.  
 Results of such tests showed that with moderately long periods and 
 rather small bases one can obtain reasonably stable action values for 
 occupied states - see Appendix \ref{app:stab}.

 For the discussed eight levels in $^{272}$Mt, the iTDSE solutions were 
 obtained with the period $T=30 \times 10^{-21}$ s. 
 The amplitudes $C_{\mu i}(\tau)$ of solutions have exponential 
 $\tau$-dependence, 
 reach very large values in the interval $[-T/2,0]$ and very small in 
 $[0,T/2]$. It is more informative to characterize solutions by 
 quasi-occupations $p_{\mu i}$ of adiabatic states for selected solutions. 
 This also makes sense from the point of view of action (\ref{eq:stot}) which 
 is built of these quantities. In Fig. \ref{fig:obsadz8lev}, quasi-occupations 
 $p_{\mu i}$ are shown for two solutions, $\phi_3$ and $\phi_5$.  
 It can be seen that at $\tau=\pm T/2$, $p_{\mu i}\cong \delta_{\mu i}$, with 
 minuscule admixtures which should vanish completely for $T=\infty$. 
  During imaginary-time evolution, $p_{\mu i}$ 
 are concentrated on the corresponding adiabatic states $\psi_{\mu=i}$, except 
 around the pseudo-crossings where a partial excitation to the 
 nearest-neighbour state occurs. Until a pseudo-crossing is isolated 
 (there is no other pseudo-crossing nearby) excitations to other states are 
 negligible. 
 If successive pseudo-crossings follow one after another, the quasi-occupations of other adiabatic levels are possible, as seen for the solution 
 $\phi_{5}$ which locally becomes a combination of $\psi_{6}$ and $\psi_{7}$, 
 and then of $\psi_{4}$ and $\psi_{6}$ - see Fig. \ref{fig:obsadz8lev}. 


  Next we discuss some properties of iTDSE solutions which seem relevant 
  for their physical interpretation and applications.

\subsection{Rise of action with the collective velocity $\dot{q}$}
\label{sec:Svsqdot}

\begin{figure}[!h]
	\centering
	\includegraphics[angle=-90, width=0.62\textwidth]{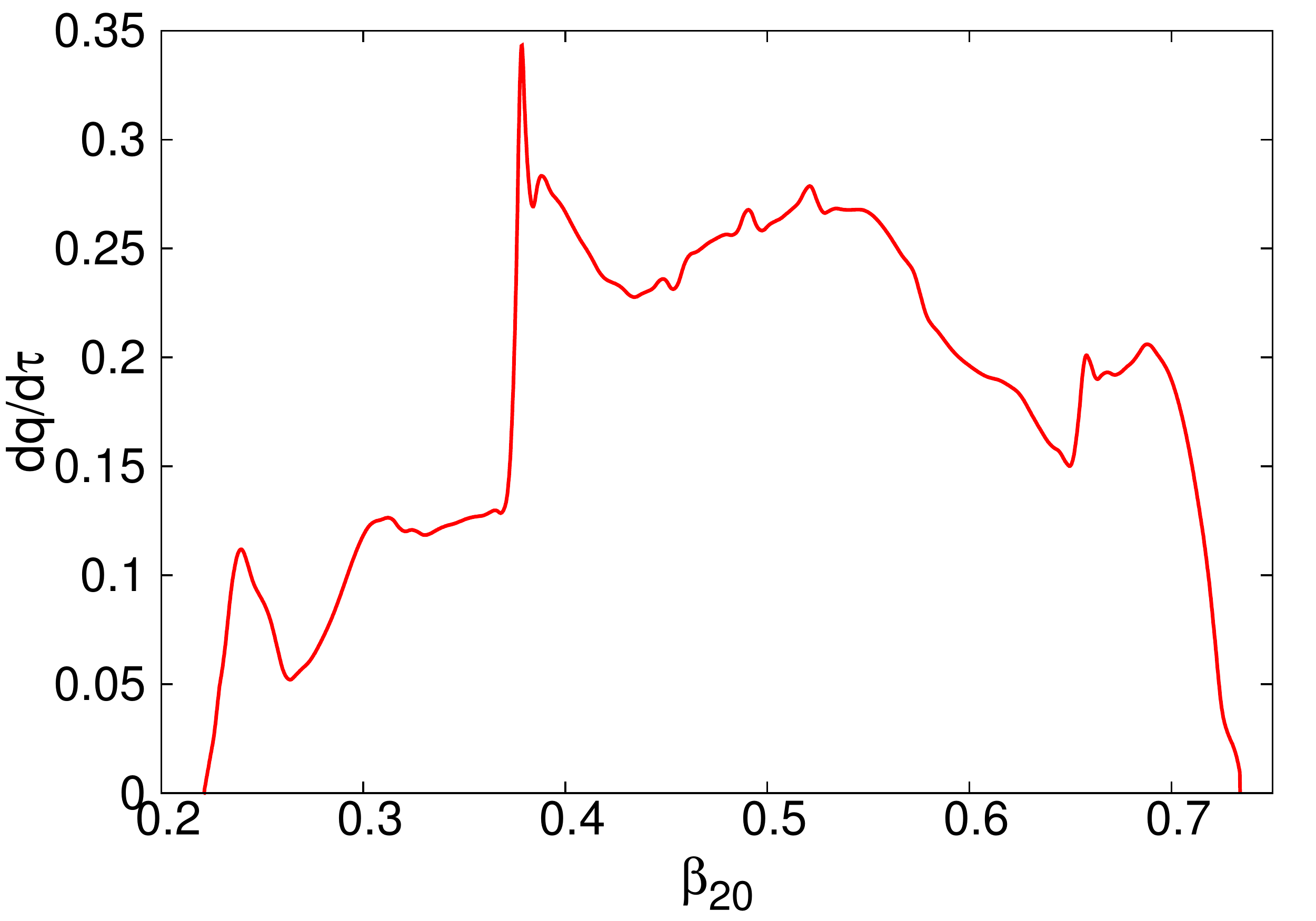} 
	\caption{Collective velocity ${\dot q}$ in units od $10^{21}$ s$^{-1}$
 calculated from (\ref{eq:qdotcoll}) for $^{272}$Mt along the path shown 
 in Fig. \ref{fig:mapaadiabmt}.} 
	\label{fig:qdot-axadiab}  
\end{figure}

 With cranking mass parameters fixed along a path, the collective 
 velocity of tunneling is proportional to $\sqrt{E(q)-E_0}$, where $E(q)-E_0$
 is a plot of the fission barrier (reduced by $E_{zp}$). 
 In a selfconsistent instanton calculation,  
 the increase in barrier height also relates to an increase in the magnitude 
 of ${\dot q}$ necessary to increase the difference between 
 $|\Phi(\tau)\rangle$ and $|\Phi(-\tau)\rangle$ in order to keep their energy 
 overlap ${\cal H}(\tau)$ constant. On the other hand, in our 
 non-selfconsistent treatment, 
 ${\dot \beta}_{20}$, i.e. our ${\dot q}(\tau)$, is simply an assumed 
 functional parameter of the solution to Eq. (\ref{equat}). However, 
 having in mind its implied physical relation 
 to the barrier height, we tested the action dependence on 
 $|{\dot \beta}_{20}|$. The collective velocity for $^{272}$Mt 
 determined from (\ref{eq:qdotcoll}) with the cranking mass parameter from the 
 neighbouring e-e nucleus (Z=108, N=168) along the path 
 depicted in Fig.  \ref{fig:mapaadiabmt} is shown in 
 Fig. \ref{fig:qdot-axadiab}. This profile was then scaled by the factors 1.3 
 and 1.6. The action calculated for all occupied neutron states of positive 
 parity for three collective velocities is given in Table 
 \ref{tab:stotvsqdot}. 
\vspace{3mm}
\vspace{-2mm}
 One can see that action indeed increases with $|{\dot q}|$, as the expected 
 relation with the barrier height would suggest. 
 Detailed outcome is dependent on the s.p. level scheme, in 
 particular, pseudocrossings close to the Fermi level.  
 In Eq. (\ref{equat}), the coupling terms causing non-adiabatic 
 transitions are $\dot{q}\langle\psi_i|\partial_q\psi_j\rangle$, so 
 the main influence on $S$ have regions in $q$ where a large $|\dot{q}|$ 
 occurs at a sharp pseudocrossing. 
\begin{table}[h]
	\renewcommand{\arraystretch}{1.3}
	\setlength{\arrayrulewidth}{0.9pt}
	\begin{center}
		\begin{tabular}{c|c}
			\hline\hline
Collective velocity & $S_{tot}=\sum_{\Omega^+}S_{\Omega^+} $ $[\hbar]$ \\
			\hline\hline
			$\dot{q}$ & 21.3465  \\
			$1.3\,\dot{q}$ & 24.6362   \\
			$1.6\,\dot{q}$ & 28.6790  \\
			\hline
		\end{tabular}
	\end{center}
	\caption{Action $S_{tot}$ for neutron states of positive parity 
 in $^{272}$Mt as a function of scaled collective velocity. 
 The profile $\dot{q}$ corresponds to the formula (\ref{eq:qdotcoll}) 
 for the path in Fig. \ref{fig:mapaadiabmt}. }
	\label{tab:stotvsqdot} 
\end{table}

\subsection{Integrand of action vs mass parameters}
\label{sec:instaBpar}

\begin{figure}
	\centering
	\includegraphics[angle=-90, width=0.48\textwidth]{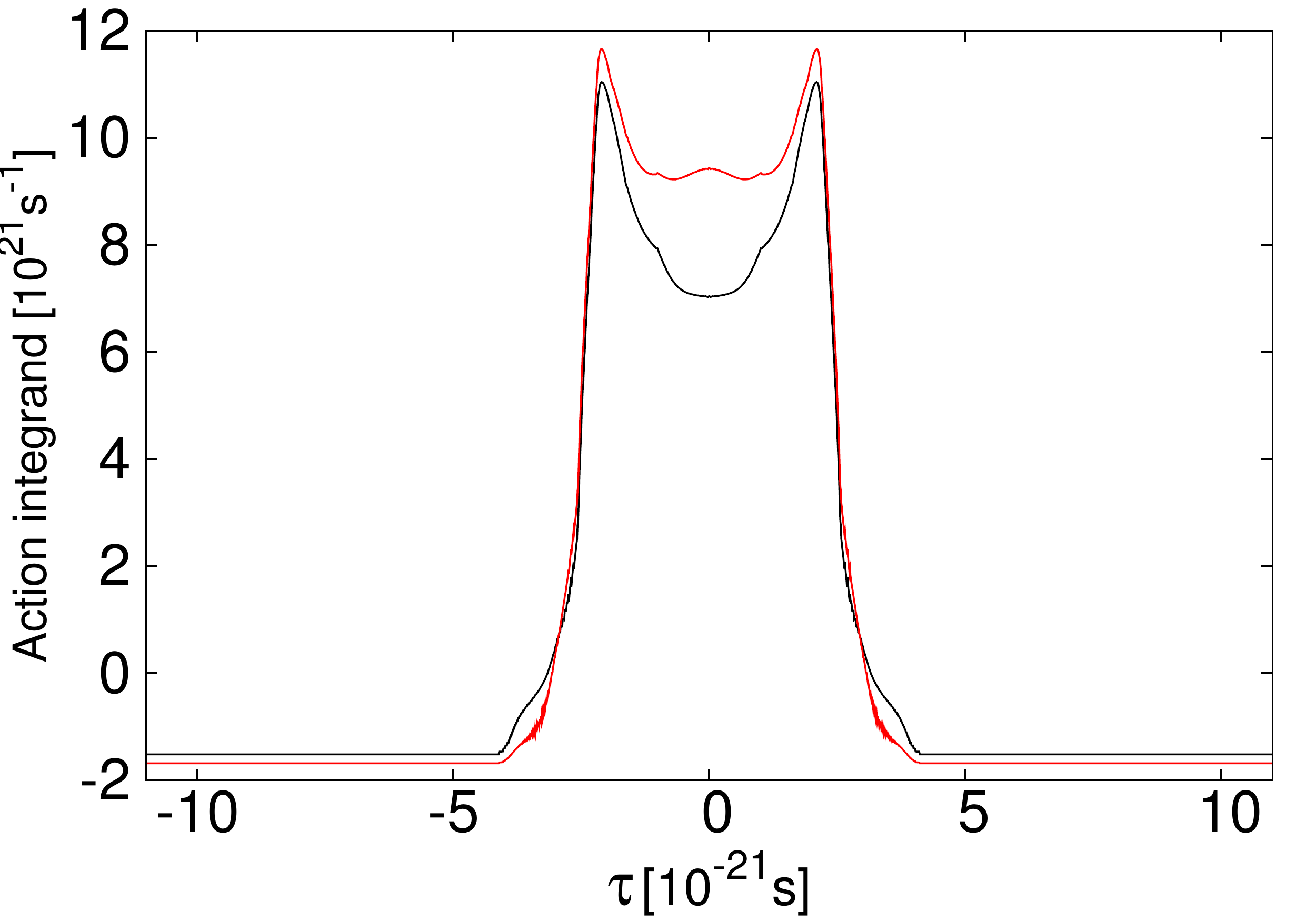}
	\includegraphics[angle=-90, width=0.48\textwidth]{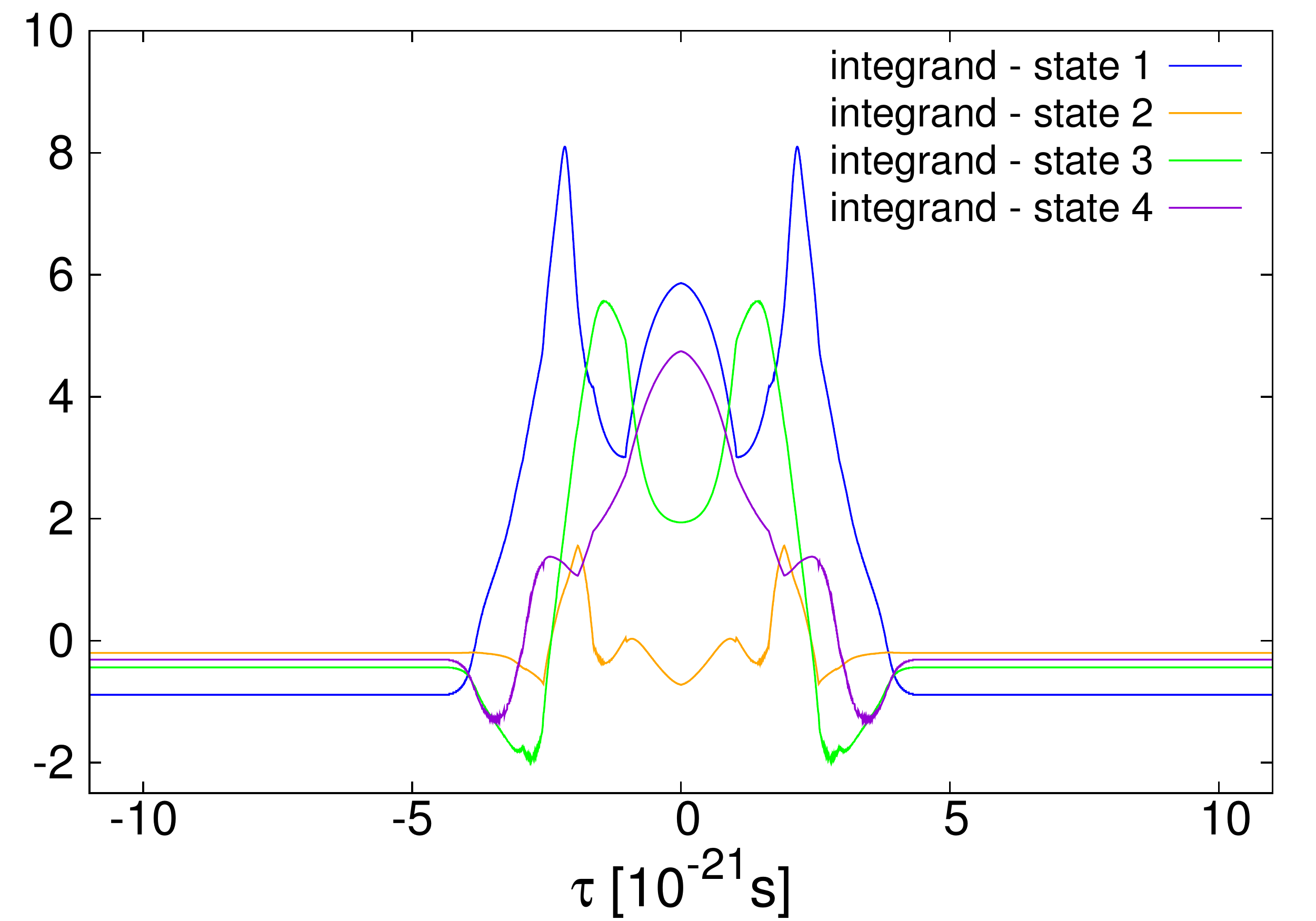} 
	\caption{ \textit{Left:} 
 The total action integrand in units $10^{21}$ s$^{-1}$ - the sum of individual
 contributions - for six (in black) and seven (in red) neutrons - 
 taken from \cite{Piaski}.
 \textit{Right:} Contributions to the integrand of action 
 from individual s.p. solutions.} 
	\label{fig:integrands}
\end{figure}

 One can ask whether it would be possible to define a mass parameter $B(q)$ 
 from the $\tau$- even action integrand in Eq. (\ref{eq:stot}) by: 
\begin{equation}
\label{Sintegrand}
\sum \limits_{i,occ}\sum \limits_{\mu=1}^{\cal N} [\zeta_i-\epsilon_{\mu}
(q(\tau))] p_{\mu i}(\tau)=B_{q q}(q)\dot{q}^2 .
\end{equation} 
 In Fig. \ref{fig:integrands} are shown contributions to the integrand of 
 action from s.p. bounce-like states and their sum for even and odd 
 number of particles (\ref{eq:stot}). Calculations were done for the same 
 $\Omega^{\pi}=1/2^+$ neutron states in $^{272}$Mt, for a path shown in 
 Fig. \ref{fig:mapaadiabmt}. 
 It can be seen that while integrands of single iTDSE solutions sometimes show 
 a rather complicated pattern, their sum is much more regular. 
 This comes from a cancellation of excitations among solutions corresponding 
 to occupied levels and only excitations to levels above the Fermi level count. 
 There is no drastic difference between the even- and odd-particle-number 
 case - it is just 
 a contribution from one singly occupied instanton-like solution, which may be 
 both positive or negative in general. This is in contrast to the cranking 
 approach, where for the odd-$A$ case, mass parameter (\ref{cranking}) and 
 the action integrand (\ref{scrank}) would show large peaks at 
  pseudocrossings of the unpaired level. 

 As seen in Fig. \ref{fig:integrands}, the integrand (\ref{Sintegrand})
 becomes negative around the endpoints $\tau\rightarrow \pm T/2$, 
 so it cannot define any mass parameter. This follows from 
  differences between the Floquet exponents $\zeta_{i}$ and s.p. energies 
 $\epsilon_i$ at the g.s. minimum, which, as stated in Sect.   
\ref{sec:metinst2}, is the artefact of using $T<\infty$ in practical 
 calculation. The same difficulty will probably remain in the selfconsitent 
  calculations. 
 
 However, even for a positive integrand of action there would be a more general 
 impediment to deriving the mass parameter. 
 The beyond-cranking treatment means that the integrand of action 
 depends on all even powers of $\dot{q}$. Thus, 
 for a given path, $B_{q q}$ of (\ref{Sintegrand}) would be dependent on 
 $|{\dot q}|$. 
   On the other hand, since a solution along the prescribed path depends on it, 
 two different paths tangent at a common point ${\bf q}$  
   (what would imply eqal effective cranking mass 
 parameters at ${\bf q}$) would have generally different integrands 
 of action at ${\bf q}$.

\subsection{Calculations along nonaxial path for neutron states in $^{272}$Mt}
\label{sec:instnonax}

\begin{figure}
	\centering
	\begin{minipage}{.55\textwidth}
		\centering
	    \includegraphics[width=1.0\textwidth]{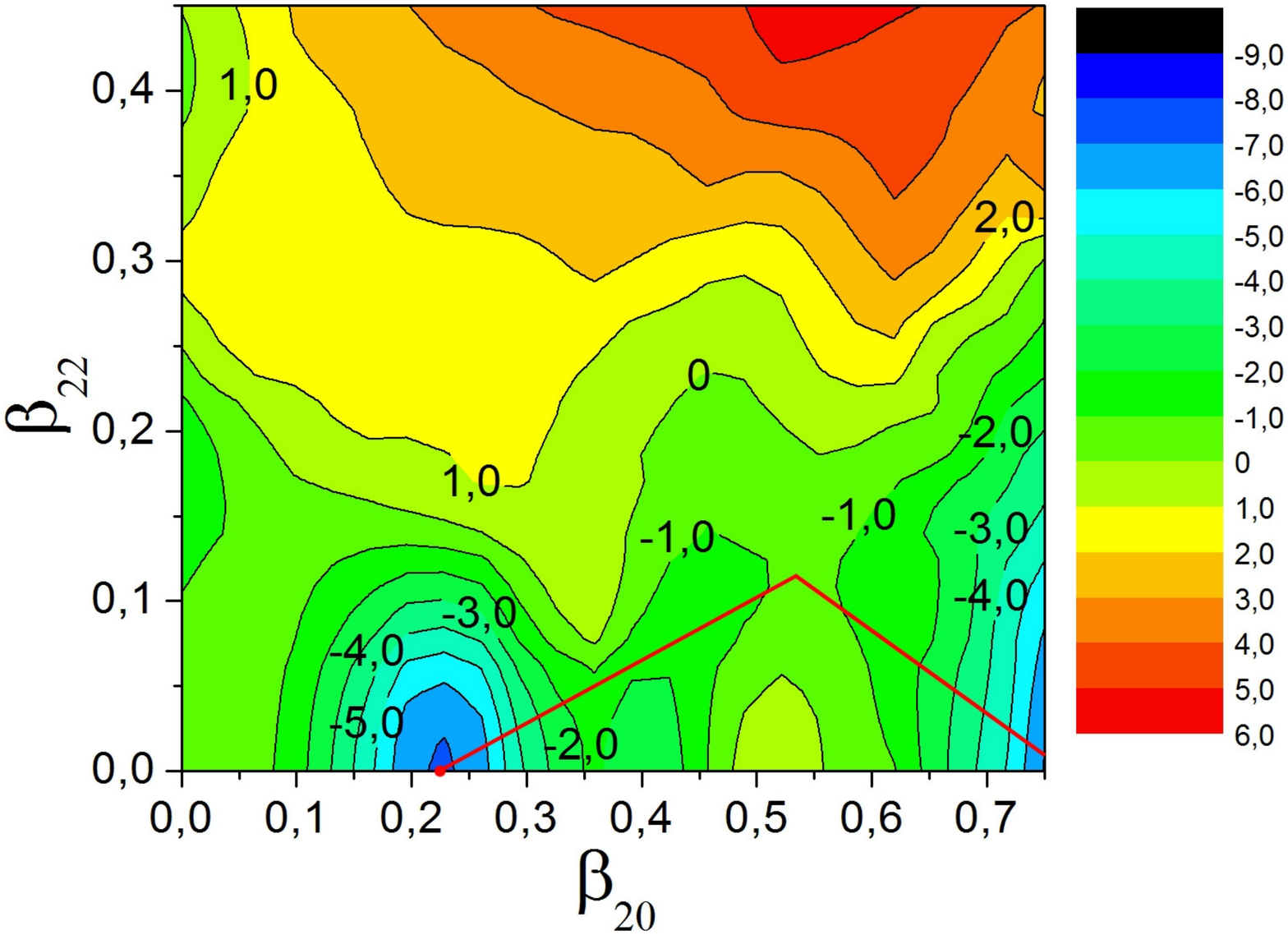}
    \end{minipage}%
    \begin{minipage}{.45\textwidth}
    	\centering
	    \includegraphics[angle=-90, width=1.0\textwidth]{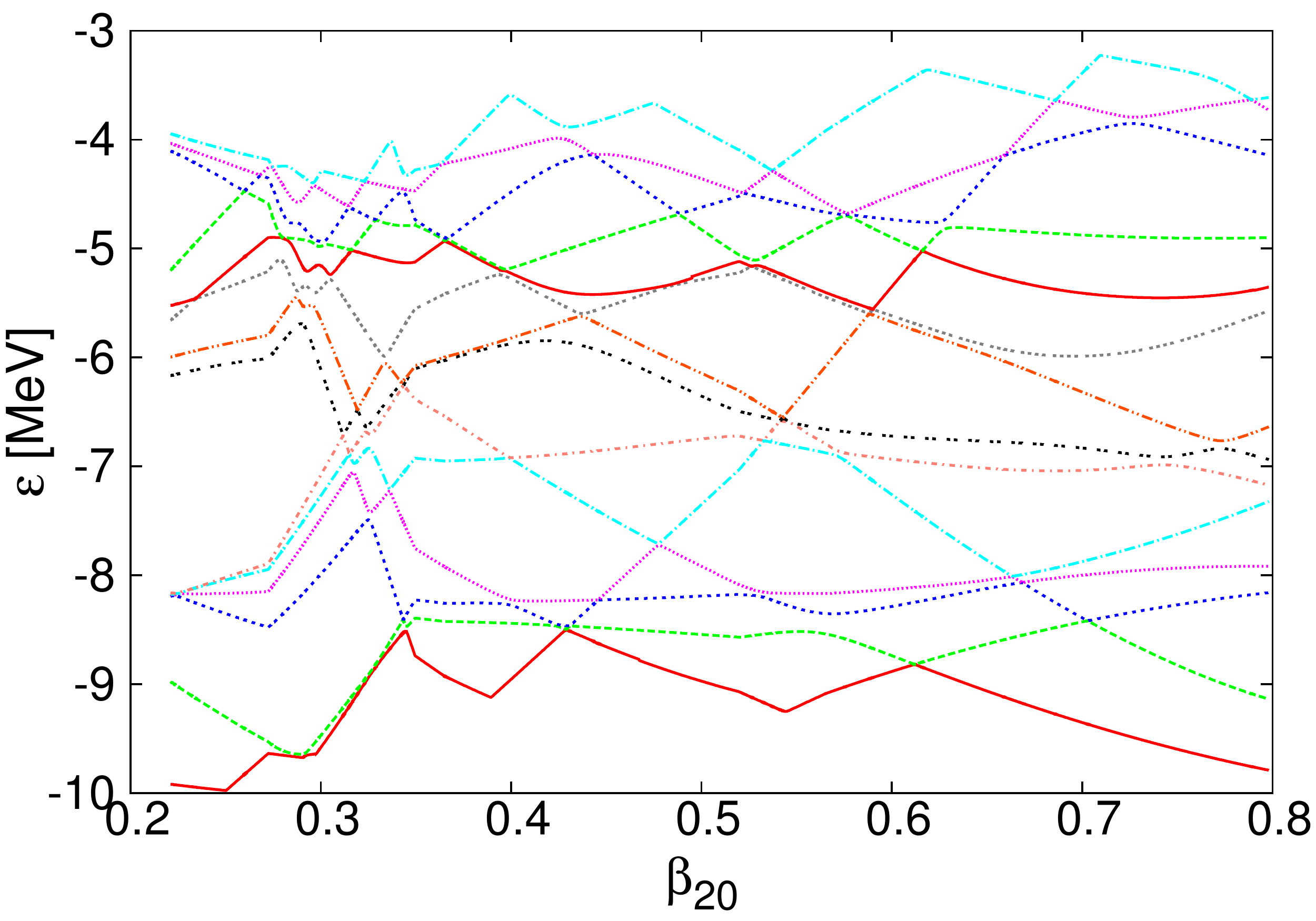}
    \end{minipage}	
	\caption{\textit{Left:} Energy landscape for $^{272}$Mt in 
 $\beta_{20}-\beta_{22}$, minimized over $\beta_{40},\beta_{60},\beta_{80}$
 with a chosen fission path (marked in red). \textit{Right:} 
 Display of 14 positive-parity neutron levels around the Fermi energy along 
 the fission path; the 7-th level from below is the last occupied. }
	\label{fig:mt-nonax}
\end{figure}

 A solution of iTDSE equations for nonaxial shapes turns out to be 
 more difficult than in the case of axial deformations considered hitherto.
 The W-S spectrum along a nonaxial fission path shows many 
 sharp pseudocrossings between levels of the same parity, some with interaction
  as small as $V\sim 10^{-5}-10^{-6}$ MeV (see Fig. \ref{fig:mt-nonax}). 
 Although for $V\rightarrow 0$ such levels would cross, the results 
 for the two-level model have shown (Sec. \ref{sec:2levmod}) that 
 this limit is subtle and depends also on the collective velocity and 
 the slopes of crossing levels. It happens that diabatic continuation, 
 i.e. assuming $V\approx 0$, may lead to large errors  in calculated action.  
 On the other hand, many pseudo-crossings with a very weak 
 interaction, leading to extremely high peaks in the matrix elements 
 which couple involved adiabatic states, are the obstacle in solving iTDSE. 
   The encountered problem and its (rather cumbersome) solution are 
  described below.  
 
 Calculations were performed along the chosen nonaxial path for $^{272}$Mt, 
 see Fig. \ref{fig:mt-nonax}, for ${\cal N}=32$ neutron states of positive 
 parity. In the first version, we used the data from the W-S code 
 along the path with a variable step, not shorter than 
 $\Delta\beta_{20}=10^{-6}$. In the second version, the minimal step was 
 smaller, $\Delta\beta_{20}=10^{-7}$. Finally, in the third version, 
 we used the procedure described in the Appendix \ref{app:nonax}, with the 
 minimal step $\Delta\beta_{20}=10^{-7}$, and the analytic model 
 (\ref{eq:uogcouplfit}) adjusted to those peaks for which the minimal stepsize 
 still did not cover their range with a sufficient precision.  
  Actions calculated for occupied instanton levels and their sum are 
 given in Table \ref{tab:actnonax}. It can be seen that actions for 
 some individual levels in the first and second versions of the calculation 
 differ widely - this means that the step $\Delta\beta_{20}=10^{-6}$ is not 
 sufficient. This is consistent with an insufficient density of points 
 for a description of particular pseudocrossings, as revealed by the inspection 
 of related coupling matrix elements.  
 In spite of this, the total action is similar in two versions of calculation. 
 This is yet another sign that action depends on pseudo - crossings 
 close to the Fermi levels - the details of crossings far above or below 
 the Fermi energy (between both occupied or both unoccupied levels) 
  do not have effect on total action.

 In the third version of the calculation, the highest peaks in the 
 coupling matrix elements were replaced by the peaks modelled analytically  
 (\ref{eq:uogcouplfit}). 
 Actions obtained within this method (in the third column in Table 
 \ref{tab:actnonax}), both for individual solutions and the total, are 
 close to those of the second version. This is probably related to the fact  
 that difficult couplings that were modelled occur at such $q$, where 
  $\dot{q}\approx 0$, so that they were suppresed in the instanton equations  
  (\ref{equat}). In general, however, the procedure of peaks modelling 
  seems indispensable for obtaining sufficiently exact actions if the instanton
  equations are to be solved in the adiabatic basis (in particular, when a very 
 large nonadiabatic coupling occurs close to the Fermi energy). 

\begin{table}
	\renewcommand{\arraystretch}{1.3}
	\setlength{\arrayrulewidth}{0.9pt}
	\begin{center}
		\begin{tabular}{c|ccc}
			\hline\hline
			Nr &	$\Delta\beta_{20}=10^{-6}$ & $\Delta\beta_{20}=10^{-7}$ & $\Delta\beta_{20}=10^{-7}$ plus fit \\
			\hline \hline		 
			1 &	3.2143 &	3.2057 & 3.1936 \\
			2 & 0.9453 & 	8.0320 & 8.0555 \\
			3 &	3.2931 & 	6.9294 & 6.9118 \\
			4 &	 3.2790 & 	-8.7864 & -8.7867 \\
			5 &	-0.0346 & 	2.1493 & 2.1684 \\
			6 &	-1.7771 &	-2.3285 & -2.3531 \\
			7 &	0.9953 &	1.1126 & 1.1129 \\
			8 &	8.8511 &	9.1817 & 9.1458 \\
			9 &	4.1217 & 	-1.3617 & -1.4455 \\
			10 & 5.5588 & 	9.6487 & 9.8299 \\
			11 &	-2.9214 &	-2.3793 & -2.3817 \\
			12 &	-4.5752 &	-4.5158 & -4.5660 \\
			13 &	-0.4160	& -0.3668 & -0.3788 \\
			14 &	6.7950 &	6.4864 & 6.4848 \\
			15 &	6.6443 &	6.4057 & 6.4033 \\
			16 &	2.8743 &	2.8123 & 2.8128 \\
			\hline
			$\mathbf{S_{tot}/\hbar}$ & \textbf{36.8479} & \textbf{36.2254} & \textbf{36.2069}
			\\
			\hline                 
		\end{tabular}
	\end{center}
	\caption{Actions for separate s.p. solutions occupied at the g.s. 
 and their sum - the total action for a nonaxial path;   
 \textit{first column:} calculations with the minimal 
 step $\Delta\beta_{20}=10^{-6}$; \textit{second column:} calculations with 
 the minimal step $\Delta\beta_{20}=10^{-7}$; \textit{third column:} 
 calculations with the minimal step $\Delta\beta_{20}=10^{-7}$ augmented 
 with the modelling of the highest peaks in the nonadiabatic couplings 
 by the formula (\ref{eq:uogcouplfit}).} 
	\label{tab:actnonax}  
\end{table} 

 We also checked the dependence of action on the dimension 
 ${\cal N}$ of the adiabatic basis. We changed ${\cal N}$ from 14 to 32, 
  always keeping the Fermi level at ${\cal N}/2$   
 (as in Appendix \ref{sec:SvsN} for the axially symmetric path). The results  
 given in Tab. \ref{tab:stotvsnnonax} indicate that the dominant  
 contribution to action comes from levels around the Fermi level.  

  Action obtained for the trajectory along nonaxial shapes was compared 
 to the one along the axially symmetric path (shown in 
 Fig. \ref{fig:mapaadiabmt}) in Table \ref{tab:axvsnonax}.  
 In both cases the same neutron levels with positive parity were included.  
 It can be seen that action along the shorter, axially symmetric path 
 is smaller in spite of the fact that the barrier is lower by $\sim 2$ MeV 
 along the nonaxial path, what in our treatment translates into a smaller 
 collective velocity ${\dot q}$. 

  It has to be emphasized that the last result cannot be treated as 
 general - it merely shows that the instanton method applied to 
 reasonably chosen paths can lead to situations similar as in 
 calculations with the cranking mass parameters.  
  Deciding whether axial or nonaxial path prevails would require 
 a minimization procedure not defined here. 
\begin{table}
	\renewcommand{\arraystretch}{1.3}
	\setlength{\arrayrulewidth}{0.9pt}
	\begin{center}
	\begin{tabular}{c|c}
		\hline \hline
		${\cal N}$ & $S_{tot}=\sum_{i=1}^{{\cal N}/2}S_{i} $ $[\hbar]$ \\
		\hline\hline
		16 & 27.0313 \\
		20 & 35.8289 \\
		24 & 35.9705 \\
		28 & 36.1187 \\
		32 & 36.2069 \\
		\hline	
	\end{tabular}
    \end{center}
\caption{Action (in $\hbar$) for neutrons of positive parity along the 
 nonaxial path for various numbers ${\cal N}$ of included adiabatic states.}  
\label{tab:stotvsnnonax} 
\end{table}

\begin{table}
	\renewcommand{\arraystretch}{1.3}
	\setlength{\arrayrulewidth}{0.9pt}
	\begin{center}
	\begin{tabular}{c|cc}
		\hline\hline
		Path & $B_{f}$ [MeV] & $S_{tot}/\hbar$ \\
		\hline \hline
		axial & 8.4 & 21.35 \\
		nonaxial & 6.5 & 36.21 \\
		\hline
	\end{tabular}
    \end{center}
\caption{ Fission barrier heights $B_f$ and actions $S_{tot}$ (in $\hbar$) for 
 neutrons of positive parity in $^{272}$Mt along the axial 
 (Fig. \ref{fig:mapaadiabmt}) and nonaxial (Fig. \ref{fig:mt-nonax}) 
 fission paths.  }
\label{tab:axvsnonax}
\end{table}

\section{Fission hindrance in odd nuclei - a study}

 Usually, the spontaneous fission hindrance factors $HF$ for odd nuclei are 
 defined as  
 $T_{sf}^{o}/T_{sf}^{ee}$, where $T_{sf}^{o}$ is the spontaneous fission 
 half-life of an odd nucleus and $T_{sf}^{ee}$ is a geometric mean of 
 the fission half-lives of its e-e neighbours \cite{Hess}. 
 Experimental facts are that 1) most of $HF$ values lie between $10^3$ to 
 $10^5$, 2) they do not display any strong dependence on the $K(=\Omega)$ 
 quantum number of the g.s. configuration \cite{Hess}.  

 Here, we will use $HF$ calculated as:   
\begin{equation}
HF=\frac{T_{sf}^{o}}{T_{sf}^{e}},
\label{HF}
\end{equation}
 where $T_{sf}^{o}$ i $T_{sf}^{e}$ are fission half-lives of an odd-A nucleus 
 and its $A-1$ e-e neighbour.  


 Experimental fission half-lives and odd-even $HF$s can be converted into 
 relations between actions for odd-$A$ and e-e neighbours by using the 
 $WKB$-motivated formula for spontaneous fission half-lives:
\begin{equation} 
\log_{10}(T_{sf}[s])=-20.54 + 0.8686\,\frac{S}{\hbar} - 
\log_{10}\left(\frac{E_{zp}}{0.5\:MeV}\right).
\label{eq:logtsf}
\end{equation}
 Here, $S$ is the minimal action chosen among all possible fission 
 paths, and $E_{zp}$ is the zero-point energy (in MeV) of vibration along the 
 fission direction around the m.s. Assuming {\it a universal 
 value} of $E_{zp}$, which is surely an approximation, one obtains: 
 \begin{equation} 
  \log_{10} (HF) \approx 0.8686 \ \frac{S_{odd} - S_{even}}{\hbar}  . 
\end{equation}

 Calculations were performend for selected superheavy nuclei with known 
 half-lives and, in some cases, known g.s. spin and parities, indicating 
 possible configurations. 
 A similar calculations for actinide nuclei would be much more involved in 
 view of their much longer and more complex barriers.

\subsection{Instanton-like action without pairing for $^{\mathbf{257}}$Rf, 
  $^{\mathbf{257}}$Rf}

  By solving iTDSE for a given path and collective velocity 
 profile ${\dot q}(\tau)$ one can calculate action for both even and odd nuclei,
 neglecting pairing. Such results would correspond to a scenario originally 
 put forward by Hill and Wheeler \cite{HillWhee}. Without pairing 
 they cannot be realistic, but allow to notice a few things, 
 among them how much fission would be hindered without pair correlations. 

 We choose the odd nucleus $^{257}$Rf as the example. Its $I^{\pi}=1/2^+$ g.s.,
 which well corresponds to the $K^{\pi}=\Omega^{\pi}=1/2^+$ configuration 
 in the W-S model, has a known spontaneous fission half-life of 
$T_{sf}^{odd}=423$ s \cite{Hess}. Also known is the experimental lower limit of 
 $T_{sf}^{odd}>490$ s \cite{Hess} for the half-life of the excited 
 $I^{\pi}=11/2^{-}$ state, corresponding 
 to the $K^{\pi}=11/2^-$ configuration in our micro-macro model. 
 The experimental spontaneous fission half-life for the e-e neighbour 
 $^{256}$Rf is $T_{sf}^{even}=6.4$ ms \cite{Hess}, which gives 
 $HF=6.6\times 10^4$ (for $K^{\pi}=1/2^+$ configuration) and 
 $HF>7.6\times 10^4$ (for $K^{\pi}=11/2^-$). 

\begin{figure}
	\centering
	\begin{minipage}{.5\textwidth}
	  \includegraphics[width=1.2\textwidth]{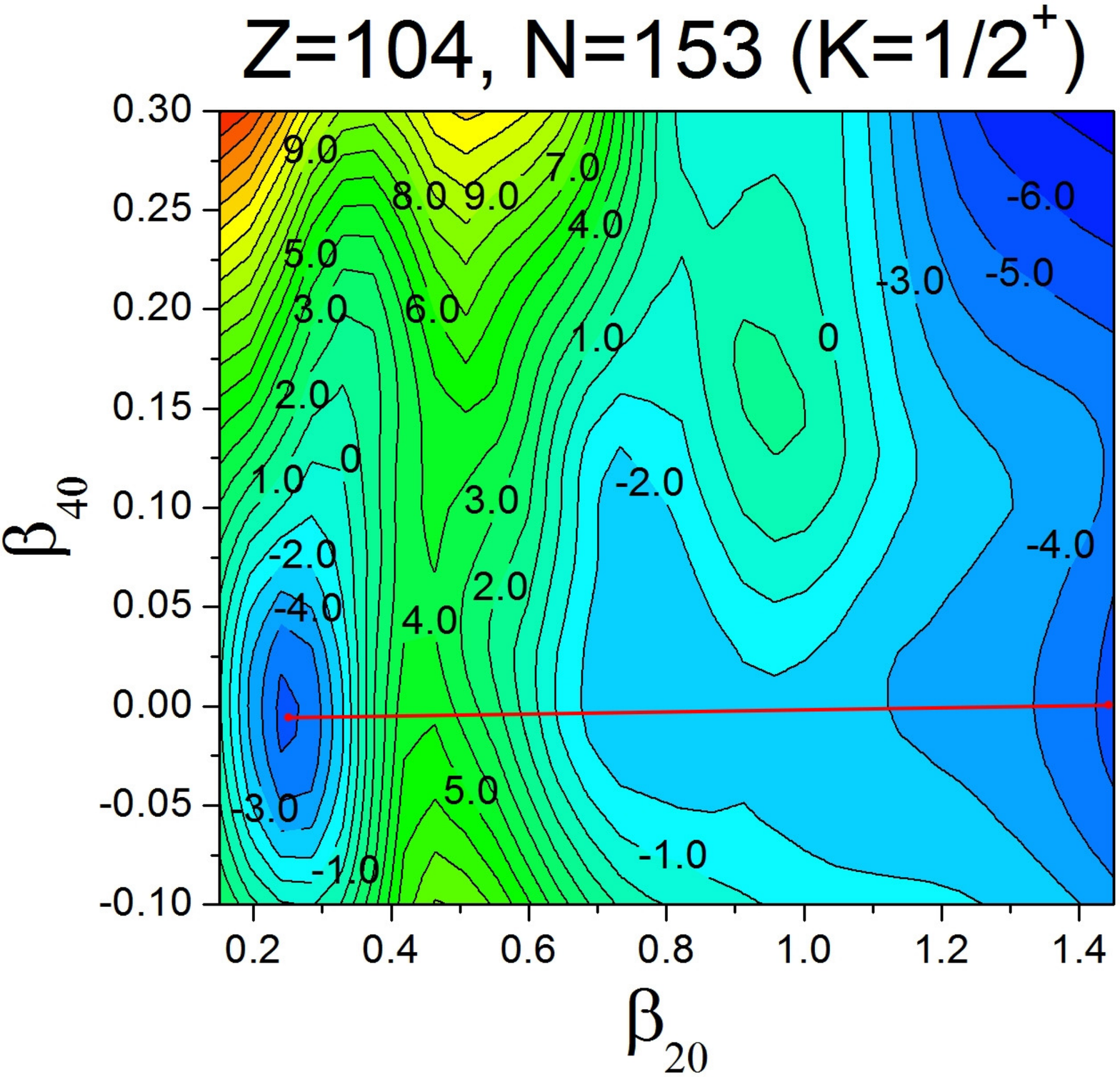}
    \end{minipage}%
    \begin{minipage}{.5\textwidth}
	  \includegraphics[width=1.2\textwidth]{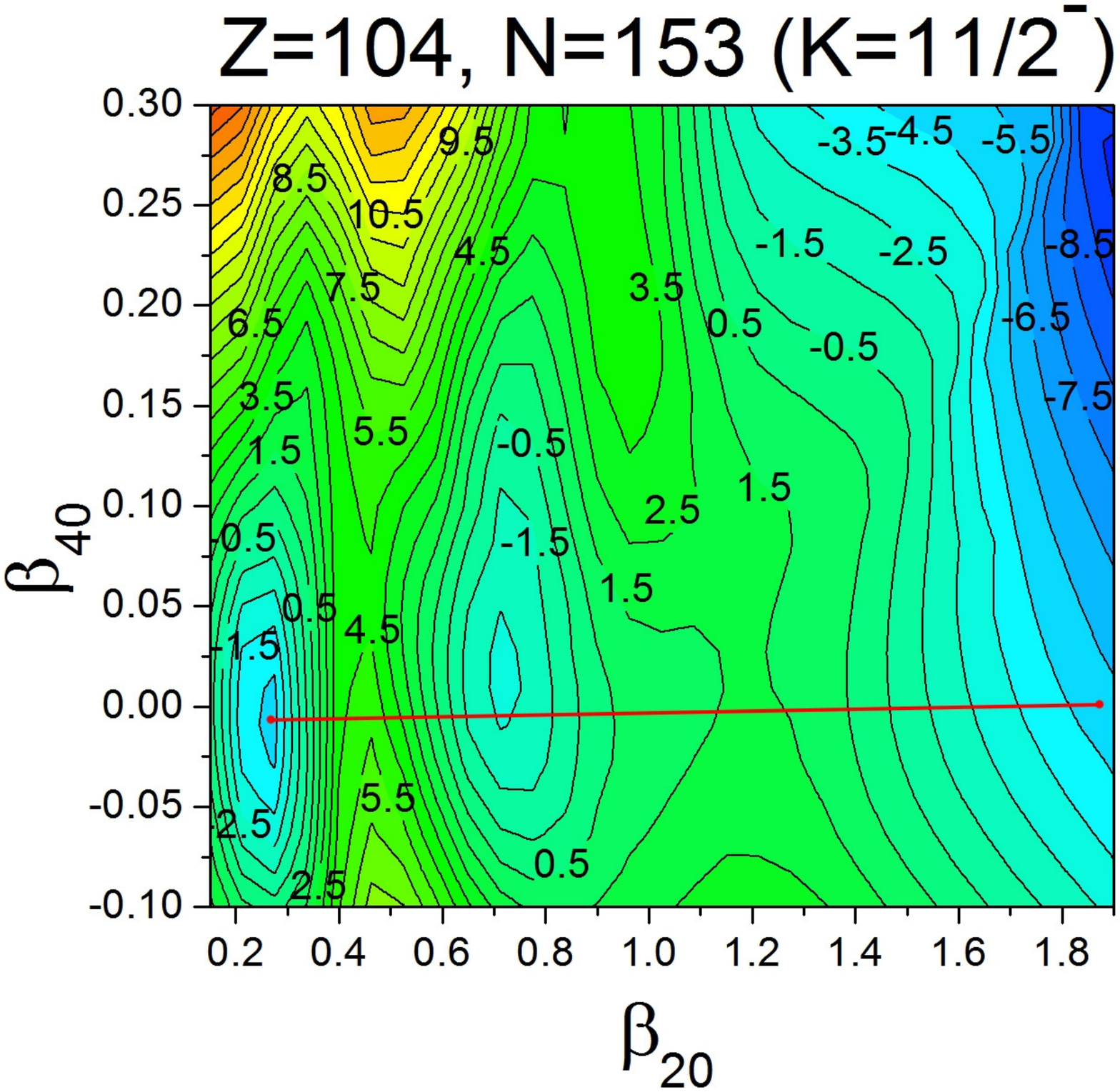}
    \end{minipage}
    \begin{minipage}{.5\textwidth}
	  \includegraphics[width=1.2\textwidth]{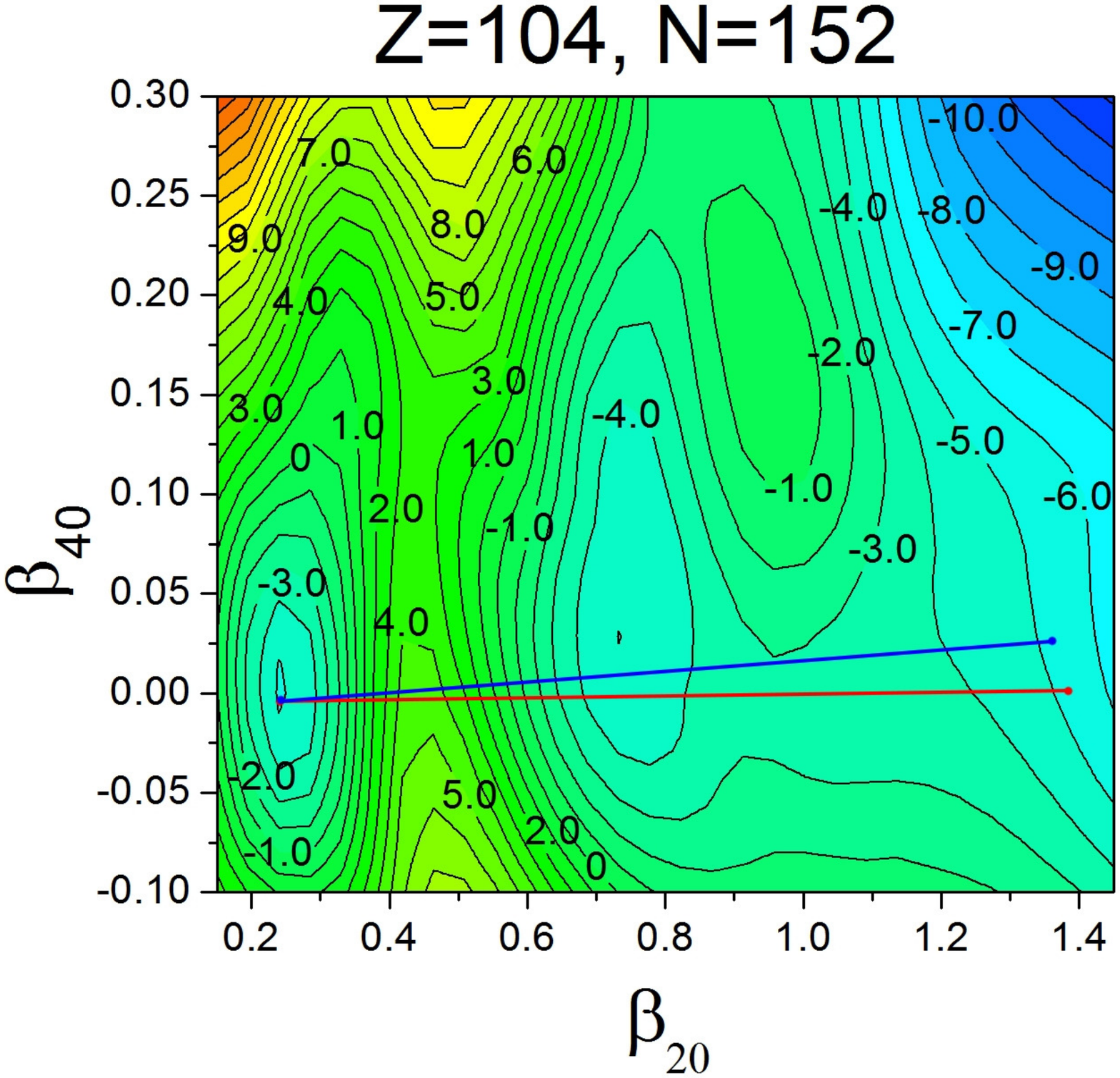}
    \end{minipage}
\caption{\textit{Upper panel:} energy landscapes for $^{257}$Rf minimized over 
 $\beta_{60},\beta_{80}$ with fixed $K^{\pi}=1/2^+$ 
(left) and $K^{\pi}=11/2^-$ (right) configuration. \textit{Lower panel:} 
 energy landscape for the neighbouring $^{256}$Rf. Chosen fission paths marked 
 in red. For e-e $^{256}$Rf also a second path (marked in blue) 
 was considered (see text). Note different ranges of $\beta_{20}$ in maps.}
\label{fig:Rf-mapy}	
\end{figure}

 The tunneling path was chosen as follows. First, micro-macro 
 energy landscapes of two nuclei were calculated by using mass-symmetric 
 axial deformations: for each $\beta_{20}-\beta_{40}$ energy was minimized 
 over $\beta_{60},\beta_{80}$, with the steps $\Delta\beta_{20}=0.05$ and 
 $\Delta\beta_{40}=0.025$. The odd nucleus configurations $K^{\pi}$ 
 were kept constrained at $K^{\pi}=1/2^+$ and $K^{\pi}=11/2^-$ for the g.s. 
 and the excited state, respectively. This means a continuation, possibly 
 non-adiabatic, of the state $\Omega^{\pi}$ occupied by the odd neutron at the 
 energy minimum. 
 A similar calculation, but without blocking, was performed for $^{256}$Rf. 
 It can be seen from the maps in Fig. \ref{fig:Rf-mapy} that 
 keeping the configuration in the odd nucleus leads to a substantial 
 increase and elongation of the barrier, especially for the excited 
 configuration $K^{\pi}=11/2^-$. 
 Taking into account the experience from action minimization calculations, 
 the fission path was chosen piecewise straight and close as possible to the 
 minimal energy, in order to keep the path short and the barrier low  
 (the path is also piecewise straight in $\beta_{60},\beta_{80}$). 
 It is depicted in red in Fig. \ref{fig:Rf-mapy} 

 Instanton-like action $S_{inst}$ was calculated by solving 
 iTDSE with the collective velocity:  
 ${\dot q}_P=\sqrt{2(E(q)-E_{m.s.})/B_P(q)}$, where $E(q)-E_{m.s.}$ 
 is deformation energy with respect to the m.s. for each 
 nucleus/configuration (i.e. with $E_{zp}$ set to zero), and $B_P(q)$ 
 is the cranking mass parameter of  
$^{256}$Rf, both {\it including pairing} and calculated along the chosen paths. 
 So, strictly speaking, ${\dot q}_P$ derives from the paired system, but 
 iTDSE is solved for the system without pairing. 
 For comparison, along the same paths we calculated actions: 
\begin{equation}
S_{cr}({\dot q}_P)=\int_{-T/2}^{T/2}d\tau\, B_{NP}(q)\,\dot{q}_P^2,
\label{eq:Sadnopair}
\end{equation}
 with the same  ${\dot q}_P(\tau)$ and the cranking mass parameter $B_{NP}(q)$  
 {\it without pairing} for each nucleus (i.e. also for  
 the odd one). The mass parameter $B_{NP}$ includes large peaks due to 
 close avoided level crossings which should considerably increase action 
 relative to $S_{inst}$. We can calculate action $S_{cr}({\dot q}_P)$ 
 accurately thanks to the large number of points - few thousands per path. 
 Both actions are given in Table \ref{tab:Rf-action}. 
\begin{table}
	\renewcommand{\arraystretch}{1.3}
	\setlength{\arrayrulewidth}{0.9pt}
	\begin{center}
		\begin{tabular}{l|cc|cc|cc}
			\hline
			\hline
		Nucleus ($K^{\pi}$)	& \multicolumn{2}{c|}{$^{257}$Rf  ($1/2^+$)} & \multicolumn{2}{c|}{$^{257}$Rf  ($11/2^-$)} & \multicolumn{2}{c}{$^{256}$Rf} \\
			\hline
		Action [$\hbar$]	& $S_{inst}$ & 
 $S_{cr}({\dot q}_P)$ & $S_{inst}$ & $S_{cr}({\dot q}_P)$ & $S_{inst}$ & 
 $S_{cr}({\dot q}_P)$ \\
			\hline
			\hline
			Neutrons ($+$) & 27.29 & 86.40 & 31.23 & 68.41 & 19.52 & 32.11 \\
			Neutrons ($-$) & 73.71 & 1378.97 & 82.06 & 1539.65 & 65.53 & 1172.65 \\
			Protons ($+$) & 46.19 & 9530.25 & 50.46 & 9754.98 & 46.09 & 9393.87 \\
			Protons ($-$) & 15.34 & 21.76 & 19.11 & 46.94 &12.87 & 16.39 \\
			\hline
			\textbf{Sum} & \textbf{162.53} & \textbf{11017.38} & \textbf{182.86} & \textbf{11409.98} & \textbf{144.01} & \textbf{10615.02} \\
			\hline			
		\end{tabular}
	\end{center}
\caption{Actions (in $\hbar$) for $^{256}$Rf and both configurations in 
 $^{257}$Rf obtained with collective velocities ${\dot q}_P$ (see text) 
 along paths shown in Fig. \ref{fig:Rf-mapy}: 
  instanton-like $S_{inst}$ and with the cranking mass parameter 
 without pairing  - $S_{cr}({\dot q}_P)$.   
 Contributions from neutrons and protons of each parity (indicated in 
 parentheses) are given separately.  }
\label{tab:Rf-action} 
\end{table}
 We also calculated cranking action without pairing $S_{crank}$, i.e. twice 
 the expression of Eq. (\ref{scrank}) with the integrand 
 $\sqrt{2B_{NP}(q)(E(q)-E_{m.s.})}$, i.e. with the mass parameter $B_{NP}(q)$ 
 and collective velocity ${\dot q}_{NP}=\sqrt{2(E(q)-E_{m.s.})/B_{NP}(q)}$.   
 
 As might be expected, $S_{cr}({\dot q}_P)$ hugely overestimates $S_{inst}$ - 
 nearly by two orders of magnitude (Tab. \ref{tab:Rf-action}), mainly because 
 of pseudo - crossings of s.p. levels close to the Fermi energy. 
 Locally, around them, $B_{NP}>>B_P$, and this results in large local 
 contributions to action $S_{cr}({\dot q}_P)$. The local bumps in $B_{NP}$,  
 capriciously dependent on details of avoided level crossings, explain 
  vastly different contributions to $S_{cr}({\dot q}_P)$ from different groups
 of levels: $\sim 90$\% of $S_{cr}({\dot q}_P)$ comes from protons of 
 positive parity, while the contributions from protons of negative parity 
 in $^{256}$Rf and 1/2$^+$ state in $^{257}$Rf are similar as those to 
 $S_{inst}$ (Tab. \ref{tab:Rf-action}).  
 Using $\dot{q}_{NP}$, which differs from ${\dot q}_P$ mainly in that it 
 is much smaller at pseudo-crossings, largely reduces action: one obtains 
  $S_{crank}=199.28\,\hbar$ for $^{256}$Rf and 
 $222.48\,\hbar$ for $^{257}$Rf ($K^{\pi}=1/2^+$), results larger than, but  
  much closer to instanton-like action $S_{inst}$.


  From (\ref{eq:logtsf}), after assuming $E_{zp}=0.5$ MeV, we obtain 
 "experimental" actions of $2 S=42.24 \ \hbar$ for $^{256}$Rf and $2 S = 
 53.34 \ \hbar$ for the g.s. of $^{257}$Rf - these doubled actions should be
  compared  to values from Tab. \ref{tab:Rf-action}. Thus, calculated 
 $S_{inst}$ are $\approx 3.5$ times bigger than the values following from 
 measured half-lives.  

  We have checked that the instanton action calculated according to 
 the given prescription very much depends on the path.   
 For the trajectory coloured in blue in Fig. \ref{fig:Rf-mapy}, 	
 we obtained for $^{256}$Rf 
 $S_{inst}({\dot q}_P)=167 \ \hbar$, larger by 23 $\hbar$ than for the 
 not very different red one. 
 Apparently, in the absence of pairing, the details of pseudo - 
 crossings have large influence on action. This shows that action 
 minimization without pairing might be very difficult and would be directing 
 into paths with more gentle crossings. 
 
 The difference between instanton-like actions $S_{odd}$ and $S_{even}$  
 comes from: 1) a collective contribution - from the differences in 
 deformation energy of the e-e and odd-$A$ nuclei, which in turn comes from:  
  a) different collective velocities and b) different lengths of the path;
  2) a contribution to action from the odd nucleon \cite{foot3}.
       
 Note that in the instanton method without pairing, the odd - even effect in 
 fission half-lives comes exclusively from different heights and lengths of 
 the fission barriers.  
 If not for these, action for odd-$A$ would lie between those of neighbouring 
 $A-1$ and $A+1$ e-e species, as it is a sum of individual 
 s.p. instanton-like actions, Eq. (\ref{eq:stot}). 

 For two configurations in $^{257}$Rf we have from Tab. \ref{tab:Rf-action}: 
 $\Delta S_{odd-even}=18.52\; \hbar$ for $K^{\pi}=1/2^+$, and 
 $\Delta S_{odd-even}=38.85\; \hbar$ for $K^{\pi}=11/2^-$. 
 This large difference of $20.33\; \hbar$  can be 
 traced to a larger ${\dot q}_P$ for the second configuration, and could 
 be predicted from their very different barriers in Fig. \ref{fig:Rf-mapy}.  
 This well illustrates the trend towards higher barriers in calculations with 
 a fixed-$K$ configurations, and those with higher $K$ values in particular. 
 Such $K$-dependence is absent in experimental half-lives (see Fig. 17 in 
 \cite{Hess}). 
 
 We note that for the relative quantities, $\Delta S_{odd-even}/S_{odd}$, 
 for the g.s. of $^{257}$Rf and $^{256}$Rf we obtain from (\ref{eq:logtsf}), 
 again assuming the same $E_{zp}$, the ratio 0.114 vs. the experimental value 
 0.21. However, the minimization of action, not attempted here, could change 
 this ratio.


\subsection{Calculations assuming collective mass parameter and 
  an odd - particle contribution \label{sec:calcHF-hybrid}}

 Without having solved Eq. (\ref{eq:itdhfb1}) with pairing, we will use 
 unpaired iTDSE solutions to study odd-even fission hindrance by  
 adopting a hybrid model which incorporates both pairing and the odd particle 
 contribution to action.    

 We {\it assume} the following scheme. Action for an e-e nucleus is taken 
 from Eq. (\ref{scrank}) with both energy and the cranking mass parameter 
 including pairing. For an odd-$A$ nucleus we assume: 
\begin{equation}
S_{odd}=S_{crank}+\frac{1}{2}S_{s.p.}^{inst}\;, 
\label{eq:Scoreiinst}
\end{equation}
 where $S_{crank}$ is the cranking action (\ref{scrank}) of the e-e core, 
 calculated with the micro-macro barrier for the odd-$A$ nucleus, 
 $E^{odd}(q)-E_0$, where $E_0=E_{m.s.}+E_{zp}$, and the 
 cranking mass parameter with pairing $B^{even}_P(q)$ of the neighbouring 
 e-e $A-1$ system, while $S_{s.p.}^{inst}$ is the contribution to 
 action from the unpaired nucleon. It can be calculated as action of the  
 instanton-like solution corresponding to the unpaired $\Omega^{\pi}$ state 
 (i.e. the one blocked in the m.s.) with the collective velocity 
 ${\dot q}_P=\sqrt{2(E^{odd}(q)-E_{m.s.})/B^{even}_P(q)}$, or as 
   the difference in actions for occupied $\Omega^{\pi}$ 
 states between the odd-$A$ and e-e $A-1$ nucleus. Both ways of calculating 
 $S_{s.p.}^{inst}$ give very similar values; we will give those by the 
 second method. 
  The factor $1/2$ in (\ref{eq:Scoreiinst}) accounts for the fact that 
 $S_{inst}$ corresponds to twice action of Eq. (\ref{scrank}).

 The rationale behind the choice of the mass parameter and, consequently, of 
 the collective velocity ${\dot q}_P$, is the assumed collectivity of  
 quantum tunneling in spontaneous fission. We reject the 
 cranking mass parameter for odd-$A$, Eq. (\ref{cranking}), as it leads to 
 huge differences between collective velocities ${\dot q}$ at the neighbouring 
 $q$ points in an odd-$A$ nucleus, and between $A$ and $A-1$ nuclei at the same 
 $q$ point. Outside regions where pseudo-crossings of the odd level take place,
 the cranking mass parameters for $A$ and $A-1$ nuclei are similar,   
 see Eq. (\ref{cranking}). Thus, eliminating huge variations from 
  the mass parameter for odd-$A$ is consistent with assuming its magnitude 
 similar as in the even-$A-1$ system, uniformly in $q$. Certainly, 
 similar does not mean equal. However, lack of arguments for any definite 
 ratio singles out the made choice as the simplest one. It means that 
 the difference in actions for $A$ and $A-1$ systems comes mainly from 
 different deformation energies. A choice of the same, or of the same 
 phenomenological formula for, mass parameters for odd-$A$ and e-e $A-1$ 
 nuclei was made in the past \cite{Molpar,LojewF}. The results of the previous 
 subsection also point out that such a choice is reasonable. 
 The quantity $S_{s.p.}^{inst}$ is the remaining difference between actions 
 for odd-$A$ and e-e $A-1$ nucleus, coming from the unpaired odd particle.  
  
 As examples of the previous subsection indicate, the important point is 
 whether deformation energy of an odd-$A$ nucleus is calculated conserving 
 the configuration $\Omega^{\pi}$ of the g.s. or releasing this requirement  
 and taking the minimal energy among various configurations at each 
 deformation. We performed calculations within our model in both ways in order 
 to compare results.  

 Included deformation parameters and the choice of fission paths were as 
 discussed in the previous subsection. 
 We selected nuclei $Z=103-112$ for which their, and their even-$A-1$ 
 neighbours fission half-lives are known, and so is the hindrance factor 
 (\ref{HF}). For most of them their g.s. spins and parities are either known 
 or attributed on the basis of phenomenological models \cite{Hess}.

\begin{figure}
	\centering
 	\begin{minipage}{.5\textwidth}
		\includegraphics[width=1.2\textwidth]{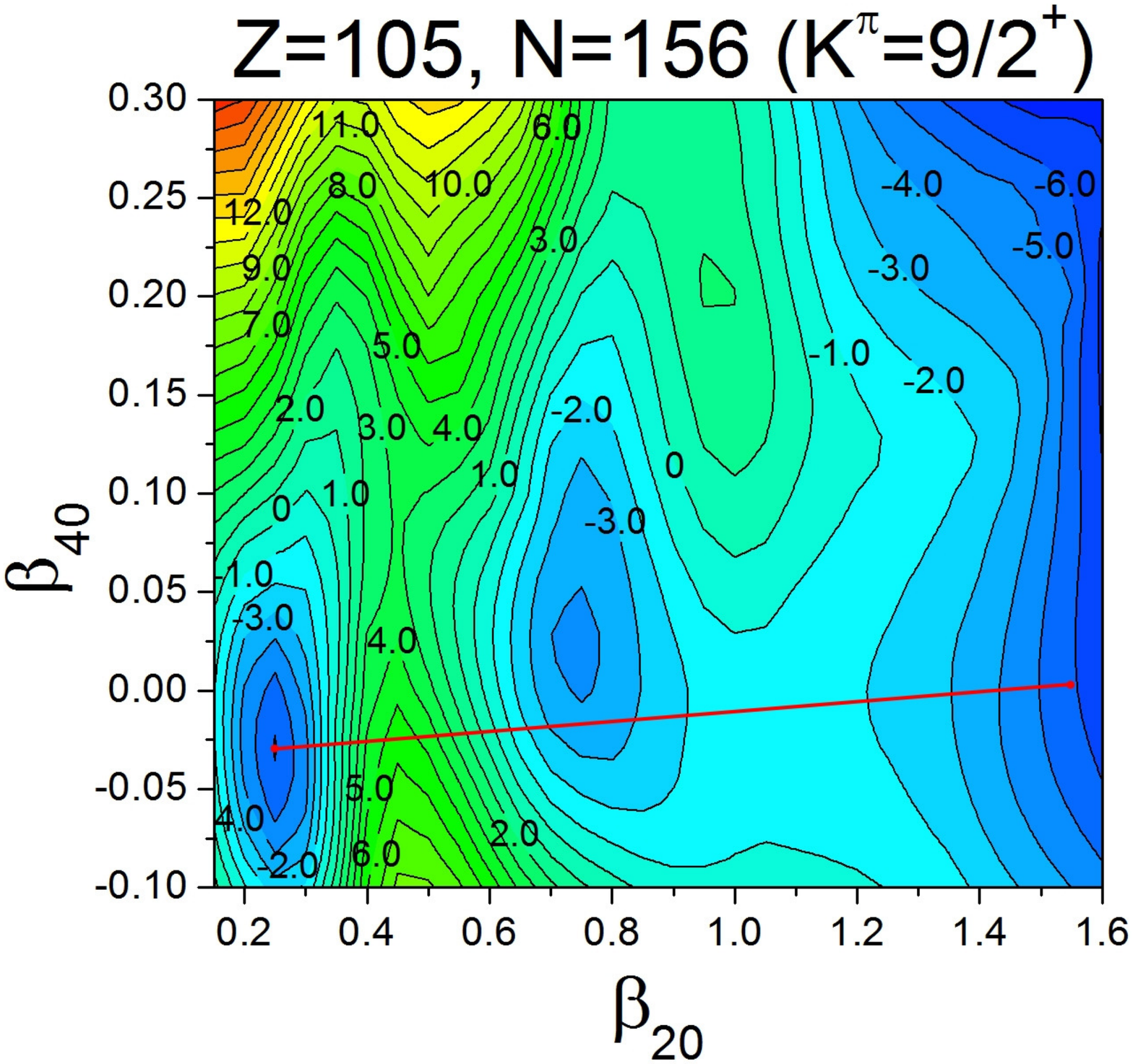}
 	\end{minipage}%
 	\begin{minipage}{.5\textwidth}
		\includegraphics[width=1.2\textwidth]{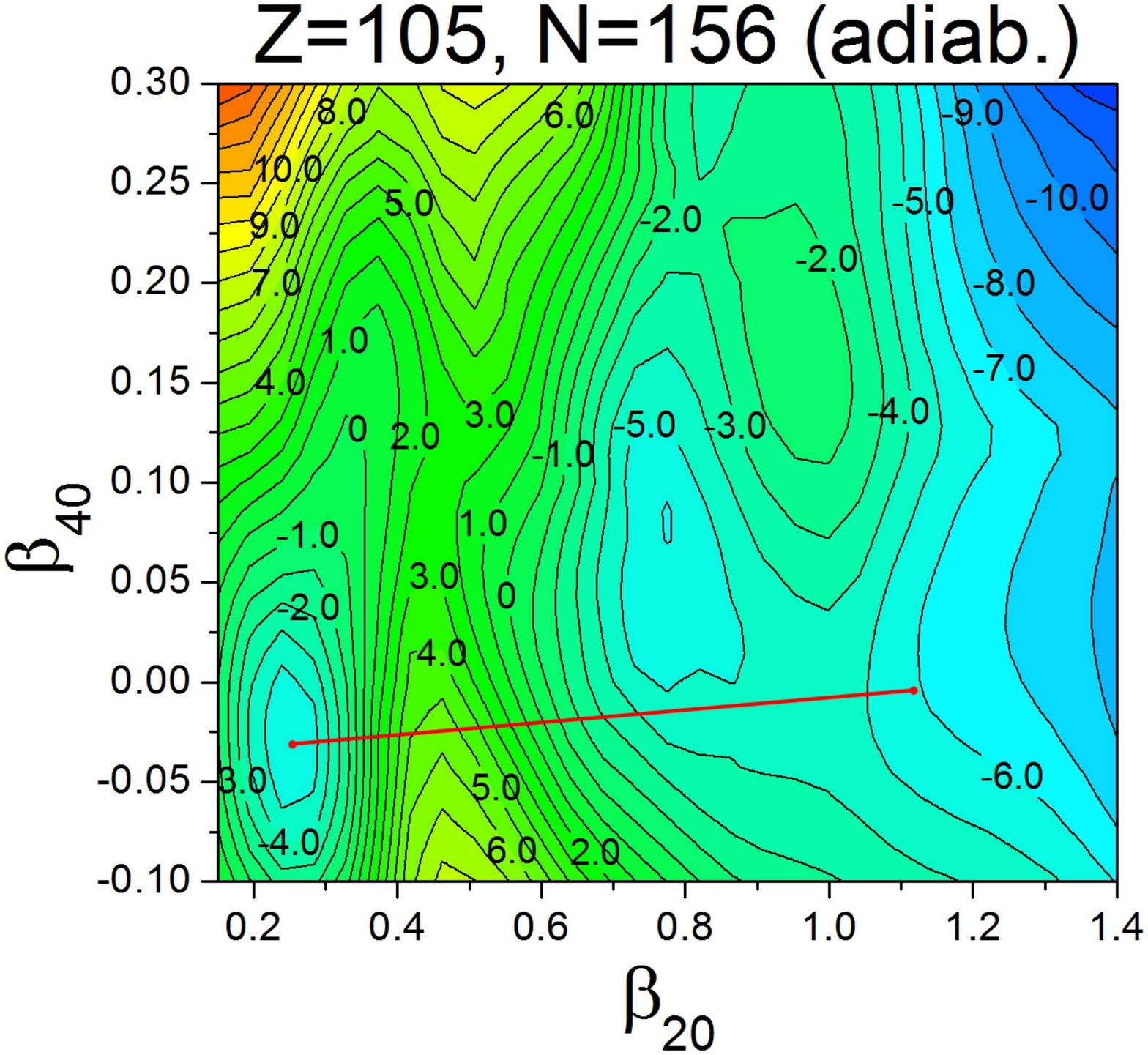}
 	\end{minipage}
 	\begin{minipage}{.5\textwidth}
		\includegraphics[width=1.2\textwidth]{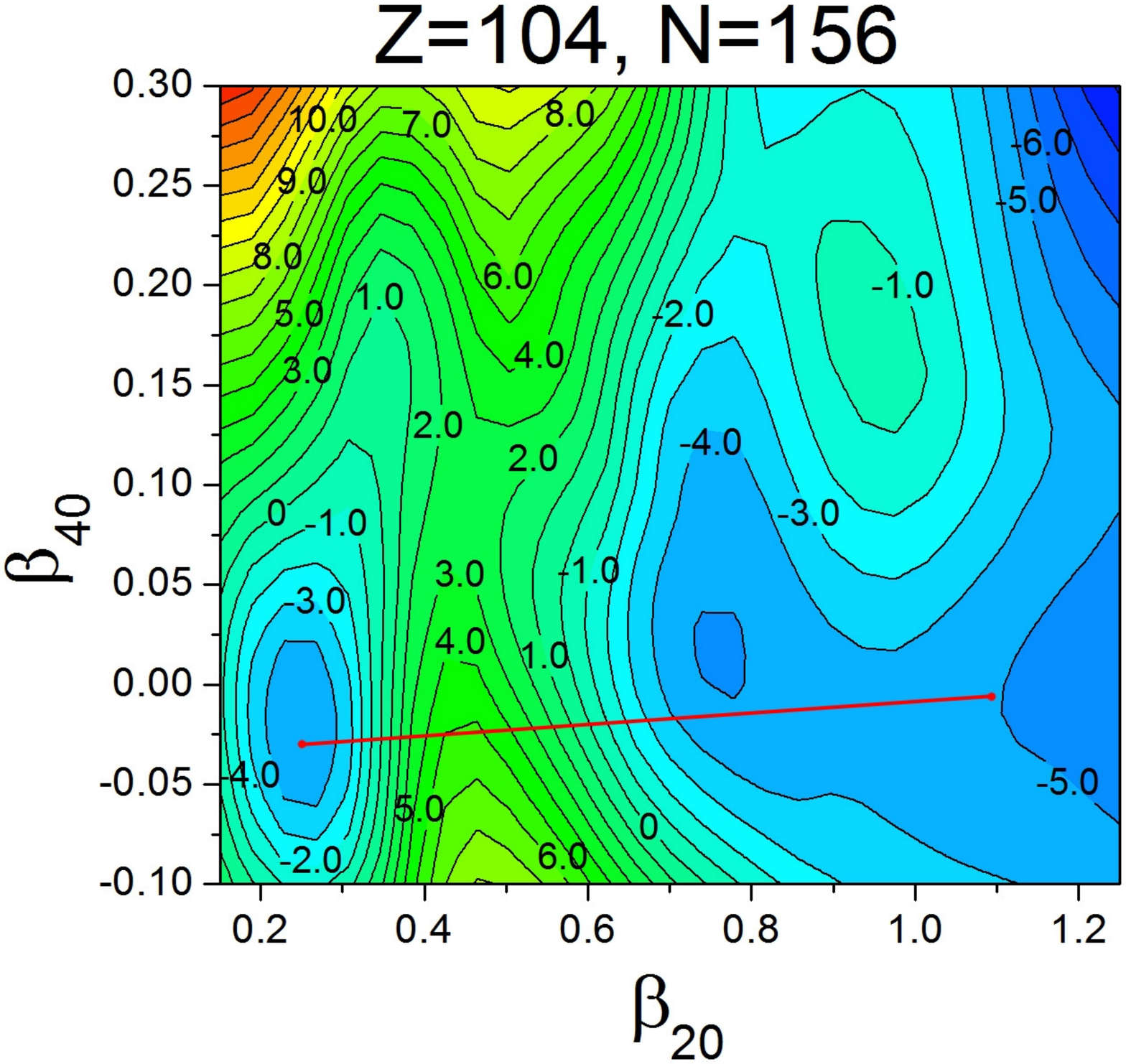}
 	\end{minipage}
	\caption{\textit{Upper panel:} energy landscapes for $^{261}$Db, 
  minimized over $\beta_{60},\beta_{80}$ with the kept g.s. configuration  
$K^{\pi}=9/2^+$ (left) and adiabatic (right). 
\textit{Lower panel:} energy landscape for $^{260}$Rf. Chosen fission 
 path marked in red. Note different range of $\beta_{20}$ in maps.}
	\label{fig:Db-mapy}	
\end{figure}

 In Fig. \ref{fig:Db-mapy}, the calculated 
 energy surfaces are shown for $^{261}$Db and its e-e neighbour $^{260}$Rf. 
  The g.s. configuration of $^{261}$Db is $K^{\pi}=9/2^+$. 
 Both surfaces for $^{261}$Db, 
 adiabatic (minimized over configurations) and constrained on the $K^{\pi}$ 
 value, are given together with chosen fission paths. 
  It can be seen that the fission barriers are double-humped, with a smaller 
 second hump.  
 A similar picture holds for other considered nuclei. 
 A clear difference between adiabatic and 
 $K^{\pi}$ - conserved surfaces can be observed 
 for $K=9/2$ in $^{261}$Db - one can notice higher and longer second 
 barrier. For smaller $K$, like e.g. the $K^{\pi}=1/2^+$ configuration in 
 $^{259}$Sg (not shown here), this difference is smaller. A large difference 
 in barriers for high-$K$ configuration was also seen for $^{257}$Rf in 
 Fig. \ref{fig:Rf-mapy}.

  At this point one has to note that our calculations do not include nonaxial 
 deformations, $\beta_{22}$, etc, which lower the first barrier, neither do 
 they account for mass asymmetric deformations lowering the second barrier. 
 Calculations which include nonaxiality    
 indicate that a path through the nonaxial saddle, lower by 1-2 MeV,  
 has a substantially greater length which moderates or even compensates 
 the effect of the lower saddle. 
 On the other hand, the mass asymmetry is lowering the second barrier and the 
 path incorporating it is not much longer (in terms of $ds=
 \sqrt{\sum_{\lambda \mu}(d\beta_{\lambda \mu}/d\beta_{2 0})^2} d\beta_{2 0}$) 
 than the one considered here because the mass-asymmetric exit from the barrier
 occurs for smaller $\beta_{2 0}$ - thus the effect of $\beta_{\lambda 0}$ with 
 odd $\lambda$ is likely to decrease the action.  

 It turns out that with realistic values of $E_{zp}$ around 0.5 - 1 MeV we 
 obtain too large actions and half-lives for e-e nuclei  
 as compared to the experimental values.   
  The reason lies in a too limited choice of nuclear shapes and in a relatively 
 small strength of the pairing interaction, dictated by the local mass fit 
 \cite{WSparmac}. 
  Indeed, we have checked for $^{256}$Rf, that with the pairing strengths and 
 $E_{zp}=0.7$ MeV used in \cite{sss95} and ignoring the second barrier hump 
  (which is largely reduced by the mass-asymmetry) we reproduce 
    the result reported there which is in good agreement with the experimental 
   value. 

\begin{table}
	\begin{center}
		\begin{tabular}{c|c|c|c}
			\hline
			\hline
			Nucleus \ & \ $S_{crank}/\hbar$	\ & \ $T_{sf}^{exp}$ [s] \  	&\ 	$T_{sf}^{calc}$ [s]\ 	\\ 
			\hline
			\hline
			$^{258}$No	&	21.60	&	1.2E-03	&	4.1E-03	\\
			$^{254}$Rf	&	18.46	&	2.3E-05	&	7.8E-06	\\
			$^{256}$Rf	&	21.91	&	6.4E-03	&	7.6E-03	\\
			$^{260}$Rf	&	22.97	&	2.2E-02	&	6.4E-02	\\
			$^{258}$Sg	&	21.92	&	2.6E-03	&	7.7E-03	\\
			$^{260}$Sg	&	23.62	&	7.0E-03	&	2.4E-01	\\
			$^{282}$Cn	&	18.82	&	9.1E-04	&	1.6E-05	\\
			\hline
			\hline
		\end{tabular}
	\end{center}	
	\caption{Calculated actions (in $\hbar$) and calculated vs experimental 
 fission half-lives (in seconds) for e-e nuclei after adjusting 
 zero-point energy $E_{zp}$ to minimize the root-mean-square error.  }
	\label{tab:even-zE0}
\end{table}	

	\begin{table}
		\begin{center}
			\begin{tabular}{cc|c|c|c|c}	
				\hline
				\hline		
		 \  		Nucleus \ &\ $K^{\pi}$\ &\  $S^{conf}_{crank}/\hbar$\ & $T_{sf}^{crank}$ [s]  &\ 	$S^{ad}_{crank}/\hbar$\ &\ $\Delta S_{crank}/\hbar$\ \\
				\hline
					$^{259}$Lr	&	 7/2-	&		33.32& 6.2E+07	& 23.44	&	9.88	\\
					$^{255}$Rf	&	 9/2-	&	        56.06& 3.5E+27 	& 25.31	&	30.75	\\
					$^{257}$Rf	&	 1/2+	&	        34.32& 4.6E+08 	& 22.58	&	11.74	\\
					$^{257}$Rf (m)	&	 11/2-	&		48.89& 2.1E+21	& 22.58	&	26.31	\\
					$^{261}$Db	&	 9/2+	&		40.79& 1.9E+14	& 26.65	&	14.14	\\
					$^{259}$Sg	&	 1/2+	&		32.44& 1.1E+07	& 23.23	& 	9.21	\\
					$^{261}$Sg	&	 3/2+	&		30.75& 3.6E+05  & 25.30	&	5.45	\\
					$^{283}$Cn	&	 5/2+	&		24.52& 1.4E+00  & 21.56	&	2.96	\\
				\hline
				\hline
			\end{tabular}
		\end{center}
		\caption{
 For odd nuclei and their $K^{\pi}$ configurations shown in columns 1 and 2 
 are given cranking actions (\ref{scrank}) calculated with the mass parameters 
 of the e-e neighbour: for a fixed $K^{\pi}$ configuration 
 $S^{conf}_{crank}$ (col. 3), for adiabatic configuration $S^{ad}_{crank}$ 
 (col. 5), their difference $\Delta S_{crank}$ (col. 6), all in $\hbar$; 
 half-lives $T_{sf}^{crank}$ (in s) resulting from $S^{conf}_{crank}$ are 
 given in col. 4. The zero point energy $E_{zp}$ was adjusted to experimental 
 fission half-lives of e-e nuclei. }
		\label{tab:Sconfvsadiab2} 
	\end{table}

 Since we focus here on fission hindrance for odd-$A$ nuclei we 
 decided to artificially change zero-vibration energy $E_{zp}$ so  
 that the mean - square deviation of fission half-lives in e-e nuclei from 
 experimental values is minimal. This happens for $E_{zp}=2.03$ MeV. 
 The fission half-lives of e-e nuclei obtained with the adjusted $E_{zp}$,   
 which will serve as the reference for the calculation of fission hindrance 
 factors in odd-$A$ nuclei, are given in Table \ref{tab:even-zE0}. They are 
 mostly of the same order of magnitude as the experimental ones, except 
 in $^{260}$Sg and $^{282}$Cn. The effect of higher $E_{zp}$ cancels the 
 contribution to action from the second barrier for $Z=102-106$. 
 This is roughly consistent with the results of \cite{sss95}, where the 
 barrier was practically reduced to the first hump.

 In Table \ref{tab:Sconfvsadiab2} we compare actions $S_{crank}$ 
 of (\ref{eq:Scoreiinst}) obtained in two ways for odd nuclei: 
 $S_{crank}^{conf}$ - by keeping the fixed configuration, and 
 $S_{crank}^{ad}$ - by using adiabatic occupation of the odd nucleon. 
 Differences between these actions, $S_{crank}^{conf}-S_{crank}^{ad}$, are 
 greater than 9 $\hbar$, except for $^{261}$Sg and $^{283}$Cn. 
 As we have checked, they remain large for a wide choice of 
 adopted $E_{zp}$ values between 0.5 and 2 MeV.  
 As for e-e nuclei, paths on the adiabatic surfaces effectively do not 
 show the second barrier. With the preserved $K^{\pi}$ configuration, 
 the contribution of the second barrier to action is substantial and strongly 
 dependent on the magnitude of $K$.  
 Fission half-lives calculated with keeping the $K^{\pi}$ configuration, 
 also given in Table \ref{tab:Sconfvsadiab2}, vastly overestimate the 
 experimental values (see col. 3 of Table \ref{tab:oddnucres2} for 
 comparison), except in $^{283}$Cn, with the largest 
 discrepancy for large $K$. Therefore, we do not include odd-particle 
 actions $S_{s.p.}^{inst}$ for them. 

	\begin{table}
		\begin{center}
			\begin{tabular}{cccc|ccccc}	
				\hline
				\hline
				\multicolumn{4}{c|}{\textbf{Nucleus data}}	 & \multicolumn{5}{c}{\textbf{Adiabatic blocking}} \\
				\hline
				\hline										
				$^{A}X$	&	$I^{\pi}$	&	$T_{sf}^{exp}$ [s]	&	$HF_{exp}$ 	&	$S_{s.p.}^{inst}/\hbar$	&	$T_{sf}^{cr}$ [s] & $T_{sf}^{cr+inst}$ [s] & $HF_{calc}^{cr}$ & $HF_{calc}^{cr+inst}$	\\
				\hline
				$^{259}$Lr	&	 7/2-	&	27.4	&	2.3E+04	&	1.02	&	0.16	& 0.45 &	3.9E+01	&	1.1E+02	\\
				$^{255}$Rf	&	 9/2-	&	3.15	&	1.4E+05	&	-1.37	&	6.83	& 1.73 &	8.8E+05	&	2.2E+05	\\
				$^{257}$Rf	&	 1/2+	&	423	&	6.6E+04	&	2.43	&	0.03	& 0.33 &	3.9E+00	&	4.34E+01	\\
				$^{257}$Rf (m)	&	 11/2-	&	$>$490	&	$>$76562.5 &	0.03	&  0.03 & 0.03	&	3.9E+00	&	3.9E+00	\\
				$^{261}$Db	&	 9/2+	&	5.6	&	2.5E+02	&	0.04	& 99.6	& 103.6 &   1.56E+03	&	1.62E+03	\\
				$^{259}$Sg	&	 1/2+	&	8	&	3.1E+03	&	1.85	&	0.11	& 0.68  &	1.43E+01	&	8.83E+01	\\
				$^{261}$Sg	&	 3/2+	&	31	&	4.4E+03	&	0.61	&	6.7	& 12.32 &  2.79E+01	&	5.13E+01	\\
				$^{283}$Cn	&	 5/2+ (*)	&	24	&	2.6E+04	&	2.76	& 0.0038 & 0.06 &  2.38E+02	&	3.75E+03	\\
				\hline
				\hline 
			\end{tabular}
		\end{center}
		\caption{For seven odd-$A$ nuclei listed in the first column 
 are given: configurations $I^{\pi}$ (experimental or from systematics), 
 experimental spontaneous fission half-lives $T^{exp}_{sf}$ 
 (after \cite{Hess}) and fission hindrance factors $HF_{exp}$ according to 
 (\ref{HF}), and calculated quantities (for the g.s. or m.s. configurations 
 $K^{\pi}=I^{\pi}$): 
 the odd nucleon instanton contribution to action $S^{inst}_{s.p.}$,
  fission half-lives and $HF$s following from the adiabatic actions 
 $S^{ad}_{crank}$ 
 for the e-e core (given in Tab. \ref{tab:Sconfvsadiab2}) 
 and the same augmented with $S^{inst}_{s.p.}$,
 $S_{crank}^{ad}+\frac{1}{2} S_{s.p.}^{inst}$. Half-lives are given in seconds, 
 actions in units of $\hbar$. Asterisk for $^{283}$Cn signals that the given 
 $T^{exp}_{sf}$ is the smaller of two conflicting experimental values and 
 spin/parity is derived from our W-S spectrum. The symbol (m) denotes the 
 excited configuration.}
		\label{tab:oddnucres2}	
	\end{table}

\begin{figure}
 	\centering
 	\includegraphics[angle=-90.,width=0.6\textwidth]{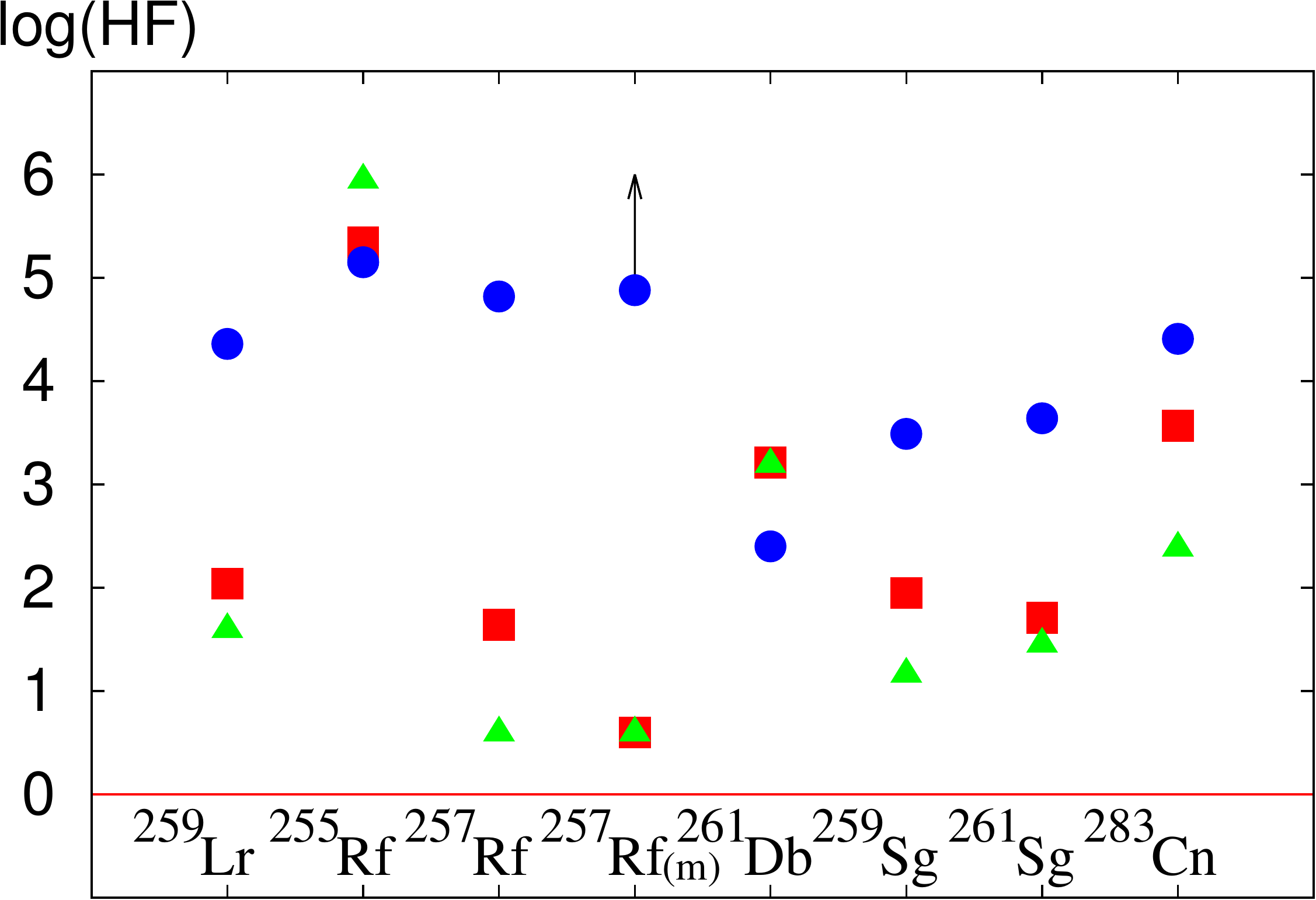} 
 	\caption{Logarithms of fission hindrance factors, $\log HF$,  
  defined by Eq. (\ref{HF}): experimental (blue circles) vs. calculated with 
  (red squares) and without (green triangles) the odd-particle instanton 
  contribution for 
  nuclei specified at the bottom of the panel; an arrow for $^{257}$Rf(m) 
  signifies that only the lower bound for $HF$ is experimentally known; 
   for further details - see text.}
 	\label{fig:logHF}
 \end{figure}

  Results pertaining to half-lives of odd-$A$ nuclei and fission hindrance 
 factors obtained with the adiabatic blocking are given in 
 Table \ref{tab:oddnucres2} and shown in Fig. \ref{fig:logHF}. 
  Here we include results obtained with $S_{crank}^{ad}$ alone and with 
  the added odd-particle contribution $S_{s.p.}^{inst}$. 
  Obtained half-lives are much closer to the experimental ones than those for 
   fixed configurations, but with no clear hindrance, i.e. 
  $HF$s are mostly underestimated (with two exceptions - $^{255}$Rf and 
 $^{261}$Db). The modification of the half-life introduced by adding 
 instanton-like action for the odd nucleon $S_{s.p.}^{inst}$  
 (\ref{eq:Scoreiinst}),
  shown in Tab. \ref{tab:oddnucres2}, moves the calculated $HF$s closer  
 to the experimental values, but the effect is still too small.  

 Odd-even fission hindrance factors calculated assuming the same collective 
 mass parameter in e-e and odd-$A$ neighbours suggest the following conclusions:
\begin{enumerate}
	\item Keeping configuration $K^{\pi}$ of the fissioning states   
 leads to the odd-even fission $HF$s larger by orders of magnitude  
 than in experiment. 
	\item Keeping the lowest configuration       
 leads mostly to (with two exceptions) too small hindrance factors.  
 
	\item Instanton-like correction for the odd nucleon added to 
 adiabatic cranking result $S^{ad}_{crank}$ (\ref{eq:Scoreiinst}) acts in the 
 right direction but is too small.  
 As a result, the obtained $HF$s are on average smaller
  than the experimental values of $10^3$ - $10^5$; they are  
 also more scattered than the latter. 
\end{enumerate}

 One can note that these conclusions concerning diffrences in $T_{sf}$ of 
 odd-$A$ and e-e closest neighbours do not seem to be much influenced by 
 the lack of the action minimization: 
 adiabatic energy landscapes of odd-$A$ nuclei and their e-e neighbours  
 are very similar, $S^{ad}_{crank}$ are relatively smooth and 
 the chosen paths are typical of realistic calculations.

\section{Summary and conclusions}

 As the cranking or ATDHF(B) approximation commonly used in calculating 
 spontaneous fission half-lives is incorrect for 
 odd-$A$ nuclei and $K$-isomers, in the present paper we tried to include 
 nonadiabatic, beyond-cranking effects in the description of quantum tunneling. 
 A treatment that avoids the adiabatic assumption is provided by the 
 method of instantons. For atomic nuclei, it takes a form of iTDHFB equations 
 non-local in time, with specific boundary condition, which seem unsolvable 
 at present. This motivated us to simplify these equations to iTDSE and study 
 actions for resulting instanton-like solutions which  
 relate to fission half-lives.  
  The rationale for taking an intermediate step before the full 
  instanton theory is also related to the question of the 
  energy overlaps (\ref{eq:tdhfener}): they are crucial in the selfconsistent 
 theory, but their proper treatment is unknown for the majority of energy 
  functionals presently used.

  The instanton equations of the selfconsistent theory were simplified to 
 iTDSE version with the phenomenological potential in the case without
  pairing, and to iTDHFB equations with the fixed potential and selfconsistent 
 pairing gap for the seniority pairing interaction. 
 The iTDSEs were solved for the phenomenological Woods-Saxon potential in a 
 number of cases. Since we do not want to relay on the cranking mass parameters
 for odd-$A$ nuclei, we had to assume the collective velocity. 
  We used for this purpose the cranking mass parameter of the neighbouring e-e 
  nucleus - a plausible, but not unique assumption.  

  The method of obtaining iTDSE solutions and actions was demonstrated 
  for axially symmetric potential. It was found that actions may be reliably 
 calculated using reasonably long periods and relatively small  
 bases of adiabatic levels, lying close to Fermi energy. 
 Compared to the cranking approximation for odd-$A$ nuclei, close avoided 
 level crossings have milder influence on instanton-like actions.  
  For collective velocities typical of e-e actinide or superheavy nuclei, the 
  quasi-occupations which characterize nonadiabatic excitations in iTDSE 
 solutions are changing mostly in the vicinity of pseudo-crossings. 
 Instanton-like action rises with the (uniformly) rising collective velocity 
 and the length of the fission path can balance the lower barrier in the 
 competition between trajectories.

 The case of triaxial potential turned out to be more demanding as a result 
 of many very weakly-interacting pseudo-crossings. The solution of iTDSE 
 in the adiabatic basis becomes difficult and an effective way of solution 
 remains to be found. One has to mention that 
  the difficulty caused by many nearly-crossing levels may be less acute 
  when one includes the antihermitean part of the mean field. This would 
 make the eigenvalues of the mean-field ${\hat h}$ complex and instanton 
 solutions less susceptible to such crossings. 
 
  In the study of odd-even fission hindrance factors we made use of 
  iTDSE solutions without pairing by combining them with the cranking 
  actions for the e-e cores. The premise of this study was that 
  effective mass parameters pertinent to spontaneous fission are the same 
 (or very similar) in neighbouring e-e and odd-$A$ nuclei.  
  The clear result obtained under this proviso is that actions calculated 
 for the fixed $K^{\pi}$ configurations along axially symmetric paths 
  hugely overestimate values from experiment. 
    The actions calculated with adiabatic energy landscapes are mostly 
   too close to those of e-e neighbours. 
  Since adiabatic energy landscapes of odd-A nuclei {\it include} the effect of 
  the pairing gap decrease due to blocking, one may say that this effect alone 
  is insufficient, while the additional effect of preserving $K$ 
  quantum number is unrealistically large.  
  The instanton-like contributions from the odd nucleon, when added to the 
  e-e core actions obtained with adiabatic landscapes, are (in most cases) 
 too small to provide for the observed hindrance factors.  
 One could say that actions for odd-A nuclei seem to be 
  closer to the scenario with unconstrained configurations what would suggest 
  changes in $K$ in tunneling, possibly related to nonaxial or more exotic 
  deformations along the fission paths.

  In the near future we plan to study the simplified iTDHFB actions including 
 pairing of Sec. \ref{sec:metinst3} in order to see how the above conclusions 
 about fission hindrance factors change. In particular, it seems interesting 
 whether one could reproduce their relatively small experimental scatter 
 of merely 2 orders of magnitude.
  We would also like to see if one can effectively use the solution method for 
 iTDSE studied here in the solution of the selfconsistent problem.  
  It would be also interesting to improve the presented micro-macro 
  instanton-like procedure. This, however, would probably require 
  some non-selfconsistent version of the antihermitean part of 
  the imaginary-time mean-field.

 \acknowledgments

 The authors would like to thank Micha\l{} Kowal for many 
 inspiring discussions and suggestions, and Piotr Jachimowicz for 
  providing energy landscapes including effects of the axial- and 
  reflection-asymmetry on fission saddles.




\appendix

\section{Cranking expressions for action \& Floquet exponents 
     \label{app:crankmet}} 

  The cranking approximation in solving the real-time Schr\"odinger equation: 
 $i\hbar\partial_t\psi(t) = {\hat h}(q) \psi(t)$, where $q=q(t)$ is a 
 collective coordinate, follows from expanding $\psi(t)$ onto adiabatic states 
 $\psi_{\mu}(q)$ (\ref{expan1}), substituting:  
 \begin{equation}
 C_{\mu}(t)=c_{\mu}(t)\exp\left(-\frac{i}{\hbar}\int_0^{t}
 \epsilon_{\mu}(t'))dt'\right)  ,
 \end{equation}
 and solving equations for $c_{\mu}(t)$:
 \begin{equation}
  \label{eqadia}
 \partial_t c_{\mu} = -{\dot q}\sum_{\nu}\langle\psi_{\mu}\mid\partial_q
 \psi_{\nu}\rangle c_{\nu}\exp\left(\frac{i}{\hbar}\int^t(\epsilon_\mu-
 \epsilon_{\nu})dt'\right) ,
 \end{equation} 
  to the leading order in ${\dot q}$, assuming that the amplitude of the 
  adiabatic ground-state dominates others: $|c_0|\approx 1$, $|c_{\mu}|<<
  1$ for $\mu>0$. For $\mu>0$, one can integrate (\ref{eqadia}) under 
 the assumption that the exponential gives the leading $t$-dependence: 
 \begin{equation}
 \label{crankcoef}
 c_{\mu} \approx i\hbar{\dot q}\frac{\langle\psi_{\mu}\mid\partial_q
 \psi_{0}\rangle}{\epsilon_{\mu}-\epsilon_0} c_0
  \exp\left(\frac{i}{\hbar}\int^t(\epsilon_\mu-
 \epsilon_0)dt'\right) ,
 \end{equation} 
 so the wave function in the cranking approximation is:
 \begin{equation}
  \label{crankwf}
 \psi(t)= c_0 \exp\left(-\frac{i}{\hbar}\int^t \epsilon_0 dt'\right) 
 \left(\psi_0 + i \hbar{\dot q} \sum_{\mu>0}\frac{
 \langle\psi_{\mu}\mid\partial_q\psi_0\rangle}{\epsilon_{\mu}-\epsilon_0}
  \psi_{\mu}\right) . 
 \end{equation} 
  This form of integration, different from the usual one for an initial 
  value problem, allows to obtain mass parameter (see below) as a function  
 {\it solely} of the coordinate $q$. Other possible integrals of (\ref{eqadia}) 
   imply dissipation of collective motion, see e.g. \cite{KHR1977} or the 
  recent \cite{Rouvel}. From (\ref{crankwf}), the initial assumption 
 $|c_{\mu}|<<1$ means: $\frac{\hbar{\dot q}}{\epsilon_{\mu}-\epsilon_0}
 \langle\psi_{\mu}\mid\partial_q \psi_{0}\rangle <<1$, that  
 {\it does not hold} in a vicinity of a sharp (avoided) level crossing, except 
 for minuscule ${\dot q}$. 

 Substituting $c_{\mu}$ of (\ref{crankcoef}) into Eq. (\ref{eqadia}) for 
 $c_0$ one obtains:
 \begin{equation}
 \partial_t c_0 \approx \frac{i}{\hbar}\left(i\hbar\langle\psi_0\mid\partial_t
 \psi_0\rangle+(\hbar{\dot q})^2\sum_{\mu>0} 
  \frac{\mid\langle\psi_{\mu}\mid\partial_q \psi_{0}\rangle\mid^2}
  {\epsilon_{\mu}-\epsilon_0}\right) c_0 ,
 \end{equation} 
 where the expression in the parenthesis is real, so $c_0$ evolves as 
 a pure phase:  
 \begin{equation}
 \label{crankc0}
  c_0 \approx \exp\left \{\frac{i}{\hbar}\int^t\left(i\hbar\langle\psi_0\mid
 \partial_t \psi_0\rangle+(\hbar{\dot q})^2\sum_{\mu>0} 
  \frac{\mid\langle\psi_{\mu}\mid\partial_q \psi_{0}\rangle\mid^2}
  {\epsilon_{\mu}-\epsilon_0}\right)dt'\right \}  ,
 \end{equation} 
  with the first term in the exponent being the topological (Berry's) phase
 \cite{Berry}. Usually, the coeficient $c_0$ is modified to 
 assure normalization of $\psi(t)$, $\sum_{\mu}|c_{\mu}|^2=1$, which 
 introduces corrections quadratic in ${\dot q}$ to $|c_0|$, but does not 
 change its phase.
 As a result, the expectation value of ${\hat h}$, 
 $\langle \psi(t)\mid {\hat h}(q)
 \mid \psi(t)\rangle \approx \epsilon_0(q)+\frac{1}{2}{\dot q}^2B_{q q}(q)$, 
 where: 
 \begin{equation}
 B_{q q}(q)=2\hbar^2\sum_{\mu>0}
 \frac{|\langle\psi_{\mu}\mid\partial_q\psi_0\rangle|^2}
 {\epsilon_{\mu}-\epsilon_0}  
 \end{equation}
 is the cranking mass parameter. 

 For a periodic hamiltonian with a period $T$, ${\hat h}(t+T)={\hat h}(t)$, 
 the cranking wave function $\psi(t)$ is quasiperiodic, with a phase augmented 
 by $-i\zeta T/\hbar$ after each period, where by 
Eq. (\ref{crankwf},\ref{crankc0}), if topological phase gives no contribution,  
 \begin{equation}
 \label{Floq1}
 \zeta = \frac{1}{T} \int_0^T[\epsilon_0(q) - \frac{1}{2}{\dot q}^2
 B_{q q}(q)]dt  .
 \end{equation}
 Thus, one can present $\psi(t)$ as: ${\tilde \psi}(t)\exp(-i\zeta t/\hbar)$, 
 where ${\tilde \psi}(t)$ is periodic with the period $T$, 
 and $\zeta$ is called the Floquet exponent. The function ${\tilde \psi}(t)$ 
 satisfies (in the cranking approximation) the equation: 
 $(i\hbar\partial_t-{\hat h}(q)){\tilde \psi}= -\zeta{\tilde \psi}$.  
 Calculating action, 
 $\int_0^T dt 
 \langle{\tilde \psi}\mid i\hbar\partial_t {\tilde \psi}\rangle$, one thus 
 obtains $\int_0^T dt (\epsilon_0+\frac{1}{2}{\dot q}^2 B_{q q}(q)-\zeta)$, 
 which from (\ref{Floq1}) equals $\int_0^T dt B_{q q}(q){\dot q}^2$. 
 This action may be used to quantize energy of collective modes, see 
 e.g. \cite{Kan}.

 The analogous solution to the equation in imaginary time $\tau=it$, 
 $\hbar\partial_{\tau} \phi+{\hat h}(q)\phi=0$, with  
 $-T/2<t<T/2$ and ${\dot q}(-\tau)=-{\dot q}(\tau)$, is:  
 \begin{equation}
  \label{crankwfi}
 \phi(\tau)= c_0 \exp\left(-\frac{1}{\hbar}\int^{\tau} \epsilon_0 d\tau'\right) 
 \left(\psi_0 - \hbar{\dot q} \sum_{\mu>0}\frac{
 \langle\psi_{\mu}\mid\partial_q\psi_0\rangle}{\epsilon_{\mu}-\epsilon_0}
  \psi_{\mu}\right) , 
 \end{equation} 
  where:
 \begin{equation}
 \label{crankc0i}
  c_0 \approx \exp\left\{-\frac{1}{\hbar}\int^{\tau}
 \left(\hbar\langle\psi_0\mid
\partial_{\tau} \psi_0\rangle+\frac{1}{2}{\dot q}^2B_{q q}(q)
 \right)d\tau'\right \} , 
 \end{equation}
 although, due to the exponential character of solutions, the range of 
 validity of the cranking approximation is probably much smaller than 
 in the real-time.  
 The corrections to $c_0$ quadratic in ${\dot q}$ which ensure the condition 
 $\langle \phi(-\tau)\mid \phi(\tau)\rangle=1$ modify the $\tau$-even part 
 of $c_0$, but not its time-odd exponent. 
 In this approximation, 
 $\langle \phi(-\tau)\mid {\hat h}(q)\mid \phi(\tau)\rangle\approx 
  \epsilon_0(q)-\frac{1}{2}{\dot q}^2B_{q q}(q)$. 
 For a periodic hamiltonian, as the one with $q(\tau)$ describing a bounce 
 solution, this wave function can be presented as 
  $\phi(\tau)={\tilde \phi}(\tau) \exp(-\zeta \tau/\hbar)$, where 
 ${\tilde \phi}(\tau)$ is periodic; the Floquet exponent here is 
 \begin{equation}
 \label{Floq2}
 \zeta = \frac{1}{T} \int_{-T/2}^{T/2}
 [\epsilon_0(q)+\frac{1}{2}{\dot q}^2B_{q q}(q)]d\tau .
 \end{equation}
  The periodic function ${\tilde \phi}$ satisfies the equation: 
 $\hbar\partial_{\tau} {\tilde \phi}=(\zeta-{\hat h}(q)){\tilde \phi}$. 
 Action defined for it by: $S=\int_{-T/2}^{T/2} d{\tau}
 \langle{\tilde \phi}(-\tau)\mid \hbar\partial_{\tau} {\tilde \phi}(\tau)
 \rangle$, can be written by using the previous relations as:
 \begin{equation}
 \label{Scrankim}
  S=\int_{-T/2}^{T/2} d{\tau}(\zeta-\epsilon_0+
 \frac{1}{2}{\dot q}^2 B_{q q}(q)) = 
 \int_{-T/2}^{T/2} d{\tau} B_{q q}(q){\dot q}^2  , 
 \end{equation}
 consistent with the cranking formula (\ref{scrank}). 



 \section{Methods applied to obtain non-selfconsistent bounce solutions 
 \label{app:solu}}
 
 The exponential behaviour of solutions to Eq. (\ref{equat}) and 
  the presence of many different exponents pose problems which require 
 special care in the numerical treatment. 
  In this section we address these difficulties and discuss 
 methods applied to obtain instanton-like solutions in this work. 

Let us first notice, that the set of equations (\ref{equat}) 
 \textit{without} the $\zeta$-term:
 \begin{equation}
\label{equat-noF}
\hbar\frac{\partial C_{\mu i}}{\partial\tau} = 
-\epsilon_{\mu}(q(\tau)) C_{\mu i} -{\dot q}
\sum_{\nu}^{\cal{N}} \langle \psi_{\mu}(q(\tau)) \mid\frac{\partial \psi_{\nu}}
{\partial q} (q(\tau))\rangle C_{\nu i}
\end{equation}
 is of the form: $\mathbf{\dot{C}}_i=\mathbf{A(\tau)}\mathbf{C}_i$, where the 
 matrix $\mathbf{A}(\tau)$ is periodic: $\mathbf{A}(-T/2)=\mathbf{A}(T/2)$, and 
 $\mathbf{C}_i$ is the column - vector of coefficients $C_{\mu i}(\tau)$ 
 of the $i$-th solution. Therefore, according to the Floquet 
 theorem, the linearly independent solutions can be written as:
 \begin{equation}
 \mathbf{C}_{i}(\tau)=\mathbf{P}_{i}(\tau)e^{-\zeta_i\tau/\hbar},
 \label{eq:solperfloq} 
 \end{equation} 
 where $\mathbf{P}_{i}(\tau)$ is a periodic function with the period $T$ while 
 $\zeta_i$ are determined by the eigenvalues $e^{-\zeta_{i} T/\hbar}$ 
 of the monodromy matrix, $\mathbf{M}=\mathbf{G}(T/2,-T/2)$, with  
 $\mathbf{G}(\tau_2,\tau_1)$ designating resolvent of (\ref{equat-noF}), 
 propagating solutions from $\tau_1$ to some other time $\tau_2$. 
 Putting (\ref{eq:solperfloq}) into (\ref{equat-noF}) we obtain equation for 
 the unknown periodic functions:
 \begin{equation}
 \label{num3}
 \mathbf{\dot{P}}_{i}=(\mathbf{I}\zeta_{i}-\mathbf{A}(\tau))
 \mathbf{P}_{i}(\tau)  , 
 \end{equation}
 with the boundary condition: $P_{ki}(-T/2)=P_{ki}(T/2)=v_{ki}$, where $v_{ki}$ 
 is the $k$-th component of the $i$-th eigenvector of $\mathbf{M}$. 
 The equation 
 above is identical to Eq. (\ref{equat}), therefore $\mathbf{P}_{i}(\tau)$ are 
 the sought bounce solutions with Floquet exponents $\zeta_i$ and boundary 
 values given by the eigenvalues and eigenvectors of the monodromy matrix.
 These considerations lead to the following scheme of solving iTDSE with 
 the instanton - like boundary conditions, which was used in the present work:
 \begin{enumerate}
 	\item Calculate the monodromy matrix $\mathbf{M}$
 of (\ref{equat-noF}) by a step-by-step forward integration along short 
 intervals of $\tau$ in the range $\tau \in \langle -T/2,T/2\rangle$, with the 
 identity matrix as the initial condition; 
 	\item Perform the eigendecomposition of $\mathbf{M}$; 
 	\item Taking the consecutive eigenvectors as initial values and their 
 corresponding eigenvalues as Floquet exponents, integrate numerically 
 Eq. (\ref{num3}) (at the final point $\tau=T/2$, according to the periodic 
 boundary condition, one should recover the initial values). In this way one 
 obtains $\cal{N}$ linearly independent bounce solutions. 
 \end{enumerate}
 In this work, Eq. (\ref{equat}) and (\ref{equat-noF}) were treated as if the 
  matrix $\mathbf{A}(\tau)$ were piecewise constant on each integration 
 interval. One step of integration of Eq. (\ref{equat-noF}) 
 consists in calculating the exponential  
 of a constant matrix and its action on the vector of coefficients  
 of the previous step:
\begin{equation}
\mathbf{C}(\tau_{i+1})=\exp\left(\mathbf{A}\cdot (\tau_{i+1}-\tau_{i})\right)
 \mathbf{C}(\tau_{i})=\mathbf{G}(\tau_{i+1},\tau_{i})\mathbf{C}(\tau_{i}).
\label{Eq:intstep}
\end{equation}
The resolvent matrix is obtained by a successive multiplication of the one-step 
 exponentials.

 The chief difficulty in applying the above procedure comes from the exponential
 behaviour of solutions. We can write them in the form with the explicit 
 exponential factor (which is an analogue of the phase factor in real-time 
 quantum mechanics) as:
\begin{equation}
C_{\mu i}(\tau)=c_{\mu i}(\tau)e^{-\frac{1}{\hbar}\int_{-T/2}^{\tau}
 \epsilon_{\mu}(q(\tau'))d\tau'}.  
\label{eq:solwithexp}
\end{equation} 
 This dependence, combined with the presence of markedly different adiabatic 
 energies $\epsilon_{\mu}(q)$, leads to the exponentially divergent numerical 
 scales. During the evolution, the coefficient associated with the lowest state 
 will be amplified relative to all others. Therefore, a simple numerical 
 multiplication of successive one-step exponentials involves a mixing of 
 elements of different orders of magnitude, which results in the loss of 
 accuracy (due to a finite numerical precision). One 
 needs a way of separating different scales at each matrix multiplication. 
 In our work we adopt the singular value decomposition (SVD) approach, described
 in \cite{Koonin}. The procedure consists of the following steps:
\begin{enumerate}
	\item SVD decomposition of the propagation matrix in the 
 first step of integration: $\mathbf{G}(\tau_{1},-T/2)=\mathbf{U}_{1}
\mathbf{\Sigma}_{1}\mathbf{V}_{1}$, where $\mathbf{U}_{1}$ i $\mathbf{V}_{1}$ 
 are orthogonal matrices, and $\mathbf{\Sigma}_{1}$ is a diagonal matrix with 
 singular values, which contain information on magnitude 
 scales present in the problem.
	\item For the successive integration steps one performs the following 
 operations:
	\begin{enumerate}
		\item Calculation of the propagation matrix over a short 
 interval $(\tau_{i-1},\tau_{i})$: $\mathbf{G}(\tau_{i},\tau_{i-1})=
 \exp(\mathbf{A}\cdot (\tau_{i}-\tau_{i-1}))$, 
		\item Multiplication of matrices in order given by the brackets
 in the expression:  $\left[\mathbf{G}(\tau_{i},\tau_{i-1})
 \mathbf{U}_{i-1}\right]\mathbf{\Sigma}_{i-1}=\mathbf{S}_{i}$,
		\item Performing the SVD decomposition of the matrix 
 $\mathbf{S}_{i}$: 
 $\mathbf{S}_{i}=\mathbf{U}_{i}\mathbf{\Sigma}_{i}\widetilde{\mathbf{V}}_{i}$,
		\item Multiplication of the $\mathbf{V}$ matrices: 
 $\mathbf{S}_{i}\mathbf{V}_{i-1}=\mathbf{U}_{i}
 \mathbf{\Sigma}_{i}(\widetilde{\mathbf{V}}_{i}\mathbf{V}_{i-1})=\mathbf{U}_{i}
 \mathbf{\Sigma}_{i}\mathbf{V}_{i}$ -- this leads to the SVD form of the 
 propagation matrix $\mathbf{G}(\tau_{i},-T/2)$  with separated numerical 
 scales stored in the diagonal elements (singular values) of the matrix 
 $\mathbf{\Sigma}_{i}$.
	\end{enumerate}
	\item Performing steps ($i=2,\dots ,N$) described above along the 
 range of integration $(-T/2,0)$ one obtains the SVD form 
 of the propagation matrix: $\mathbf{G}(0,-T/2)=\mathbf{U}_{N} 
 \mathbf{\Sigma}_{N}\mathbf{V}_{N}$. 
\end{enumerate}

\noindent
 The monodromy matrix has the form: $\mathbf{M}= \mathbf{G}(T/2,-T/2)= 
 \mathbf{G}(T/2,0)\mathbf{G}(0,-T/2)$. Due to the property: 
 $\mathbf{A}(\tau)=\mathbf{A}^{\dag}(-\tau)$, fulfilled by the 
 matrix of Eq. (\ref{equat-noF}),  $\mathbf{G}(T/2,0)=
\mathbf{G}^{\dag}(0,-T/2)$, and: $\mathbf{M}=\mathbf{G}^{\dagger}(0,-T/2)
\mathbf{G}(0,-T/2)$. Thus, the monodromy matrix is hermitean and 
 positive-definite: $\mathbf{M}=\mathbf{V}^{\dagger}_{N}
 \mathbf{\Sigma}^{\dagger}_N \mathbf{\Sigma}_N \mathbf{V}_N$, and   
   the products: $\sigma^*_i \sigma_i$, with $\sigma_i$ the $i$-th singular 
 value of $\mathbf{\Sigma}_N$, are equal to the eigenvalues 
 $e^{-\zeta_{i} T/\hbar}$ of the monodromy matrix. 
 It is thus sufficient to integrate Eq. (\ref{equat-noF}) over a half of  
 period, i.e. in the range $\left(-T/2,0\right)$, to obtain the monodromy 
 matrix; we make use of this property in our calculations.

 Another issue that requires some attention is the instability 
 of instanton - like
 solutions with $\zeta_j>\zeta_1$ (where $\zeta_1$ -- the lowest $\zeta$). 
 From Eq. (\ref{eq:itdhf}) and its counterpart for $\phi_i^*(-\tau)$ one 
 obtains:
\begin{equation}
\langle\phi_{i}(-\tau)|\phi_{j}(\tau)\rangle = \langle\phi_{i}(-\tau_0)|\phi_{j}(\tau_0)\rangle e^{\frac{1}{\hbar}(\zeta_j-\zeta_i)(\tau-\tau_0)}.
\label{eq:ovlevol} 
\end{equation}
This means that if at some $\tau_0$ the overlap 
$\langle\phi_{i}(-\tau_0)|\phi_{j}(\tau_0)\rangle\ne 0$ (which is inevitable 
 due to a limited numerical precision), the evolution causes its exponential 
rise and spoils $\phi_j$ solution by increasing admixtures of $\phi_i$ with 
 lower $\zeta_i$  to it. To eliminate this effect, the orthogonalisation of 
 $\phi_j$ with respect to all solutions with $\zeta_i<\zeta_j$ 
 was performed after each integration step.   

  The accuracy of the applied method of solution was tested by comparing 
 the results with the ones of the algorithm with a finer imaginary time-step 
 (and thus more densly calculated adiabatic Woods-Saxon energies and wave 
 functions) and by running the code in quadrupole precision. 
 The other tests, of more physical significance, are described in 
 Appendix \ref{app:stab}.


 \section{Stability of solutions with respect to period and the size of 
  the adiabatic basis \label{app:stab} }

 The stability of iTDSE solutions, in particular their actions, with respect 
 to the assumed period $T$ and basis dimension ${\cal N}$ was checked on a 
 few examples. Here we give results obtained for the $\Omega^{\pi}=1/2^+$ 
 neutron levels in $^{272}$Mt, discussed in Section IV A.

 \subsection{Stability of action with respect to the period } 

\begin{table}
	\renewcommand{\arraystretch}{1.3}
	\setlength{\arrayrulewidth}{0.9pt}
	\begin{center}
		\begin{tabular}{c|cccccc}
			\hline \hline	      
			Nr & T=20 &	T=25 &	T=30 &	T=35 &	T=40 &	T=45 \\
			\hline \hline
			1 &	0.2893 &	0.2953 &	0.2970 &	0.2976 &	0.2978 &	0.2983 \\
			2	&   0.6306 &	0.6368 &	0.6399 &	0.6399 &	0.6401 &	0.6402 \\
			3 &	1.5633 &	1.5813 &	1.5854 &	1.5870 &	1.5874 &	1.5875 \\
			4 &	-0.0210 &	-0.0093 &	-0.0051 &	-0.0038 &	-0.0034 &	-0.0033 \\
			\hline
		\end{tabular}
	\end{center}
	\caption{Action values (in $\hbar$) calculated for the four lowest 
 iTDSE solutions for various assumed periods $T$ (in $10^{-21}$ s).}
	\label{tab:actzT}
\end{table}
\begin{table}
	\renewcommand{\arraystretch}{1.3}
	\setlength{\arrayrulewidth}{0.9pt}
	\begin{center}
		\begin{tabular}{c|cccccc|c|c}
			\hline\hline
			Nr &	T=20 &	T=25 &	T=30 &	T=35 &	T=40 &	T=45 & $\mathbf{\zeta_{T\rightarrow\infty}}	$ &	$\mathbf{\epsilon_{g.s}}$ \\
			\hline\hline
			1 &	-9.906 &	-9.750 &	-9.631 &	-9.544 &	-9.477 &	-9.424 &	\textbf{-9.044} & 	\textbf{-8.990} \\
			2 &	-8.514 &	-8.424 &	-8.363 &	-8.319 &	-8.287 &	-8.262 &	\textbf{-8.059} & 	\textbf{-8.061} \\
			3 &	-6.288 &	-6.148 &	-6.054 &	-5.988 &	-5.938 &	-5.900 &	\textbf{-5.588} & 	\textbf{-5.600} \\
			4 &	-4.930 &	-4.776 &	-4.660 &	-4.576 &	-4.511 &	-4.460 &	\textbf{-4.089} &	\textbf{-4.037} \\
			\hline   		
		\end{tabular}
	\end{center}
	\caption{Floquet exponents $\zeta_i$ [MeV] for the four lowest 
 instanton-like iTDSE solutions, for increasing values of the period 
 $T$ [$10^{-21}$ s], and the limiting value $\zeta_i(T\rightarrow\infty)$ 
 [MeV], estimated from the formula in the text, 
 vs s.p. energies $\epsilon_i$ [MeV] at the g.s. deformation.}
	\label{tab:floquetzT}   
\end{table}
 The values of actions $S_i$ and Floquet exponents $\zeta_{i}$ of solutions 
 $\phi_i$ change with increasing period $T$. As the instanton-like 
 solution would correspond to $T=\infty$, it is of relevance that $S_i$ 
 and $\zeta_{i}$ should stabilize above some $T$.  
 It is indeed the case: actions $S_i$, shown in Tab. \ref{tab:actzT}, 
 change not more than $\sim$ 3\% except 
 the very small ones, whose contribution is negligible anyway. 
 The convergence of the Floquet exponents to the eigenenergies at 
 the initial (and final) state can be well approximated by the formula:  
 $\zeta_i(T)=A_i+B_i/T$ with constant $A_i$ and $B_i$, and in calculations  
 the relation $\zeta_i(\infty)=A_i\approx \epsilon_i$, although not axact,  
 is approximated reasonably well - see Tab. \ref{tab:floquetzT}.

\subsection{Stability of action with respect to the dimension $\cal{N}$ of 
 the adiabatic basis }

\label{sec:SvsN}
\begin{table}[b]
	\renewcommand{\arraystretch}{1.3}
	\setlength{\arrayrulewidth}{0.9pt}
	\begin{center}
		\begin{tabular}{c|c}
			\hline\hline
			${\cal N}$ & $S_{tot}=\sum_{i=1}^{{\cal N}/2}S_{i} $ $[\hbar]$ \\
			\hline\hline
			8 & 2.5172  \\
			10 & 2.5388   \\
			12 & 2.5657  \\
			14 & 2.5779   \\
			\hline
		\end{tabular}
	\end{center}
	\caption{Total action $S_{tot}$ for the lower half of the iTDSE 
 solutions (i.e. occupied instanton-like states) as a function of the number $\cal{N}$ of adiabatic basis states 
 included in calculations.}
	\label{tab:stotvsn} 
\end{table}
 We also tested the change of the total action $S_{tot}$ (\ref{eq:stot}) 
 with increasing number of adiabatic basis states $\cal{N}$ included 
 {\it symetrically below and above the Fermi level}. 
 Intuition would suggest that the main contribution to action should come 
 from states lying close to the Fermi level. 
 For trajectory depicted in Fig. \ref{fig:mapaadiabmt}, action values for 
 increasing ${\cal N}$ are presented in Tab. \ref{tab:stotvsn}. One can see  
 that for larger ${\cal N}$ changes in action become negligible.

\vspace{3mm}
\begin{figure}[!h]
	\centering
	\includegraphics[angle=-90, width=0.59\textwidth]{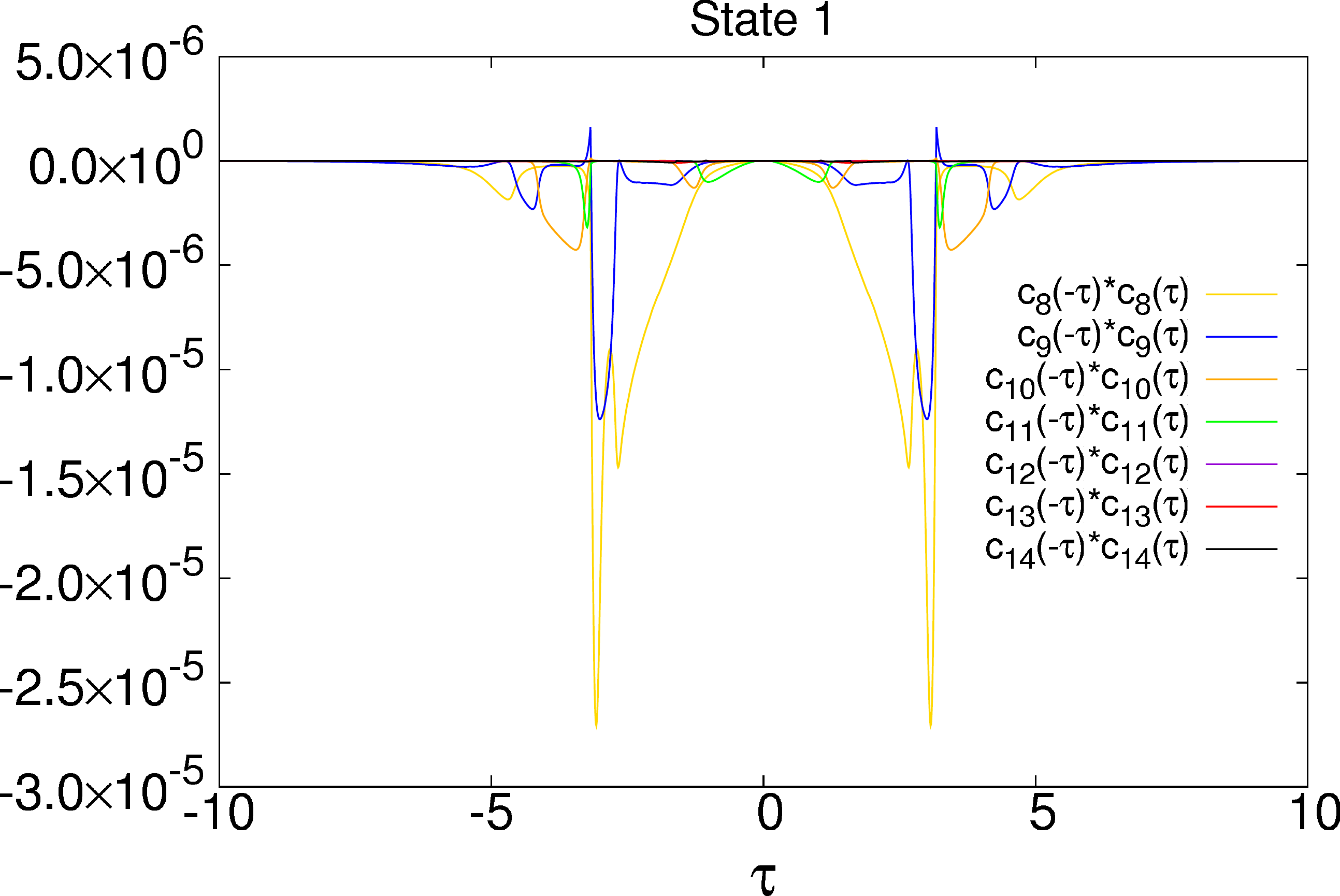} 
	\includegraphics[angle=0, width=0.59\textwidth]{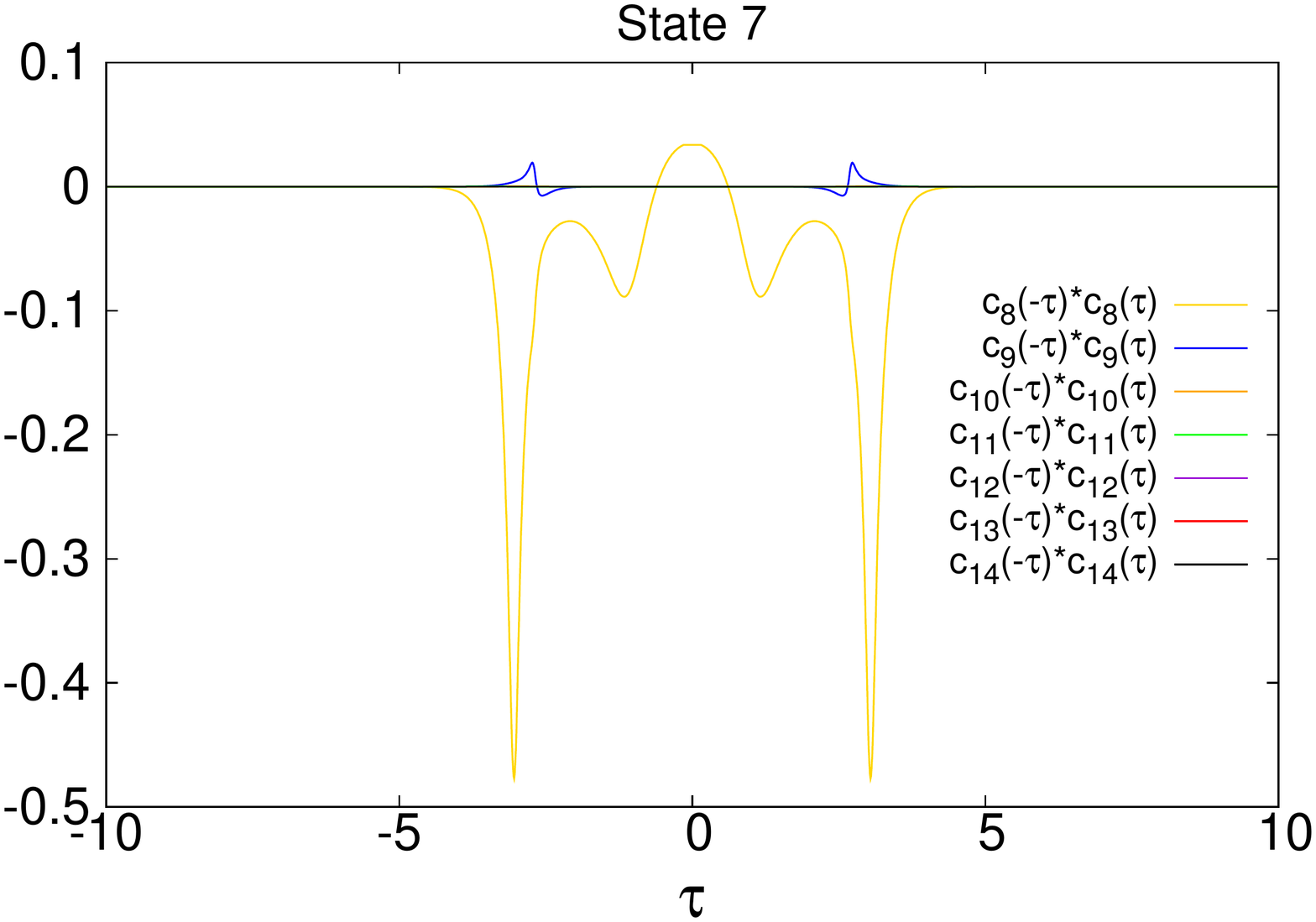}
	\caption{Quasi-occupations of seven upper adiabatic states for the 
 lowest (\textit{upper panel}) and the seventh (i.e. last occupied; \textit{lower panel})  
 instanton solution for ${\cal N}=14$. Note $\sim$ 4 orders of magnitude 
 difference in vertical scales in both panels. } 
	\label{fig:1-7pseudoobs}  
\end{figure}
\vspace{-2mm}
 
 For the case of ${\cal N}=14$ basis states, in the upper panel of 
 Fig. \ref{fig:1-7pseudoobs}, we show quasi-occupations of 
 adiabatic states above the Fermi energy,  $\epsilon > \epsilon_{F}$, 
 in the lowest 
 iTDSE solution $\phi_1$. It can be seen that excitations to adiabatic 
 states above the Fermi level are marginal and nearly do not contribute 
 to action. In the lower panel of Fig. \ref{fig:1-7pseudoobs}, are shown 
 quasi-occupations of 
 the same adiabatic states in the highest occupied instanton-like state  
 $\phi_7$. It can be seen that transitions occur mainly to the adiabatic 
 states closest in energy. These results indicate that adiabatic states 
 in a wide enough energetic window around the Fermi level suffice to calculate 
 instanton - like action.


\section{Treating sharp pseudocrossings along nonaxial fission paths
       \label{app:nonax}}

\begin{figure}[!b]
	\centering
	\includegraphics[width=0.49\linewidth]{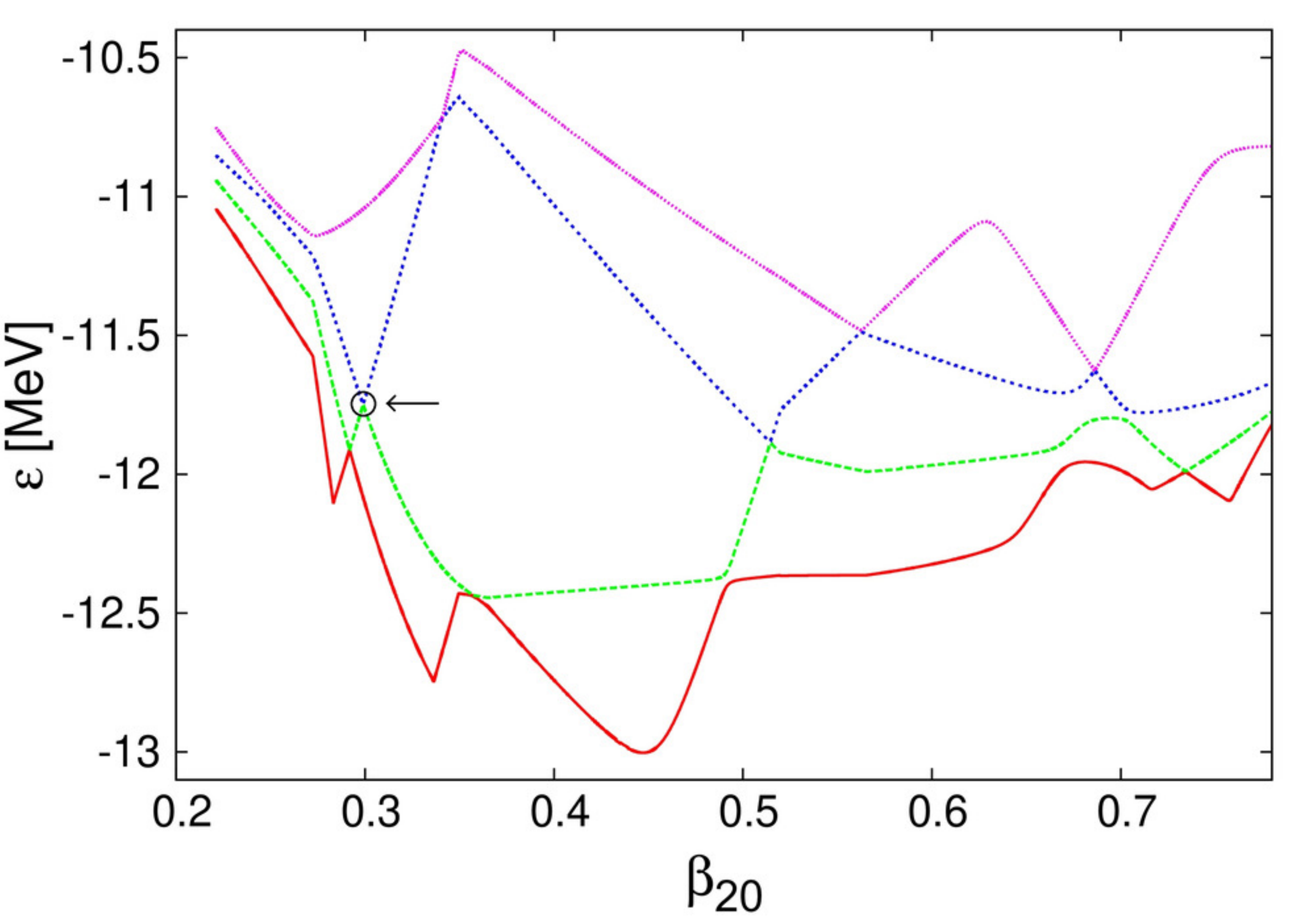}
	\includegraphics[width=0.49\linewidth]{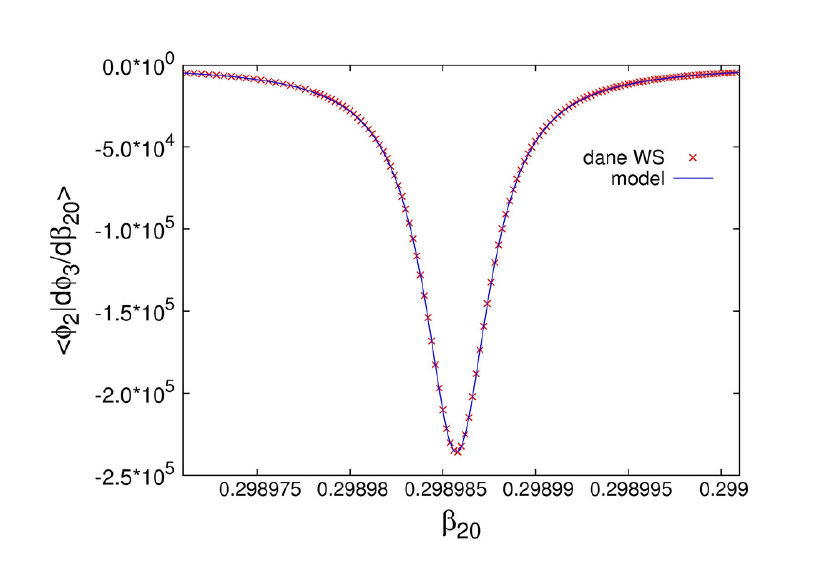}
	\caption{\textit{Left panel:} 
 s.p. energies in the Woods-Saxon potential along the chosen nonaxial path 
 (parametrized by $\beta_{20}$, Fig. \ref{fig:mt-nonax}); a sharp 
 pseudocrossing is marked by a circle. \textit{Right panel:} adiabatic coupling 
 between the two levels in the vicinity of crossing and its fit by a model 
 curve (\ref{eq:2levcoupl}). }
	\label{fig:enbet-testfit2}
\end{figure}

 Sharp pseudocrossings in the s.p. spectrum for nonaxial shapes generate very 
 narrow (in $q$) and large peaks 
 in the matrix elements of the adiabatic coupling; an example is shown in Fig. 
  \ref{fig:enbet-testfit2}. Those present an obvious impediment to an 
 effective solution of iTDSE. 

 A rapid change of adiabatic states with $q$ at sharp pseudocrossings 
 suggests the unsuitability of the adiabatic basis.  
  In chemistry, there were many trials in such situations to find a suitable 
 quasi-diabatic basis with smaller and regular coupling between crossing 
 states  \cite{Baer,SunLin,Pacher}. 
 The diabatic basis, like $\{|\chi_{i}\rangle\}$ in the two-level model 
 (sect. \ref{sec:2levmod}), might seem a good candidate. It is related 
 to the adiabatic basis via  $\theta$  angle, being a function of 
 $\alpha=V/E$ and $q-q_{0}$, where $q_0$ is the crossing point. 
 One can locally fit these parameters to each crossing and define a 
 new basis by means of the angle $\theta$, while leaving all not crossing 
 levels unchanged. This is an approximation, so the resulting basis is not 
 strictly diabatic (with $\langle\chi_{i}|\partial_{q}\chi_{j}\rangle = 0$),
 but quasi-diabatic ($\langle\chi_{i}|\partial_{q}\chi_{j}\rangle << 
 \langle\phi_{i}|\partial_{q}\phi_{j}\rangle$). One can show that in 
 the general case of many levels and many deformations $q_{i}$ 
 a strictly diabatic basis does not exist \cite{Mead}. 

 The calculations have shown that the quasi-diabatic basis found by this 
 procedure does not bring any advantage in comparison with the adiabatic one: 
 the density of points necessary to probe the neighbourhood of a crossing 
 in order to ensure an approximately correct action value is the same for 
 both bases (very dense mesh is needed in both cases). 
  
 An alternative solution would be solving instanton equations using a large 
 basis, smoothly changing with deformation (like that of the harmonic 
 oscillator), 
 without resorting to the adiabatic basis. Then the problem of sharp 
 crossings would be avoided, however, not without a cost: large basis would be 
 needed that probably would lead to the necessity of using quadruple 
 precision and more time-consuming calculations.  

 We kept the adiabatic basis. In order to integrate Eq. (\ref{equat})  
 we used a changing step in $\beta_{20}$ for calculating input data, i.e. 
 energies and adiabatic couplings along the path.   
  The step $\Delta \beta_{20}$ was diminished when a change in any of the 
 couplings was above 10\% of its preceding value. It was necessary to impose 
 the minimal step value, $\Delta\beta_{20}=10^{-7}$ (with $\beta_{20}$ 
 as the parameter of the path). Such 
 a probing was dense enough for a nearly exact integration for most of the 
 peaks. However there were a 
 few narrow and high peaks which were still not well rendered. In those 
 cases, the shape of such peak was modelled by the formula (\ref{eq:2levcoupl}) 
 (with parameters $\alpha$ and $q_{0}$) using the least squares fit to the 
 calculated points. Next, for each such modelled crossing, a $2\times 2$ 
 transition matrix $\mathbf{G}(\tau_{fin}, \tau_{ini})$ for the two crossing 
 levels was integrated [defined by the Eq. (\ref{Eq:intstep})], 
 where $\tau_{ini}, \tau_{fin}$ means the beginning and end of the peak. 
 The integration of a model peak is simple due to its analytic formula which 
 makes many Woods-Saxon calculations unnecessary.  
  Then the propagation matrix $\mathbf{\widetilde{G}}(\tau_{fin} , \tau_{ini})$
 for all ${\cal N}$ levels is calculated as follows: propagation of the 
 ${\cal N}-2$ not crossing levels is done in a standard way while 
 for two crossing levels one substitutes the matrix $\mathbf{G}$ calculated 
 for the fitted model. Denoting the index of the lower crossing level $i$,  
 one can schematically write the matrix $\mathbf{\widetilde{G}}$: 
\begin{equation}
\bordermatrix{ & 1 & 2 & \ldots &  i &  i+1 & \ldots & {\cal N} \cr
	1 & \widetilde{G}_{11} & \widetilde{G}_{12} & \ldots & 0 &  0 & \ldots & \widetilde{G}_{1{\cal N}} \cr 
	2 & \widetilde{G}_{21} & \widetilde{G}_{22} & \ldots &  0 & \ 0 & \ldots & \widetilde{G}_{2{\cal N}} \cr 
	\vdots & \vdots & \vdots & \ddots &  \vdots & \vdots & \ddots & \vdots \cr 
	i &  0 & 0 & \ldots &   G_{i\ i} &  G_{i\ i+1} & \ldots & 0 \cr 
	i+1 &  0 & 0 & \ldots & G_{i+1\ i} &  G_{i+1\ i+1} & \ldots &  0 \cr
	\vdots & \vdots & \vdots & \ddots & \vdots & \vdots & \ddots & \vdots \cr 
{\cal N} & \widetilde{G}_{{\cal N}1} & \widetilde{G}_{{\cal N}2} & \ldots & 0 & 0 & \ldots & \widetilde{G}_{{\cal N}{\cal N}} \cr
}
\label{eq:mprop}
\end{equation}

\noindent Thus, we neglect the cross terms, setting: 
 $\mathbf{\widetilde{G}}_{\alpha l}=\mathbf{\widetilde{G}}_{l \alpha}=0$, 
 where $\alpha\neq i,\,i+1$ and $l=i,\,i+1$. It means we treat the crossing 
 of two levels as isolated: the evolution of $c_{i}, c_{i+1}$ is dominated 
 by the coupling between them, 
 $\langle\phi_{i}|\partial_{q}\phi_{i+1}\rangle$, while the effect of other 
 states $c_{\alpha\neq i,i+1}$ on crossing levels and the effect of the pair on those other 
 states can be neglected in the vicinity of crossing. 
  
 This procedure was tested in few cases in which the vicinity of the crossing 
 could be probed dense enough for the solution without any fit 
 to be exact. Then the solutions for smaller density of calculated points but 
 with the modelled adiabatic coupling in the vicinity of crossing 
 was compared to the exact one. It turned out that for the desired accuracy 
 the model for the coupling should include independent parameters for the 
  height and half-width: 
\begin{equation}
\left\langle\phi_{1}\bigg|\frac{d\phi_{2}}{dq}\right\rangle=
\frac{1}{2}\frac{\alpha}{(q-q_{0})^{2}+\sigma^{2}}.
\label{eq:uogcouplfit}  
\end{equation}
 With this model, the calculated actions differed less than 1\% from the 
 reference results, except for very small actions, for which the difference 
 was of no consequence anyway.


\begin{thebibliography}{99}
\bibitem{foot2}
 One can show this e.g. by using the expression for action in terms of the 
 operator ${\hat {\cal S}}$ of (\ref{Soper}), given by the formula (44) 
 in \cite{JS}, and expanding instantons $\phi_i(\tau)$ onto eigenvectors 
 of ${\hat {\cal S}}(\tau)$.
\bibitem{foot3} 
 We omit here a small difference in W-S potentials for nuclei $A$ and $A-1$.
\end{thebibliography}
\end{document}